\documentclass[prc,nofootinbib]{revtex4}
 \usepackage{epsfig}
 \usepackage{graphicx}
 \usepackage{dcolumn}
 \usepackage{amsfonts}
 \usepackage{amsmath}
 \usepackage{amssymb}
 \usepackage{color}
 \usepackage{mathrsfs}
 \newcommand{\beq}{\begin{equation}}
 \newcommand{\eeq}{\end{equation}}
 \newcommand{\ba}{\begin{array}}
 \newcommand{\ea}{\end{array}}
 \newcommand{\beqa}{\begin{eqnarray}}
 \newcommand{\eeqa}{\end{eqnarray}}
 \newcommand{\bal}{\begin{align}}

 \def\Rb{\mathbb{R}}
 
 \DeclareRobustCommand{\scrI}{\mathscr{I}}
  \DeclareRobustCommand{\scrD}{\mathscr{D}}

\newcommand{\blue}[1]{{ \color{blue} #1 }}

\newcommand{\mc}[1]{\mathcal{#1}}

\DeclareRobustCommand{\scrD}{\mathscr{D}}
\DeclareRobustCommand{\scrI}{\mathscr{I}}

\usepackage{docs}%
\usepackage{bm}%
\expandafter\ifx\csname package@font\endcsname\relax\else
 \expandafter\expandafter
 \expandafter\usepackage
 \expandafter\expandafter
 \expandafter{\csname package@font\endcsname}%
\fi

\begin{document}

\title{The evolving many-nucleon theory of nuclear rotations}

 \author{David J.~Rowe$^1$}
 \affiliation{$^1$Department of Physics, University of Toronto, Toronto, ON M5S 1A7, Canada}

\date{Sept., 2017}

\begin{abstract}  
The many  approaches that have been pursued in seeking an understanding of nuclear rotational dynamics are reviewed and reassessed with a view to their development in the light of  recent progress  and the research tools that are now available.
A  motivation for this review is the widespread observation
of nuclear shape coexistence and  sequences of rotational states in all regions of the nuclear periodic table combined with the recognition that 
the study of the rotational dynamics of quantum fluids  
has led to significant advances in the quantum theory of many-boson systems.
Recent experimental investigations of the rotational  dynamics of a low-temperature $^6$Li gas  indicate that its slow rotational flows  are likewise the irrotational flows of a superfluid.  
In this context, the dynamics of rotating nuclei are  of fundamental interest because the nucleus is a unique  zero-temperature finite many-fermion quantum system. 
A  promising approach is provided by algebraic mean-field theory which, as its name suggests, is a combination of algebraic and mean-field methods.
Static mean-field theories play a central role in many-body theory  by defining optimal independent-particle and independent quasi-particle   basis states for the quantum mechanics of  many-fermion systems.
Their  time-dependent extensions also lead,  in the small-amplitude random-phase approximation, to the quantisation of the classical normal-mode vibrations of many-fermion systems about their static equilibrium states.
This review shows that mean-field methods become significantly more powerful when combined with algebraic methods and an appropriate coupling scheme for the nuclear shell model.

\end{abstract}
 \maketitle

 \tableofcontents

\section{Introduction}
Following the discovery of the nucleus \cite{GeigerM09,Rutherford11}
and the development of the Bohr-Sommerfeld model 
\cite{Bohr1913,Sommerfeld1916} of the atom, the modern era of nuclear physics was initiated by the observations of patterns in the properties of nuclei which indicated the emergence of simple structures where chaotic behaviour had been expected. 
Nobel prizes appropriately rewarded the early model interpretations of these properties, one of which  
\cite{Mayer49,HaxelJS49,MayerJ55}
led to the nuclear shell model 
\cite{Flowers52,Kurath56,Elliott58ab,FrenchHMW69}
and another \cite{Rainwater50,Bohr52,BohrM52} 
to the collective model and the Bohr-Mottelson-Nilsson unified model
\cite{Bohr54,BohrM53,Nilsson55,MottelsonN59}.
Reviews of the many developments that followed can be found in three recent edited volumes 
\cite{Dudek16,WoodH16,Launey17}.

A comparison of  the dynamics of a rotating nucleus  with that of a macroscopic quantum superfluid proves to be instructive.
For a quantum fluid, such as liquid helium below a critical temperature, it can happen that, 
after reaching an equilibrium steady state in a slowly rotating non-symmetrical container, 
the circulation of the fluid's current flow vanishes and the flow becomes irrotational.  
Recent experimental investigations \cite{ClancyLT07} also find that the slow rotational dynamics 
of a  low-temperature $^6$Li gas  are likewise the irrotational flows of a superfuid.  
At higher angular velocities,
it can become energetically favourable for a rotating quantum fluid  to lower its kinetic energy at the expense of increasing its potential energy  by the creation of quantised vortices.
At temperatures above the critical temperature,  the viscosity of the fluid  ceases to be zero and the equilibrium rotations  become those of a normal fluid with rigid-body moments of inertia and perturbations due to the Coriolis and centrifugal forces.

Experimental observations show that the moments of inertia required to describe the observed rotational states of well-deformed nuclei are typically 4-5 times  those for irrotational flow and approximately half those of a rigid body  \cite{NilssonP61}.  
Thus, given that  nuclei are unique  zero-temperature finite many-fermion quantum systems, it is  of fundamental interest to understand the nature of their rotational dynamics. 
The objective of this review is to identify realistic but practical methods for describing the dynamics of nuclear rotations  in deformed nuclei and the emergent phenomenon of  shape coexistence \cite{HeydeIWWM83,WoodHNHvD92,HeydeW11} 
in terms of many-nucleon quantum mechanics. 
It starts with an assessment of what has been achieved and proceeds by considering promising strategies  for further progress in  light of the powerful resources that are now available.

It is useful to begin by considering the  differences between translations and rotations of a nucleus. A fundamental difference is that, whereas the Hamiltonian of an isolated nucleus is invariant under both translations and rotations, there is no rotational analog of Galilean invariance; 
 rotational motions are coupled to other degrees of freedom by inertial Coriolis and centrifugal forces.
One might nevertheless expect that  perturbations by the inertial  forces of the intrinsic structure  of a stable  and well-deformed rotational nucleus could be negligible in a state of small angular momentum.
This would be the case for an adiabatically rotating nucleus in classical mechanics.
However, in quantum mechanics, the angular momentum is quantised and the smallest non-zero value for an even-mass nucleus might already be too large for a rotational state of a well-deformed nucleus  to be considered adiabatic.

A possible interpretation of nuclear rotations is that the low-energy states of a rotational band 
already contain quantised vortices which increase in number continuously with increasing angular momentum with the result that the kinetic energy of the rotations decreases and the potential energy, associated with the creation of vortices, increases such that the two components of the energy combine to give the near  $L(L+1)$ energy-levels observed in rotational nuclei.

A first interpretation of nuclear rotations, in terms of many-nucleon quantum mechanics, was given by Elliott's SU(3) model \cite{Elliott58ab} which,
with an effective charge and an effective interaction,   has many of the observable properties of a rotor model.
This  is  now understood from the observation that the SU(3) Lie algebra is the projected image of a rigid-rotor algebra onto a shell-model subspace 
and, as a result, retains many of the properties of a genuine rotor model as an effective shell-model of nuclear rotations, 
as discussed in  Section \ref{sect:AlgModel}.
Thus, for example, whereas the rotational states of a rigid rotor model \cite{Ui70} all have  identical potential energies, which implies that its rotational energies are purely kinetic, the states of an SU(3) model are exactly the opposite; they have identical kinetic energies and their energies are purely potential.
A realistic theory of nuclear rotations evidently lies between the rigid rotor and SU(3) limits.

In pursuing practical approaches to many-nucleon quantum mechanics,
it is profitable to begin with an expression of the Hilbert space of a nucleus as an ordered sum of subspaces, each of which is the  space for an irrep (irreducible representation) of some dynamical  group of unitary transformations.
Such a decomposition and classification of many-nucleon states is said to define a coupling scheme.  
Expressing the infinite-dimensional many-nucleon Hilbert space of a nucleus as an ordered sum of subspaces  is necessary, in practical calculations, in order that it may be restricted to a finite-dimensional subspace  in a meaningful way.
An ideal ordering is such that calculations of low-energy eigenstates of a given Hamiltonian can be  implemented  in a sequence of spaces of increasing dimensions until convergence of the results is achieved to some acceptable level of accuracy.
However, it is important to be aware, 
as illustrated dramatically in Sect.\ \ref{sect:BahriRmodel},
that states of much lower energy could still occur if  a differently ordered basis were chosen.
This characteristic of an infinite-dimensional Hilbert space undoubtely underlies the phenomenon of shape coexistence \cite{WoodH16} in which states  show up in the low-energy spectra of nuclei that, in a standard shell-model coupling scheme, are expected to lie high in energy.
Thus, it is appropriate to choose a coupling scheme  for a microscopic  calculation that is based on the unitary irreps of the Lie algebra of a dynamical model which to some level of approximation describes the properties of the low-energy states of interest.
A given nuclear Hilbert space can have many coupling schemes
\cite{Racah43,Racah49,MayerJ55,Jahn50I,%
Jahn51II,Flowers52,Kurath56,Elliott58ab,deShalitT63,FlowersS64b,%
FrenchHMW69}.
In accord with the above philosophy,  we  focus here on  schemes associated with algebraic models aimed at understanding the emergence of  collective properties in nuclei.

This review  starts with outlines of the several approaches 
to  questions  concerning the deformations of nuclei and the dynamics of their collective motions.   
These questions have been addressed primarily in two  ways: one based on mean-field theory and the other on algebraic methods.
Mean-field approaches are natural because the  highly successful phenomenological  unified model  is  based on the correlation of the many-nucleon dynamics within the common mean field in which the nucleons move.
On the other hand, an algebraic approach is appropriate because the
quantum mechanics of any system is fundamentally an algebraic model 
\cite{Dirac26,Weyl50}.
Moreover, the standard model of a nucleus, 
given by  non-relativistic many-nucleon quantum mechanics, 
is a particularly simple algebraic model with a uniquely defined Hilbert space, 
namely that of a totally antisymmetric unitary irrep of the Lie group of all one-body unitary transformations. 
This is  significant because of the wealth of elegant and highly developed mathematical tools available for the study of such systems.
It implies, for example, that the Hilbert space of an $A$-fermion nucleus is spanned by Slater determinants made up of $A$ single-fermion wave functions.
As a result,   algebraic and symmetry-based methods provide relationships between the many different models and the possibility of exploiting their various insights.
A primary objective  of this review is to expose the advantages  of combining algebraic and mean-field methods.
Also  discussed, will be  the relationships this provides between the classical and quantum mechanical treatments of a dynamical system.

It must be acknowledged at the outset that the many historical developments  mentioned in this review  are  a  subset of  those that merit recognition.
Those referenced are  representative of the many that have been influential in the evolution of the models and microscopic theory of nuclear collective structure outlined in this review. 
For example,  important developments that are not discussed in this review are the  significant density-functional methods \cite{DrutFP10,NazarewiczRSV13,Dobaczewski16} which  are  aimed primarily at  obtaining reliable characterisations of nuclear properties in terms of their sizes, density distributions, binding energies, etc., for the many practical purposes  in which such information is needed.
Density-functional theory clearly has a vested interest in the foundations of mean field theory
\cite{Rosensteel17} and has some overlaps with the considerations of this review.
Also omitted are the IBM (Interacting Boson Model) \cite{IachelloA87} methods  which effectively illustrate the power of algebraic methods and to which some of what is discussed in this review  applies.

A particularly significant omission in this review is the important extension of the
mean-field theories based on independent-particle approximations 
to the independent quasi-particle theories, based on the BCS theory \cite{BardeenCS57} of superconductivity, which were brought into nuclear physics by Bohr, Mottelson and Pines \cite{BohrMP58}, Belyaev \cite{Belyaev59}, Baranger \cite{Baranger60}, and others.
Algebraic mean-field (AMF) methods can no doubt be extended to include the  symmetry-breaking correlations brought about by pairing interactions.
However, an alternative and promising approach, within the framework of the proposed 
AMF theory, is suggested which exploits the so-called quasi-dynamical symmetries that are observed in situations in which the dominant dynamics are adiabatic
\cite{RoweRR88,Rowe04,Rowe04b,RoweCancun04,RosensteelR05} and which avoids 
violating nucleon-number conservation.

More complete reviews of most of the separate topics that feature in this paper can be found in the literature; see, for example, 
Refs.\ \cite{Rowe85,Rowe96,WoodH16,Dudek16,Launey17} and some standard texts \cite{Rowebook70,BohrM75,RingS80,Talmi93,RoweWood10}.

 \section{The cranking model} \label{sect.Inglis}
 
 An early model of nuclear rotational dynamics was the  Inglis cranking model \cite{inglis54,Inglis55} for deriving moments of inertia.
The basic cranking model considers a wave function for a deformed nucleus in a body-fixed  frame that  is rotating about a fixed axis with a small angular velocity $\omega$, 
and  seeks to determine the rotational component of its energy which it equates with 
$\frac12 \scrI \omega^2$, 
where $\scrI$ is the moment of inertia for rotation about the chosen axis.  
  This model has  been remarkably influential and has contributed to subsequent developments in mean-field theory, e.g., the Thouless-Valatin model (see Sect.\ \ref{sect:MFrotor}).

If $\hat H_0$ is an independent-particle model Hamiltonian for the intrinsic states of a deformed rotational  nucleus, as given, for example, by the Nilsson model \cite{Nilsson55}, then the wave function for the nucleus, when rotating with some small angular velocity $\omega$, is assumed in the cranking model to be a solution of the time-dependent Schr\"odinger equation
 \beq
 \hat H_0 |\phi_\omega (t)\rangle    
 = {\rm i} \hbar \frac{\partial}{\partial t}  |\phi_\omega (t)\rangle. 
 \eeq
However, if  $|\phi_\omega (t)\rangle$ is transformed to the rotating frame of reference
 \beq 
|\phi_\omega (t)\rangle \to |\tilde\varphi_\omega (t) \rangle  
= e^{ -{\rm i}\omega t \hat J_3}     |\phi_\omega(t) \rangle ,
\eeq
where $\hat J_3$ is the 3-component of the angular-momentum operator,
it becomes a stationary state  of the form
\beq 
|\tilde\varphi_\omega (t)\rangle 
=  e^{-\frac{{\rm i}}{\hbar} E(\omega)t}  |\tilde\varphi_\omega \rangle
\eeq
 It  follows that the state $|\tilde\varphi_\omega \rangle$ is a solution of the time-independent Schr\"odinger equation
\beq
(\hat H_0 -\hbar\omega \hat J_3) |\tilde\varphi_\omega \rangle
= E(\omega) |\tilde\varphi_\omega \rangle 
\eeq
and is given, for an arbitrarily small value of $\omega$, 
to first-order in perturbation theory by
\beq
|\tilde \varphi_\omega\rangle = |0\rangle +
\hbar\omega \sum_\nu |\nu\rangle \frac{\langle \nu | \hat J_3 |0\rangle }{E_\nu-E_0},
\eeq
where $|0\rangle = |\tilde \varphi_0\rangle$ is the ground state  of $\hat H_0$ and $ |\nu\rangle$ is an excited state of excitation energy $E_\nu-E_0$.
The corresponding moment of inertia of the cranking model 
\beq \label{eq:CMformula}
\scrI_3 = 2\hbar^2 \sum_\nu 
\frac{ \big| \langle \nu |\hat J_3 |0\rangle \big|^2}{E_\nu-E_0},
\eeq
is then obtained by equating the energy increase with the  rotational energy                in the equation
\beq
\langle \tilde \phi_\omega | \hat H_0 | \tilde\phi_\omega\rangle 
=E_0 + \tfrac12 \scrI_3 \omega^2 .
\eeq

In a simple application to an even-even nucleus \cite{BohrM55},  $\hat H_0$ was taken to be the triaxial harmonic-oscillator Hamiltonian
\beq
\hat H_0 = \tfrac12 \sum_{ni} 
\hbar\omega_i  \big(b^\dag_{ni} b_{ni} + b_{ni} b^\dag_{ni}\big) ,
\eeq
in which $n$ is summed over the nucleon number, $i$ is summed over the three coordinate axes, and $b^\dag_{ni}$ and $ b_{ni}$ are harmonic oscillator raising and lowering operators that obey the boson commutation relations
\beq [b_{ni}, b^\dag_{mj}] = \delta_{m,n} \delta_{i,j} . \eeq
The spins of the nucleons were considered to be coupled to zero, so that $\hat  J_3$ could be replaced by the orbital angular momentum 
\beq
\hbar\hat L_3 =\sum_n (\hat x_{n1} \hat p_{n2} - \hat x_{n2} \hat p_{n1}),
\eeq
and expressed in terms of the raising and lowering operators by the standard relationships
\beq \label{eq:xp.coords}
 \hat x_{ni} = \frac{1}{\sqrt{2} a_i} \big(b^\dag_{ni} + b_{ni}\big), \quad \hat p_{ni} = {\rm i}\hbar \frac{a_i}{\sqrt{2}} \big(b^\dag_{ni} - b_{ni}\big) ,
 \eeq
with $a_i = \sqrt{M\omega_i / \hbar}$\,.
The cranking model then gives the moment of inertia
\bal \label{eq:CM.MofI}
 \scrI_3 = \frac{\hbar}{2\omega_1\omega_2}
  \Big[ \frac{(\omega_2-\omega_1)^2}{\omega_2+\omega_1} (\sigma_1+\sigma_2) 
  +\frac{(\omega_2+\omega_1)^2}{\omega_2-\omega_1} (\sigma_1-\sigma_2)  \Big] ,
\end{align}
with
\beq  \sigma_i =\tfrac12 \sum_n  
 \langle 0| b^\dag_{ni} b_{ni} +b_{ni} b_{ni}^\dag |0\rangle .
\eeq
This result was also derived without the use of perturbation theory by Valatin \cite{Valatin56}. 
It is remarkable, as noted by Bohr and Mottelson \cite{BohrM55, BohrM75}, because
if $\sigma_1= \sigma_2=\sigma_3$ but the values of $\omega_i$ are not necessarily equal,
the cranking-model moments of inertia are precisely those of an irrotational-flow model.  And, if $\sigma_1$, $\sigma_2 $, and $\sigma_3$ are not all equal,
but the harmonic oscillator frequencies
satisfy the shape-consistency condition (cf.\ Section  \ref{sect:CMspaces})
\beq \label{eq:shape.consistency.a}
\sigma_1 \omega_1=\sigma_2 \omega_2=\sigma_3 \omega_3,
\eeq
then the moments of inertia are those of a rigid-body.

Two other results are worth noting.
The first  is when $\sigma_1 > \sigma_2$ and $\sigma_2=\sigma_3$ or
$\sigma_1 = \sigma_2$ and $\sigma_2>\sigma_3$ and the shape-consistency condition 
(\ref{eq:shape.consistency.a}) is satisfied.
The nucleus then has an axis of symmetry and its wave function
 is invariant under rotations about this symmetry axis.
It  then has no collective degree of freedom corresponding to rotations about this symmetry axis, 
and its  cranking model moment of inertia  for rotations about  axes orthogonal to its symmetry axis is  that of a rigid rotor.
The second is that the cranking-model moments of inertia are generally consistent with the expectations of the 
Bohr-Mottelson unified model \cite{BohrM53}.
As a quantum liquid-drop model, the moments of inertia of  the Bohr model \cite{Bohr52} in deformed rotational states have the irrotational-flow values of a quantum fluid.
However, in the  unified model, the closed-shell-core component of a nucleus,  when polarised to a non-spherical shape by extra-core nucleons, 
was expected to contribute an irrotational flow component to the moments of inertia 
while the extra-core nucleons were expected to make a much larger  contribution to the total moments of inertia; in the Inglis  model \cite{BohrM55}, the extra-core nucleons bring the combined moment of inertia up to that of a rigid body.

In fact,  the  rigid-body moments of inertia of the simple cranking model are much larger than those observed \cite{NilssonP61} which lie midway between  irrotational and rigid-body values.
It is also known that  closer agreement with observed moments of inertia can be obtained if more realistic rotational wave functions are used in the cranking model.
Thus, Belyaev \cite{Belyaev59} obtained an expression which gave much closer agreement with experiment  by including the nucleon spins and replacing the independent-particle intrinsic state of the Inglis model with an independent quasi-particle state.
He thereby included the effects of pairing correlations and, as expected from  the BCS theory of superconductivity \cite{BardeenCS57}, obtained cranking model moments of inertia closer to those of a superfluid.
Other extensions have also been made to the cranking model; cf., for example, Refs.\ 
\cite{Skyrme57,Migdal59,Levinson63,KelsonL64,WongTT68, DudekDRS80}.
Particularly influential were the self-consistent field methods of Thouless and Valatin \cite{Thouless60,ThoulessV62}, which are discussed briefly in  Section \ref{sect:MFrotor}.

A possible interpretation of the extended cranking model results
 is that, with the inclusion of pairing interactions for example, 
the ratio $\omega_2/\omega_3$ in the cranking model formula (\ref{eq:CM.MofI}) is effectively changed from the shape-consistent value of mean-field theory 
without pairing, such that Eqn.\ (\ref{eq:CM.MofI})  does give the experimentally 
determined moments of inertia.
The moments of inertia  of the cranking model  then have an interpretation 
\cite{Rosensteel02} as those  of a  rotating  Riemann ellipsoid 
\cite{Chandrasekhar69} in which the rotations are linear combinations of irrotational and rigid flows.

Such an  interpretation of nuclear moments of inertia  in terms of rigid and irrotational  current flows implicitly assumes that  rotational energies are kinetic energies.
 This would appear to be highly reasonable given that the underlying concept of a rotational model is that its energies are kinetic energies with potential energy contributions only arising, for example, due to centrifugal stretching and Coriolis perturbations.
It is nevertheless important to ascertain the extent to which the so-called rotational energies arising in model calculations are consistent with this presumption.
It is also remarkable that the cranking model leads to such simple and elegant results
for what is essentially a complex problem; it is known, for example, that pure rigid-flow is only attainable in an unphysical limit in the quantum mechanics of a many-fermion system \cite{Rowe70,GulshaniR78,GulshaniR78b}.

\section{Mean-field theory as an interface between classical and quantum mechanics}  
\label{sect:MFtheory} 
From an algebraic perspective, mean-field theory is a coherent-state theory
and, as such, it can be understood in terms of both classical and quantum mechanics.
It was characterised as a semi-classical theory long ago as a way of explaining the identical equations obtained in Time-Dependent Hartree-Fock theory 
\cite{Ferrell57,GoldstoneG59,Thouless72}, and in the Random Phase Approximation \cite{BohmP52,Thouless72,Sawicki61}.
As reviewed in this section, small-amplitude TDHF theory has a quantal interpretation that is equivalent to that of the RPA \cite{Rowe66a,Rowe66b} and, more generally
\cite{RoweRR80}, it provides an  interface between classical and quantum mechanics.
These properties are important because, as shown in Section \ref{sect:AMFtheory},
standard Hartree-Fock mean-field theory can be extended to a large class of algebraic models  
to produce powerful new results.

It is first useful to recall  the prototype coherent-state relationships between the classical and quantal representations of the Heisenberg-Weyl algebra \cite{Bargmann61}.

\subsection{Quantisation and  dequantisation of a 
Heisenberg-Weyl algebra} 
\label{sect:3.A}
The Dirac quantisation \cite{Dirac26,Dirac67,Weyl50}  of the dynamics of a particle
is achieved by mapping the position and momentum coordinates,
$\{ x_{i},  p_{i};  i = 1,2,3\}$,  of its classical phase space to operators
$\{ \hat x_{i}, \hat p_{i}; i = 1,2,3\}$ on a Hilbert space of square-integrable wave functions of the nucleon coordinates such that
\beq 
\hat x_{i} \psi(x) = x_{i}\psi (x) , \quad
\hat p_{i}  \psi(x)= -{\rm i} \hbar \frac{\partial}{\partial x_{i}} \psi(x).
\eeq
This corresponds to constructing a unitary irrep of the Heisenberg-Weyl algebra in which the  coordinates satisfy the commutation relations
\beq \label{eq:HeisenbergCR1}
 [\hat x_{i},\hat p_{j} ]
={\rm i}\hbar \delta_{i,j} \hat I \ , \quad
  [\hat x_{i},\hat x_{j} ] 
=[\hat p_{i},\hat p_{j} ] =0 ,
\eeq 
where $\hat I$ is the identity operator.

Conversely,   classical mechanics is regained from quantum mechanics by  coherent-state methods \cite{Schrodinger26,Glauber63b,Perelomov85,KlauderB-S85}. 
Let $|0\rangle$ denote a minimum-uncertainty state for the position and momentum observables of the particle given by the ground state of a harmonic-oscillator Hamiltonian
$ \hat H_{\rm HO}
=\sum_{i}  \big( \alpha_{i} \hat x_{i}^2 + \beta_{i} \hat p_{i}^2\big)$.
Then
\beq
\langle 0| \hat x_{i} |0\rangle = \langle 0| \hat p_{i}|0\rangle =0,
\quad \text{for all  $i$} ,
\eeq
and it follows that the coherent states
 \beq \label{eq;3.|x,p>}
 |x,p\rangle = \exp \big[ \frac{\rm i}{\hbar} 
 \sum_{i}\big( p_{i} \hat x_{i}- x_{i}\hat p_{i}\big) \big] |0\rangle  ,
\eeq
with real values of $x_{i}$ and $p_{i}$, have the property that
\beq
\langle x,p |\hat x_{i}|x,p\rangle = x_{i}, \quad
\langle x,p |\hat p_{i}|x,p\rangle = p_{i}.
\eeq
Other classical  observable are similarly defined as functions of the  $x_i, p_i$ phase-space coordiinates by their quantum mechanical expectation values.  For example, if $\hat H$ is the quantum mechanical 
Hamiltonian, the corresponding classical Hamiltonian is given by
\beq
\mathcal{H}(x,p) = \langle x,p |\hat H |x,p\rangle.
\eeq

 From the identities 
\bal \begin{split}
& \frac{\partial}{\partial x_{i}}|x,p\rangle =  - \frac{\rm i}{\hbar}\, \hat p_i |x,p\rangle , \\
 &  \frac{\partial}{\partial p_{i}}|x,p\rangle =   \frac{\rm i}{\hbar}\, \hat x_i |x,p\rangle ,
\end{split}
\end{align}
it then follows that 
\bal 
\begin{split}
&\frac{\partial\mathcal{H}}{\partial x_i} = -\frac{\rm i}{\hbar} \langle x,p | [\hat H, \hat p_i]|x,p\rangle ,
\\
&\frac{\partial\mathcal{H}}{\partial p_i} = \frac{\rm i}{\hbar} \langle x,p | [\hat H, \hat x_i]|x,p\rangle .
\end{split}
\end{align}
The time-dependent Schr\"odinger equation
\bal \begin{split}
&{\rm i} \hbar\frac{\partial x_i}{\partial t} = \langle x,p | [\hat x_i,\hat H]|x,p\rangle, \\
&{\rm i} \hbar\frac{\partial p_i}{\partial t} = \langle x,p | [\hat p_i,\hat H]|x,p\rangle, 
\end{split} \end{align}
then leads to the classical equations of motion
\beq 
 \frac{\partial x_i}{\partial t} = \frac{\partial\mathcal{H}}{\partial p_i} , \quad
 \frac{\partial p_i}{\partial t} = - \frac{\partial\mathcal{H}}{\partial x_i} .
\eeq

These relationships reveal an embedding of the classical phase space for a system of spin-free particles as a smooth manifold in the Hilbert space of its quantum mechanics
in which a classical point with  coordinates $x= \{ x_{ni}\}$ and $p= \{ p_{ni}\}$ is identified with a quantum-mechanical  coherent state $|x,p\rangle$.

For a system of particles with intrinsic spins, the relationship between its classical and its quantum mechanics would appear to be much more complicated.
In fact, for nucleons, which are spin-half fermions, it is actually  simple
because the Hilbert space of a many-fermion system contains only fully anti-symmetric states; 
this restricts them to a single irrep and imposes a huge restriction on the possible many-fermion states.

\subsection{Mean-field theory as a classical representation of a Lie algebra}
\label{sect:HFtheory}
A remarkable observation is that a coherent-state theory analogous to that used above  to obtain an interface between the classical and quantum mechanics of the Heisenberg-Weyl Lie algebra applies, with some generalisation, to any Lie algebra whose irreps have lowest- (and/or highest-) weight states.
The best-known example is given by Hartree-Fock mean-field theory.

The Hartree-Fock  theory of a many-fermion system  starts from the observation that its Hilbert space is that of a fully anti-symmetric unitary irrep of a Lie algebra of one-body operators.
In the language of second quantisation, this irrep is expressed in terms of creation and annihilation operators, $a^\dag_\mu$ and $a^\mu$, of single-nucleon states as  linear combinations of the one-body operators
\beq
\hat C_{\mu}^{\nu} = a^\dag_\mu a^\nu,
\eeq
for which the commutation relations 
\beq
[\hat C_{\mu}^{\nu}, \hat C_{\mu'}^{\nu'}] =
\delta^{\nu}_{\mu'} \hat C_{\mu}^{\nu'} - \delta^{\nu'}_{\mu} \hat C_{\mu'}^{\nu} ,
\eeq
are obtained from the fermion anti-commutation relations
\beq
 \{ a^\mu , a^\dag_\nu\}  =  a^\mu  a^\dag_\nu + a^\mu a^\dag_\nu
 =\delta^\mu_\nu , \quad
 \{ a^\mu , a^\nu\} = \{ a^\dag_\mu , a^\dag_\nu\} = 0 .
\eeq

Following an Einstein convention,   creation operators $\{ a^\dag_\nu\}$, with lower indices, 
are components of a covariant tensor and   annihilation operators $\{ a^\nu\}$,
with upper indices, are  components of  a contravariant tensor.
They are defined such that a scalar product of these operators is  obtained by summing over common upper and lower indices; e.g., if 
the indices are SU(2) angular-momentum quantum numbers $\nu = jm$, for a given value of $j$,
a scalar number operator is defined by the sum
\beq
\hat N_j = \sum_m a^\dag_{jm} a^{jm} .
\eeq
The standard coupling of covariant SU(2) tensors of the same angular momentum to form a scalar, is given by
\beqa
\big[ a^\dag_j \otimes a_j \big]_0 &=&
\textstyle \sum_m (j,-m, j,m |0,0)\, a^\dag_{jm} a_{j,-m} \nonumber \\
&=& \sum_m \frac{(-1)^{j+m}}{\sqrt{2j+1}} \, a^\dag_{jm} a_{j,-m} .
 \eeqa
Thus, SU(2) tensors with upper and lower indices can be related by defining
 \beq \label{eq:a_barm}
a_{j\bar m} \equiv  a^{jm} =  (-1)^{j+m} a_{j,-m} .
\eeq 
With this definition, it then  follows that
\beq
\hat N_j = \sqrt{2j+1}\, \big[ a^\dag_j \otimes a_j \big]_0 \,,
\eeq 
and that pairs of covariant fermion operators  satisfy the anti-commutation relationships
\beq 
 \{ a_{j \bar m} , a^\dag_{j' m'}\}  \equiv \{ a^{j m} , a^\dag_{j' m'}\}
= \delta^j_{j'}\delta^m_{m'} .
 \eeq
It is important to note, however, that the relationship between the covariant and contravariant components of a tensor depends on the nature of the tensor.  It is given here for the standard labelling of SU(2) and SO(3) tensors.  For Cartesian tensors, which have a positive-definite metric,
 there is no need to make a distinction; ie., in terms of an orthonormal basis, $x^i \equiv x_i$ and  $x^2 = \sum_i x^2_i$.

The Hilbert space of a nucleus is now determined by standard Lie algebra methods in terms of a lowest-weight state and sets of raising and lowering operators, defined as follows.
For an $A$-nucleon nucleus, a  lowest-weight  state for an irrep of the Lie algebra of one-body operators is a state whose wave function is a Slater determinant of the wave functions  of  occupied single-nucleon states, as explained below. 
These states might be selected, for example, on the basis of independent-particle shell model considerations. 
However, the choice is, in principle, arbitrary.
For any given choice,  raising and lowering operators are  defined as follows.

Let indices $h$ (for hole) and  $p$ (for particle) label  single-particle states that are, respectively, occupied and unoccupied in a lowest-weight state $|\phi\rangle$.
The operators
\beq
\hat \eta^h_p = a^\dag_p a^h \quad \text{and} \quad
\hat \eta^p_h = a^\dag_h a^p ,
\eeq
are then, respectively,  particle-hole creation and annihilation operators and the lowest-weight state is a so-called particle-hole vacuum state.
Starting from such a state, a basis for the many-nucleon Hilbert space of a nucleus is  obtained by exciting the vacuum state to many particle-hole  states generated by the  actions on it of the particle-hole creation operators as raising operators.

It is now apparent that any Slater-determinant state is a particle-hole vacuum state with respect to a corresponding definition of particle and hole operators and can serve as a lowest-weight state for the construction of the quantum-mechanical Hilbert space of a nucleus.
Moreover, any normalised Slater determinant  is related to any other such Slater determinant by a unitary transformation of its occupied single-particle states.
Thus, the Slater determinants of Hartree-Fock theory are said to lie on an orbit of the group  of unitary one-body transformations and  are, by definition, coherent states of this group \cite{Perelomov72, Perelomov85,KlauderB-S85}.
As a result \cite{RoweRR80,RosensteelR81b}, the Slater determinants of Hartree-Fock theory form a manifold of lowest-weight states that is in one-to-one correspondence with a classical phase space on which a classical Hamiltonian dynamics is defined in parallel with that for the Heisenberg algebra.
Such properties of Lie group orbits are well known in mathematics 
\cite{Kirillov76} and feature in the theory of geometric quantisation
\cite{Souriau70,Kostant70}.

The identification of Hartree-Fock mean-field states (normalised Slater determinants) with points on a classical phase space is  obtained by first defining local position and momentum operators \beq \label{eq:lcoords1}
\hat x_{ph} = \frac{1}{\sqrt{2}}\big(\hat\eta^p_h + \hat\eta^h_p \big),
\quad  \hat p_{ph} = \frac{{\rm i} \hbar}{\sqrt{2}}
\big(\hat\eta^p_h - \hat\eta^h_p \big) ,
\eeq
for a neighbourhood of  an arbitrary particle-hole vacuum state $|\phi\rangle$.
These operators  satisfy the equations 
\bal \label{eq:HeisenbergCR2}
\begin{split}
&\langle \phi | [\hat x_{ph},\hat p_{p'h'} ] |\phi\rangle
={\rm i}\hbar \delta_{p,p'} \delta_{h,h'} , \\
& \langle \phi | [\hat x_{ph},\hat x_{p'h'} ] |\phi\rangle
=\phi | [\hat p_{ph},\hat p_{p'h'} ] |\phi\rangle =0 ,
\end{split}
\end{align} 
and can be used to define mean-field states with local phase space coordinates  
$(x_{ph},p_{ph})$ coordinates  defined  by 
\beq \label{eq:lcoords3}
 |\phi (x,p)\rangle = \exp \big[ \frac{\rm i}{\hbar} 
 \sum_{ph}\big( p_{ph} \hat x_{ph}- x_{ph}\hat p_{ph}\big) \big] |\phi\rangle  .
\eeq

Such $(x_{ph}, p_{ph})$ coordinates can be defined for a finite neighbourhood of any point on the manifold of lowest-weight states \cite{AbrahamM87}.
This is always possible for  a classical phase space for which a neigbourhood of any point can be put into one-to-one correspondence with a neighbourhood of a flat Euclidian space.
However,  a single global $(x,p)$ coordinate chart, 
for which the equations
 (\ref{eq:HeisenbergCR2}) hold at every point,
would not be possible on a phase space which, for example, had the topology of a sphere.
This now a problem, however, because there is no requirement that a classical phase space should be Euclidean.
The important observation is that for a given Hamiltonian $\hat H$, the classical equations of motion
\beq \label{eq:CurvedSpaceHEqns}
\frac{d x_{ph}}{dt} = \frac{\partial\mathcal{H} }{\partial p_{ph}} , \quad
\frac{d p_{ph}}{dt} = -\frac{\partial\mathcal{H} }{\partial x_{ph}} ,
\eeq
with
\beq 
\mathcal{H}(x,p) = \langle \phi(x,p)| \hat H | \phi(x,p)\rangle ,
\eeq
are determine to apply precisely at any point of a smooth classical phase-space .
This follows, as for the Heisenberg-Weyl Lie algebra,  because any  point of the space can be considered as the origin of a local cordinate chart for a neighbourhood of that point.
 Thus, the infinitesimal generators $\hat x_{ph}$ and $\hat p_{ph}$ can be chosen at any point to satisfy the Eqns.\  (\ref{eq:HeisenbergCR2}) exactly.

The Hartree-Fock  $|\phi_0\rangle$ state is now selected as the  independent-particle state on the manifold of lowest-weight states for which the energy 
$\langle\phi_0 |\hat H |\phi_0\rangle$ is a minimum and which therefore
satisfies the equations
\beq \label{eq:variational.eq}
\langle \phi_0 | [\hat H, a^\dag_p a^h]|\phi_0\rangle 
=\langle \phi_0 | [\hat H, a^\dag_h a^p]|\phi_0\rangle= 0 ,\quad\text{for all $p$ and $h$}.
\eeq
Thus, if $(x,p)$ coordinates are defined about such a state $|\phi_0\rangle$ 
as a point of phase space, the classical Hamiltonian $\mathcal{H}$  satisfies the equations 
\beq
\left(\frac{\partial \mathcal{H}}{\partial x_{ph}}\right)_{x=p=0}
=\left( \frac{\partial \mathcal{H}}{\partial p_{ph}}\right)_{x=p=0} = 0, 
\quad\text{for all $p$ and $h$}.
\eeq
The standard Hartree-Fock procedure for identifying such a minimal energy  independent-particle state  is to start from some reasonable first guess 
$|\phi\rangle$ and define  an independent-particle Hamiltonian 
\beq
\hat h = \sum_{\mu,\nu} 
\langle \phi |\{ a^\mu , [\hat H , a^\dag_\nu]\} |\phi \rangle a^\dag_\mu a^\nu .
\eeq
Diagonalisation of this Hamiltonian then gives a new lower-energy state as its lowest-energy eigenstate \cite{RoweR76}. 
This process is iterated until it converges to a  state $|\phi_0\rangle$ that satisfies the variational equation (\ref{eq:variational.eq}).

It is now observed that  Hartree-Fock theory is much more than  a means to determine an approximation for the ground state of a quantum-mechanical system.
It  provides an explicit map of the quantum mechanics of a many-fermion system to a corresponding classical mechanics in which the Hartree-Fock Hamiltonian
\beq
\hat h(x,p) = \sum_{\mu,\nu} 
\langle \phi(x,p) |\{ a^\mu , [\hat H , a^\dag_\nu]\} 
|\phi(x,p) \rangle a^\dag_\mu a^\nu ,
\eeq
defined in terms of local $(x,p)$ coordinates at any point of a phase space,
determines the classical dynamics at that point.
Thus, a TDHF dynamics is defined by the classical equations of motion on this phase space, without the need of an adiabatic approximation as has often  been considered  necessary for such purposes.
An adiabatic approximation may nevertheless be useful to restrict the large-amplltude mean-field dynamics to the valley floor of the classical phase space,  as defined in \cite{RoweR82b};
cf. also \cite{MatsuyanagiMNHS10,MatsuyanagiMNYHS16}.
Such large-amplitude TDHF dynamics can be relevant, for example, to the study of heavy-ion reactions and nuclear fission.

The  relationship between the classical and quantum representations 
of the Lie algebra of one-body operators of a many-fermion system
was shown in nuclear physics \cite{HolzwarthY74,KermanK76,RoweB76} and used by many \cite{RoweB76,MoyaDeGuerraVillars77,Marumori77,%
BarangerV78,ReinhardG78,GoekeRR81}
to explore the topography of the classical space of mean-field energies in terms of
valleys,  fall lines, peaks, ridges and passes.
Applications of some of these methods  to the theory of nuclear structure were reviewed
 in terms of generator-coordinate methods by Reinhard and Goeke \cite{ReinhardG87}.  
 Other methods are currently being developed by Matsuyanagi and colleagues
\cite{MatsuyanagiMNHS10,MatsuyanagiMNYHS16}.
It should be remembered, however,
in considering the topographical maps obtained in mean-field theory,
that distance scales on the map will generally vary from point-to-point and can give a distorted view of the situation as shown in a  rigorous coordinate-free treatment \cite{RoweR82b,RoweR82}.  
Distance scales and their directional dependence, defined by the metric on the multi-dimensional mean-field manifold, can be determined at a point by a so-called local harmonic-oscillator treatment, corresponding to constrained RPA solutions \cite{RoweB76,Marumori77}.
One result is that the path which follows the bottom of a valley will only, in general, be identical to a fall line, i.e., a line of steepest descent \cite{RoweR82}, if neither is curved.
Such distortions are familiar in topographical maps of landscapes on the curved surface of  Earth.

\subsection{Classical normal-mode vibrations and the random phase approximation} \label{sect:NModes}

It will be taken as understood  that the nuclear Hamiltonian $\hat H$ is rotationally invariant.
For the purposes of this section, it will also be understood that the minimum-energy HF state is uniquely defined.
This being the case, implies that it also rotationally invariant and consequently it is a state of zero angular momentum.
For, if it were not, it would  be one of a set of minimum-energy states generated by its rotations. 
This more general case, is of special interest and is considered in depth in Section
\ref{sect:MFrotor}.

The dual classical/quantal interpretation of TDHF theory relates the classical small-amplitude normal-mode vibrations of a many-fermion system about its lowest-energy equilibrium state to the quantum mechanics of the elementary excitations of a nucleus as given in the RPA (random-phase approximation).

The small-amplitude normal-mode vibrations of a classical system 
about an equilibrium state are obtained by  expanding the Hamiltonian 
$\mathcal {H}$ to quadratic terms in the amplitudes of the position and momentum coordinates which take zero values when the system is at equilibrium.
For position and momentum coordinates
$\{x_\alpha\}$ and $\{ p_\alpha\}$ that are canonical at the equilibrium point, 
the Hamiltonian is then of the form
\beq 
\mathcal{H}(x,p) 
= E_0 +\sum_{\alpha,\beta} \frac{1}{2B_{\alpha\beta}} p_\alpha p_\beta
+\sum_{\alpha,\beta} \frac{1}{2} C_{\alpha\beta}\, x_\alpha x_\beta  + \cdots ,
\eeq
where  $E_0$ is the equilibrium energy of the lowest-energy state.
The Hamilton equations of motion (\ref{eq:CurvedSpaceHEqns})
then give harmonic small-amplitude vibrational solutions with
\beq 
\frac{dx_\alpha}{dt} = \sum_\beta \frac{1}{B_{\alpha\beta}} p_\beta ,
\quad 
\frac{dp_\alpha}{dt} = - \sum_\beta  C_{\alpha\beta} x_\beta .
\eeq

The corresponding mean-field equations are obtained as follows.
With the notation that $m$ and $n$  label particle states 
and $i$ and $j$  label hole states,
a small-amplitude time-dependent mean-field vibrational state is  expressed as
\beq
|\phi(t) \rangle = 
e^{-\frac{{\rm i}}{\hbar} E_0 t} e^{\epsilon \hat X (t)} |\phi_0\rangle,
\eeq
where  $\epsilon$ is a small parameter and $\hat X(t)$ is  the skew-Hermitian operator
\beq
\hat X(t) = \sum_{mi}
\big[ X_{mi}(t) a^\dag_m a^i - X^*_{mi} (t) a^\dag_i a^m \big] .
\eeq

From the identities
\bal 
& \langle \phi (t) | a^\dag_i a^m |   \phi (t)\rangle
=    \epsilon X_{mi}(t) + 0(\epsilon^3), \\ 
&  
{\rm i} \hbar \frac{d}{d t}\langle \phi(t) |a^\dag_i a^m |\phi(t)\rangle = 
\langle \phi(t) |[a^\dag_i a^m, \hat H ] |\phi(t)\rangle 
= \epsilon \langle \phi_0 | [[a^\dag_i a^m ,\hat H] , \hat X(t) ] |\phi_0\rangle
+ O(\epsilon^3)  ,
\end{align}
it  follows that
 \beq \label{eq:tdhf1}
 {\rm i}\hbar \frac{d}{d t} X_{mi}(t) = 
  \langle \phi_0 | [[a^\dag_i a^m ,\hat H] , \hat X(t) ] |\phi_0\rangle 
    + O(\epsilon^2) .
    \eeq
Thus, for harmonic small-amplitude  normal-mode vibrations about the Hartree-Fock  minimum-energy state, for which
   \beq 
   X_{mi}(t) = Y_{mi} e^{-{\rm i}\omega t} +  Z^*_{mi} e^{{\rm i}\omega t} ,
   \eeq
Eqn.\ (\ref{eq:tdhf1})  separates into a set of coupled eigenvalue equations:
    \beq  \begin{array}{c} \displaystyle
   \sum_{nj} \big[ A_{minj} Y_{nj} +  B_{minj} Z_{nj} \big]
    = \hbar \omega Y_{mi} \,,\\ \displaystyle
   \sum_{nj} \big[ A_{minj} Z^*_{nj} +  B_{minj}  Y^*_{nj}] = -\hbar \omega Z^*_{mi} \,,
   \end{array} \label{eq:6.tdhf2}
  \eeq
 where $A$ is a Hermitian matrix and $B$ is a symmetric matrix with components
\bal \begin{split}
&A_{minj} := 
  \langle \phi |[ a^\dag_i a^m,  [\hat H , a^\dag_n a^j ]] |\phi\rangle    
 \\
&B_{minj} := - \langle \phi |[ a^\dag_i a^m, [\hat H , a^\dag_j a^n ]] 
  |\phi\rangle  .
\end{split}\end{align}
After taking the complex conjugate of the second equation, these equations are expressed in the matrix form 
  \beq \begin{pmatrix} A&B\\B^* & A^* \end{pmatrix}
  \begin{pmatrix} Y\\ Z \end{pmatrix} =
  \hbar\omega \begin{pmatrix} Y\\ -Z \end{pmatrix} . \label{eq:6.TDHFeqn}
  \eeq
Thus, the solutions of  this matrix equation for the $\{ Y_{mi}\}$ and $\{Z_{mi}\}$ coefficients and the corresponding values of $\omega$ define the classical small-amplitude normal-modes  of a nucleus and their vibrational frequencies.

\subsection{The  quantum-mechanical Random Phase Approximation} \label{sect:RPA}
The RPA, regarded as a quantum-mechanical coherent-state version of  classical normal-mode vibrations, is now obtained \cite{Rowe66a,Rowe66b}
by identifying the small-amplitude TDHF vibrational wave functions with harmonic-oscillator coherent states.
This is  achieved by recognising that the operators
\bal \label{eq:53}
 \begin{split}
&O^\dag_\lambda := \sum_{mi} \big( Y_{mi}(\lambda) a^\dag_m a^i 
  - Z_{mi}(\lambda) a^\dag_i a^m \big)  , \\ 
&  O^\lambda :=  \sum_{mi} \big( Y^*_{mi}(\lambda) a^\dag_i a^m 
  - Z^*_{mi}(\lambda) a^\dag_m a^i  \big) , 
  \end{split}
\end{align}
for which the $Y_{mi}(\lambda)$ and $Z_{mi}(\lambda)$ coefficients are solutions of 
Eqn.\ (\ref{eq:6.tdhf2}) with $\omega_\lambda = \omega$,  satisfy the equations
 \bal \label{eq:8.38} 
 \begin{split}
&\langle \phi |[ \hat X, [\hat H, O^\dag_\lambda ]] |\phi\rangle =
 \hbar\omega_\lambda 
 \langle \phi | [\hat X, O^\dag_\lambda ] |\phi\rangle  , \\
&\langle  \phi | [\hat X, [\hat H, O^\lambda ]] |\phi\rangle = 
-\hbar\omega_\lambda  
\langle  \phi | \hat X, O^\lambda]  |\phi\rangle , 
\end{split}  
\end{align}
for any one-body operator $\hat X$ 
and can be normalised  to satisfy the orthogonality relationships 
\bal \label{eq:56}
 \begin{split}
&\langle \phi | [O^\kappa, O^\dag_\lambda ] |\phi\rangle   
    = \delta^\kappa_\lambda,\\
&\langle  \phi | [O^\kappa, O^\lambda ] |\phi\rangle =
 \langle  \phi | [O^\dag_\kappa, O^\dag_\lambda ] |\phi\rangle = 0 .
\end{split}
\end{align}
The operators $O^\dag_\lambda$ and  $O^\kappa$ are then interpreted in the RPA as  excitation and de-excitation operators of one-phonon vibrational excitations of the nuclear ground state 
$|0\rangle$
Thus, the ground state $|0\rangle$ of the nucleus  is implicitly defined in the RPA as a state that is annihilated by the $\{ O^\lambda\}$ lowering operators, i.e., $O^\lambda |0\rangle = 0 $, 
for all $\lambda$, and  states of excitation energy $E_\lambda - E_0 = \hbar\omega_\lambda$
are given by $ |\lambda\rangle =O^\dag_\lambda |0\rangle$.

A significant result is that  Eqns. (\ref{eq:53}) - (\ref{eq:56}) imply the presence of vibrational correlations in the quantum mechanical ground state $|0\rangle$.
Such correlations are implied  
when the lowering operators  $\{ O^\lambda\}$, 
which should annihilate the ground state, contain  non-zero $Z^*_{mi}$ terms.
However, the
properties of the vibrational excited states of the RPA  are obtained algebraically  without the need to derive explicit expression for either the correlated ground state or its excited states.
In particular,   matrix elements of a one-body transition operator,
  \beq \hat Q = \sum_{\mu\nu} Q_{\mu\nu} a^\dag_\mu a^{\nu}, \eeq
between the ground and excited one-phonon states,  
\beq |\lambda\rangle := O^\dag_\lambda |0\rangle, \quad \forall\, 
O^\dag_\lambda \; \text{for which}\; \omega_\lambda >0 ,\eeq
are evaluated from the expression
 \beq \langle 0|\hat Q |\lambda\rangle 
  = \langle 0| [\hat Q, O^\dag_\lambda ] |0\rangle ,\eeq
and are given, within the harmonic-oscillator approximation,  by
  \bal 
  \langle 0|\hat Q |\lambda\rangle 
  = \langle \phi| [\hat Q, O^\dag_\lambda ] |\phi\rangle 
 = \sum_{mi} \big( Y_{mi}(\lambda) Q_{im} + Z_{mi}(\lambda)Q_{mi} \big) .
\end{align}

The above double-commutator equations-of-motion approach, 
developed in Refs.\ \cite{Rowe68,Rowebook70},
 provides the simplest and now standard expression of the RPA,
as reviewed, for example, in the book of Ring and Schuck \cite{RingS80}.

\section{The emergence of rotational states in mean-field theory}
\label{sect:MFrotor}
If the minimum-energy Hartree-Fock state for an even-even nucleus is rotationally invariant,   it is a state of zero angular momentum and is considered to be an 
approximation to the ground state of the nucleus.
Moreover, if the nucleus under consideration has a low-energy excited state of some angular momentum $J$ that decays strongly, relative to a single-particle transition, by an electromagnetic $J$-pole transition to the ground state,  it will naturally be expected that such a state can be interpreted as a one-phonon collective vibrational state with an excitation operator given to a good approximation by an RPA calculation.
The low-energy $3^-$ excited states of $^{16}$O and $^{40}$Ca at  6.13 MeV and 3.73 MeV, respectively,  are good examples of this and have been successfully treated as such in Refs.\ \cite{BrownET61,GilletS64,GilletS67}.
Numerous other examples are referenced, for example, 
in \cite{Rowebook70} and \cite {RingS80}.
Thus, the TDHF-RPA theory is understood to provide a good first-order many-nucleon description of one-phonon vibrational excitations of spherical nuclei.
However, rotationally invariant spherical Hartree-Fock minimum-energy states are relatively uncommon.  
They occur for doubly closed-shell nuclei.
But, even then, it is frequently observed, consistent with experimental observations \cite{HeydeW71,WoodH16},  that there are  strongly deformed Hartree-Fock solutions with only a little more energy.

\subsection{Broken symmetry and rotational states of nuclei}
Most  frequently, it transpires that the lowest-energy mean-field state for a rotationally invariant nuclear Hamiltonian, is  neither rotationally invariant nor a state of good angular momentum.
Its  coupling to rotated lowest-energy mean-field states is then far from negligible, but is hidden  in mean-field theory because  such  states  have  energy expectation values that remain unchanged under rotations.
Consequently,  the TDHF equations have zero-frequency (Nambu-Goldstone  \cite{NambuJL61,GoldstoneSW62}) normal-mode solutions corresponding to rotations of the broken-symmetry mean-field state for which there are no restoring forces.
 Such a broken-symmetry  state was interpreted  by 
Peierls and Yoccoz \cite{PeierlsY57} as a semi-classical state of a rotor model 
with zero angular momentum and a specific orientation.
This led  to the  Thouless-Valatin \cite{Thouless60,ThoulessV62} generalisation of the Inglis cranking model in which the Hartree-Fock mean-field equations were solved for a Hamiltonian
$\hat H - \hbar\omega \hat J_1$.
An expression for the moment of inertia $\scrI_1$ was then determined from the energy increase
\beq 
\tfrac12 \scrI_1\omega^2 = \Delta E .
\eeq
Unlike the Inglis model, the Thouless-Valatin procedure,  takes full account of  the self-consistent adjustment of the mean-field Hamiltonian for a rotating state of the nucleus.
It can similiarly be extended to include pairing interactions.

\subsection{Angular-momentum projection} \label{sect:AMprojection}
A  precise expression of the Peierls-Yoccoz \cite{PeierlsY57} observation,
 within the framework of the  generator-coordinate theory of Hill, Wheeler and Griffin
\cite{HillW53,GriffinW57},  
is that when the lowest-energy mean-field state is not rotationally invariant
a set of low-energy states of the nucleus with good angular-momentum quantum numbers can be obtained as linear combinations of the equal-energy states generated by its rotations. 
This leads to a microscopic many-nucleon model of nuclear rotations in which the nuclear Hamiltonian is diagonalised in the Hilbert space spanned by  these rotated states. 
A basis of angular-momentum states generated in this way is defined by 
angular-momentum projection methods.
Many applications of these methods have recently  been reviewed by Sun \cite{Sun16}.
Here we briefly review, with some adjustment, the elegant approach of Lee and Cusson \cite{LeeC72,CussonL73} as it applies to doubly even nuclei.

A Slater determinant, considered as an intrinsic state for a system of rotational states, can  be expressed as  a sum of states
\beq \label{eq:intr_state}
|\Phi\rangle = \sum_{JK} n_{JK} |\phi_{KJ}\rangle \equiv \sum_{JK} n_{JK} |\phi_{KJK}\rangle
\eeq
of angular momentum $J$ and component of angular momentum $K$ relative to a so-called body-fixed axes.
For a given intrinsic state $|\Phi\rangle$, the objective is then  to identify the states 
$|\phi_{KJ}\rangle$ and the corresponding sets of states 
$\{ |\phi_{KJM}\rangle, M= -J, \dots,J\}$ generated by rotating them.
In standard mean-field theory, the state $|\Phi\rangle$ is the minimum-energy Slater determinant.
More generally, it  could  be determined separately for each angular-momentum state by variation-after-projection methods \cite{CussonL73}.   
Other possibilities, in which $|\Phi\rangle$ could 
be one of several optimally chosen intrinsic states  for the irreps of a microscopic  collective model, are discussed in Sections \ref{sect:GCM1} and \ref{sect:GCT}. 

Body-fixed axes, for an intrinsic state $|\Phi\rangle$, are appropriately chosen to be principal axes of its quadrupole mass tensor.
However,  principal axes of the quadrupole mass tensor are only defined to within the subgroup of rotations that leave the quadrupole mass tensor invariant.  
This subgroup is the so-called \emph{vierergruppe} group $D_2$; i.e., the group of rotations through multiples of angle $\pi$ about each of the principle axes.
A significant characteristic of Hartree-Fock mean-field theory is that, although it
frequently breaks the  rotational invariance of a nuclear Hamiltonian, 
it commonly retains its $D_2$ invariance \cite{Bar-TouvK65}.
It may be noted that $D_2$ invariance is  conserved in the asymmetric-top models of nuclear and molecular rotations and  emerges  in Elliott's SU(3)  model \cite{Elliott58ab} and Ui's  rotor model \cite{Ui70} from their algebraic structures.
When $D_2$ symmetry is conserved, the intrinsic state 
belongs to a one-dimensional  irrep of $D_2$ and the $K$ quantum number in Eqn.\ (\ref{eq:intr_state})
 is restricted to either even- or odd-integer values.
If the mean field for the minimum-energy state also has an axis of symmetry, then the intrinsic state has the larger symmetry of the group $D_\infty$ which includes all rotations about the symmetry axis plus rotations through multiples of  angle 
$\pi$ about axes perpendicular to the symmetry axis.
It then has one-dimensional irreps with a single basis state given by  combinations of states of angular momentum $\pm K$ relative to the axis of symmetry.
However, just as all even-even nuclei are observed to have $J=0$ ground states,  we expect the lowest-energy states of an axially symmetric rotor to have $K=0$ for an even-even nucleus.
In general, there will be a large range of values of $J$ for each value of $K$. 
For an odd nucleus, the intrinsic symmetry group will have a spinor irrep.

Basis states for the space generated by rotations of the  state $|\Phi\rangle$ are the  states
 $\{|\phi_{KJM}\rangle\}$  in the expansion of the rotated  states
\beq \label{eq:R(Omega)Phi_expansion}
\hat R(\Omega) |\Phi\rangle 
=  \sum_{JK} n_{JK} \hat R(\Omega) |\phi_{KJK}\rangle  \nonumber\\
= \sum_{JKM} n_{JK} |\phi_{KJM}\rangle   \scrD^J_{MK} (\Omega) ,
\eeq
where
\beq
  \scrD^J_{MK} (\Omega)  = \langle KJM |\hat R(\Omega) |KJK\rangle 
\eeq
is a standard rotation matrix.
Thus, the basis states $\{ |\phi_{KJM}\rangle\}$  for the rotational model Hilbert space with good angular-momentum quantum numbers are  obtained in the form
\beq
 |\phi_{KJM}\rangle = \frac{1}{n_{KJ}} 
\frac{2J+1}{8\pi^2}\int \hat R(\Omega)|\Phi\rangle  \scrD^{J*}_{MK} (\Omega)\, d\Omega
\eeq
with the norm factors  
\beq
|n_{KJ}|^2 = \frac{2J+1}{8\pi^2}
\int \scrD^{J*}_{KK}(\Omega) \langle \Phi | \hat R(\Omega ) |\Phi \rangle \, d\Omega .
\eeq

Matrix elements of the Hamiltonian and a multipole transition operator 
$\hat W^\lambda_\mu$ are similarly expressed in terms of integrals over rotational angles of products of $\scrD^{J*}_{MK} (\Omega)$ functions and intrinsic matrix elements $\langle\Phi | \hat W^\lambda_\mu  \hat R(\Omega ) |\Phi \rangle$.
Such integrals  can be evaluated precisely by the quadrature expressions \cite{MilanovicCS08} 
\bal
&\frac{1}{2\pi}\int_0^{2\pi}\! f(x ) \,dx, 
               = \frac{1}{n}\sum_{r=0}^{n-1}f\left(\frac{r\pi}{n}\right) , 
               \quad x= \text{$\alpha$ or $\gamma$}, \label{eq:intalpha}\\
&\int_{-1}^{1} f(x)\, dx  
               = \sum_{i=1}^N f(x_i) w_i, \qquad x= \cos \beta , \label{eq:intbeta}
\end{align}         
for any values of $n$ and $N$ larger than the number of values of $K$ and $J$, respectively, in Eqn.\ (\ref{eq:intr_state}) for which $n_{JK}$ is 
 considered to be non-negligible.
 Equation (\ref{eq:intbeta}) is the standard Gauss-Legendre expansion for which the values of $x_i$ and the  weights $w_i$ for a given $N$ are listed, for example, in Table 25.4 of Abramowitz and Stegun  \cite{AbramowitzS72} (see \verb|en.wikipedia.org/wiki/Gaussian_quadrature| for details of the
method). 
 This expression is exact for polynomial functions of $\cos\beta$ of degree no larger than $N$.
 Equation (\ref{eq:intalpha}) is a corresponding expansion for functions that are periodic over $2\pi$ intervals.  
In practice, one can increase the numbers $n$ and $N$ until convergence is obtained to the desired accuracy of the calculations.
 
To determine the norms $|n_{JK}|^2$, the energies of  projected states, and transition matrix elements,
it is necessary to calculate the overlaps $\langle \Phi' |\Phi\rangle$ and
matrix elements $\langle \Phi'|\hat X |\Phi\rangle$,
where $\Phi' = \hat R(\Omega) |\Phi\rangle$ and $\hat X$ is 
 a one- or  two-body operator.
A limitation of the following derivations of the matrix elements is that they apply only to pairs of Slater determinants, $|\Phi\rangle$ and $|\Phi'\rangle$, for which 
 $\langle \Phi |\Phi' \rangle \not= 0$.
 This is sufficient for angular-momentum projection from a single Slater determinant.
 However, for deriving the matrix elements of one- and two-body-operators between states projected from other states and from orthogonal Slater determinants,
 more general projection methods are required (such as those reviewed by Sun \cite{Sun16}).

 If $|A\rangle$ and $|B\rangle$ are the  $N$-fermion Slater determinants
 \bal
& |A\rangle = a^\dag_1 a^\dag_2 \cdots a^\dag_N |-\rangle ,\\
 & |B\rangle = b^\dag_1 b^\dag_2 \cdots a^\dag_N |-\rangle ,
\end{align}
where $|-\rangle$ is the bare-fermion vacuum state and $\{a^\dag_i\}$ and $\{b^\dag_i\}$ are each sets of orthogonal single-nucleon creation operators, 
their overlap is  the determinant
\beq 
\langle B |A \rangle = D= \det (d), 
\eeq
in which $d$ is the $N\times N$ matrix whose elements, 
\beq \label{eq:dmatrix}
d^\nu_\mu = \langle b^\nu a^\dag_\mu\rangle = \langle-| b^\nu a^\dag_\mu |-\rangle ,
\eeq
are the overlaps of single-particle states; 
recall that $b^\nu$ is the Hermitian adjoint of $b^\dagger_\nu$.

If $\{ c^\dag_\alpha, c^\alpha\}$ is another set of single-fermion creation and annihilation operators, then the  matrix element 
\beq
 \langle B | c^\dag_\beta c^\alpha | A \rangle  
 = \langle -|  b^N  \cdots b^2 b^1 c^\beta c^\dag_\alpha  
 a^\dag_1 a^\dag_2 \cdots a^\dag_N |-\rangle 
 \eeq
is expressed as the sum of terms obtained by replacing the single-particle overlaps 
$\langle b^\nu a^\dag_\mu\rangle$ in det$(d)$, one at a time, by
$\langle b^\nu c^\dag_\beta c^\alpha a^\dag_\mu\rangle$.
The result is that
\beq 
\langle B | c^\dag_\beta c^\alpha | A \rangle =
\sum_{\mu\nu} \langle b^\nu c^\dag_\beta c^\alpha a^\dag_\mu\rangle M^\mu_\nu
= \sum_{\mu\nu}  \langle b^\nu c^\dag_\beta\rangle
M^\mu_\nu \langle c^\alpha a^\dag_\mu\rangle,
\eeq
where $M^\mu_\nu$ is the  cofactor of the element 
$d^\nu_\mu=\langle b^\nu a^\dag_\mu\rangle$ in the determinant $D= \det(d)$.
The matrix $M$ with elements $M^\mu_\nu$, known as the adjugate of the matrix $d$, 
is simply related to the inverse of the matrix $d$ by  Cramer's rule \cite{Cramer-rule} which states that, 
provided $D\not= 0$, the matrix $d$ has an inverse given by
\beq 
d^{-1} = \frac{1}{D} M .
\eeq
It follows that $M= Dd^{-1}$ and that
\beq 
\langle B | c^\dag_\beta c^\alpha | A \rangle =
D \sum_{\mu\nu}  \langle b^\nu c^\dag_\beta\rangle
(d^{-1})^\mu_\nu \langle c^\alpha a^\dag_\mu\rangle,
\eeq
as determined by Cusson and Lee \cite{CussonL73}.

Two-body matrix elements can be derived in a similar way.
A matrix element 
\beq
X^{\gamma\delta}_{\alpha\beta} = 
\langle B| c^\dag_\alpha c^\dag_\beta c^\delta c^\gamma |A\rangle 
 = \langle   b^N  \cdots  b^1 c^\dag_\alpha c^\dag_\beta  
c^\delta c^\gamma a^\dag_1 \cdots a^\dag_N \rangle 
 \eeq
is expressed as a sum of terms obtained by replacing the two-particle overlaps
\beq 
f^{\nu'\nu}_{\mu'\mu} 
= \langle b^\nu b^{\nu'} a^\dag_{\mu'} a^\dag_{\mu}\rangle 
=\langle b^\nu a^\dag_{\mu}\rangle\langle b^{\nu'} a^\dag_{\mu'}\rangle
- \langle b^\nu a^\dag_{\mu '}\rangle\langle b^{\nu'} a^\dag_{\mu}\rangle ,
\eeq
in det$(d)$, one pair at a time, with 
$\mathcal{B}^{\nu'\nu}_{\alpha\beta}  \mathcal{A}^{\gamma\delta}_{\mu' \mu}$ ,
where
\beq
 \mathcal{B}^{\nu'\nu}_{\alpha\beta} = 
 \langle b^\nu b^{\nu'}c^\dag_\alpha c^\dag_\beta\rangle , \quad
  \mathcal{A}^{\gamma\delta}_{\mu'\mu} = 
 \langle c^\delta c^\gamma a^\dag_{\mu'} a^\dag_{\mu}\rangle .
 \eeq
  Then, if $M^{\mu'\mu}_{\nu'\nu}$ is the cofactor of $f^{\nu'\nu}_{\mu'\mu} $
in the determinant $\det(d)$, so that 
\beq 
\det (d) = \sum_{\mu' < \mu, \nu'<\nu}
f^{\nu'\nu}_{\mu'\mu} M^{\mu'\mu}_{\nu'\nu} ,
\eeq
the two-particle matrix element $X^{\gamma\delta}_{\alpha\beta}$
is given by
\beq
X^{\gamma\delta}_{\alpha\beta} = 
\sum_{\mu' < \mu, \nu'<\nu} \mathcal{B}^{\nu'\nu}_{\alpha\beta} 
M^{\mu'\mu}_{\nu'\nu} \mathcal{A}^{\gamma\delta}_{\mu'\mu} .
\eeq
 
Several other approaches have been reviewed by Sun \cite{Sun16}.
Most commonly, an approximation devised by Kamlah \cite{Kamlah68} has been used in practical calculations.

\section{Kinetic energy considerations} \label{sect:KE}
Numerous studies have been made of the decomposition of the many-nucleon kinetic energy into its collective and intrinsic components \cite{Villars57,Cusson68,%
ScheidG68,Ui70,Rowe70,VillarsC70,Zickendraht71,%
DzyublikOSF72,FilippovOS72,GulshaniR76,WeaverCB76,BuckBC79}.
These studies were motivated by the  expectation that the rotational energies of  strongly deformed nuclei should be predominantly kinetic energies, with corrections due to centrifugal and Coriolis perturbations, 
and the recognition that, in quantum mechanics, the kinetic energy of a many-nucleon nucleus is proportional to the Laplacian operator on its Hilbert space.
The decomposition to emerge trom these many studies was  summarised in a  simple 
but precise expression of the Laplacian on functions of the $3A$ many-nucleon 
coordinates  \cite{RoweR79} and gave the nuclear kinetic energy as a sum of three terms
\beq 
\hat T = \hat T_{\rm cm} +\hat T_{\rm coll}+\hat T_{\rm intr} :
\eeq
a centre-of-mass kinetic energy $\hat T_{\rm cm}$, 
a collective kinetic energy $\hat T_{\rm coll}$, 
and an intrinsic kinetic energy $\hat T_{\rm intr}$.

The collective component $\hat T_{\rm coll}$ of this  decomposition is expressed in terms of  three kinds of  momentum and corresponding mass parameters that are simple functions of the nuclear monopole and quadrupole moments.
Each of the three momenta has three components
and  together they are the infinitesimal generators of a group GL$(3,\Rb)$
of general-linear transformations.
Three components are those of the orbital angular momentum of the nucleus which are  infinitesimal generators of rigid-body rotations.
A second three are  infinitesimal generators of vortex rotations defined as circulations of the nuclear matter  in  the intrinsic frame of the nucleus that leave its quadrupole moments invariant.
The third three components are  infinitesimal generators of quadrupole deformation, likewise relative to the intrinsic axes, for which the quadrupole moments remain diagonal.
Thus, as discussed in the following section, the above decomposition contributed to the emergence of an algebraic theory of nuclear collective states \cite{WeaverB72}
with quadrupole vibration, rigid rotation and vorticity degrees of freedom.
The irreps of this model \cite{RosensteelR76,WeaverCB76} were shown to have both standard and vortex angular-momentum quantum numbers.

\section{Many-nucleon algebraic models} \label{sect:AlgModel}
As mentioned in the introduction, the  simplicity of many-nucleon quantum mechanics, relative to what it might have been,
 follows from the fact that it is an algebraic model with a Hilbert space given for each nucleus by a fully anti-symmetric tensor product of single-nucleon spaces.
As a result, the Hilbert space of a nucleus has a basis of fully antisymmetric independent-particle states;
this  makes it possible, to exploit the facility of large computers to work with huge bases of many-fermion states defined in a binary occupation-number representation
\cite{Whitehead72,WhiteheadWCM77}.
A nuclear Hilbert space is also a  product of two subspaces: a subspace of centre-of-mass states and a complementary subspace of states with no centre-of-mass degrees of freedom.
Another important property is that the nuclear Hilbert space is  invariant and
irreducible under all one-body unitary transformations of its basis states.
Consequently, essentially all algebraic models of nuclei, that have dynamical groups expressed in terms of one-body unitary transformations, have irreps on subspaces of  many-nucleon Hilbert spaces.

The most  useful algebraic models  define physically meaningful basis states for the many-nucleon Hilbert space of a nucleus that are classified by the quantum numbers of their irreps  and those of their submodels.  
A desirable property of an algebraic model of a nucleus is also that it has unitary irreps which leave the centre-of-mass states of the nucleus invariant.
Such models define coupling schemes for more complete calculations in spaces spanned by a number of its irreps,  and thereby provide physical interpretations of the results of such calculations.

\subsection{First attempts}
The search for an algebraic  collective model as a submodel of many-nucleon quantum mechanics was initiated in 1955 by Tomonaga \cite{Tomonaga55}
in two dimensions and extended by Miyazima and Tamura \cite{MiyazimaT56} to three. 
These searches identified quadrupole-moment operators as infinitesimal generators of irrotational flows and prepared the way for subsequent developments.
However, they did not succeed in identifying a closed subalgebra of the many-nucleon algebra of one-body operators for an algebraic model of nuclear collective states.

\subsection{The Elliott SU(3) model} \label{sect:SU3model}
The  many-nucleon  algebraic model of nuclear rotational dynamics,
given by  Elliott's SU(3) model \cite{Elliott58ab}, 
has been particularly influential in demonstrating the value of  a relevant coupling scheme for identifying collective subdynamics  of the nuclear shell model.
It also provided an alternative to the shell-model coupling schemes of  Flowers and Edmonds \cite{Flowers52,EdmondsF52a,EdmondsF52b} and showed how states with rotational properties could emerge within the framework of the nuclear shell model.
However, it was not the algebraic model of nuclear rotational states that was being sought.
For, although the SU(3) Lie algebra contains angular momentum and quadrupole moment operators expressed in terms of nucleon coordinates and momenta,
its quadrupole moments are not the physical quadrupole moments of nuclei; 
they are  only their  spherical harmonic-oscillator energy-conserving components.  
Nevertheless, the SU(3) model provides an important effective shell model of rotations in light nuclei.

In retrospect, the remarkable successes of the SU(3) model in obtaining shell-model states with properties close to those of a rotor model can be attributed  to the fact that an SU(3) model Hilbert space is the projected image of a map from a rigid-rotor model space onto that of a single spherical harmonic-oscillator shell.
This follows from the observation that the SU(3) quadrupole moments are the restrictions of physical quadrupole moments to the space of a single spherical harmonic-oscillator shell.
Thus, it fits  into the Lee-Suzuki construction \cite{LeeS80}
of an effective shell model of nuclear rotations.
However,  as  an effective shell model of a rotor, the SU(3) model is unable to give any information about the dynamics of physical nuclear rotations.
It has nevertheless proved to be an important sub-model of the desired algebraic collective model that was being sought.
Most significantly, it provides a coupling scheme for large shell model calculations 
\cite{DytrychSBDV08,DytrychLDMVSCSLC13,DreyfussLDDB13}
in spaces that include many SU(3) irreps.

\subsection{The Ui Rot(3)  model}
An algebraic model similar to the SU(3) model, but which contains  the physical quadrupole moments in addition to the angular momentum operators, was  introduced by Ui  \cite{Ui70}.
It is  a genuine rotor model and is  useful because  its representation theory provides a rigorous and systematic procedure for constructing a rigid-rotor model with intrinsic symmetries.  
It was also  a  vital step towards the objective of understanding  nuclear rotational dynamics. 
In particular, it defines an algebraic rotor model for which 
Elliott's SU(3) model is its projection onto an algebra that leaves a spherical harmonic-oscillator Hamiltonian invariant.
Its limitation is that its irreps contain eigenstates of the quadrupole-moment operators which are delta-functions and can only be realised, in the many-nucleon Hilbert space, as non-normalisable limits.

\subsection{The Weaver-Biedenharn-Cusson GCM(3) model  }
A next major step  was the construction  by Weaver, Biedenharn and Cusson \cite{WeaverB70,WeaverBC73} of an algebraic collective model with both rotational and vibrational degrees of freedom.
The unitary irreps of this model can be seen in retrospect as many-nucleon versions of an algebraic expression  \cite{Rowe04a,RoweWC09,RoweWood10,WelshR16} of the Bohr model \cite{Bohr52}
and its extension to include quantised vorticity.

First observe that  a many-nucleon version of the Bohr model 
starts with the replacement of  its surface shape coordinates $\{ \alpha_\nu\}$  by microscopic Cartesian quadrupole moments 
\beq Q_{ij} = \sum_{n=1}^A x_{ni} x_{nj} , \quad i,j = 1,2,3, \;\; n=1,\dots, A. \eeq
(The original CM(3)  model of Weaver \emph{et al.}\  did not include the monopole moment.
However, its inclusion is natural and avoids the necessity of assuming the nuclear fluid to be incompressible.)
Time derivatives of the quadrupole moments and corresponding momentum observables are   given by
\bal 
&\dot Q_{ij} = \frac{dQ_{ij}}{dt} =\sum_n (\dot x_{ni} x_{nj} + x_{ni} \dot x_{nj}) ,  \\
&P_{ij} = M \dot Q_{ij} =\sum_n ( p_{ni} x_{nj} + x_{ni} p_{nj}), \label{eq:7.P&Q}\end{align}
where $M$ is the nucleon mass.
An appropriate quantisation of the Bohr model with these observables is 
then obtained in the standard way by replacing the nucleon
coordinates, $x_{ni}$ and $p_{ni}$,  by operators
$ \hat x_{ni}$ and $\hat p_{ni}$ with commutation relations
$[\hat x_{ni}, \hat p_{mj}] = {\rm i}\hbar \delta_{i,j} \delta_{m,n} $, to give  quantal shape and momentum observables
\beq \hat Q_{ij} :=\sum_n \hat x_{ni} \hat x_{nj} , \quad
\hat P_{ij} =  \sum_n ( \hat p_{ni}\hat x_{nj} +\hat x_{ni} \hat p_{nj}), \label{eq:7.hatQhatP}
\eeq
which satisfy commutation relations
\beq [\hat  Q_{ij}, \hat P_{kl}] =  {\rm i}\hbar \big( \delta_{il}\hat Q_{jk} +  \delta_{ik}  \hat Q_{jl} + \delta_{jl} \hat  Q_{ik} + \delta_{jk}  \hat Q_{il}  \big) . \label{eq:QPcr} \eeq
Together with the angular-momentum operators
\beq \hat L_{ij} = \sum_n \big(\hat x_{ni}\hat p_{nj}  -\hat x_{nj}\hat p_{ni}\big) , \eeq
 these operators  span the so-called GCM(3) Lie algebra of a generalised collective model.
 
  An interesting result that emerges is that whereas  the standard  Bohr model 
\cite{Bohr52}  and  its later algebraic collective model version \cite{RoweT05,RoweWC09,WelshR16} have single unitary irreps, the many-nucleon GCM(3) model has many.
They  were derived by Rosensteel \cite{RosensteelR76} and again by Weaver \emph{et al.}\  \cite{WeaverCB76} and shown to be characterised by quantised vorticity.
Thus, only the zero-vorticity irrep corresponds to the  irrotational-flow Bohr model.

In addition to being a microscopic version of an extended Bohr model and, to a large extent the Bohr-Mottelson \cite{BohrM53,BohrM75} unified model,
the GCM(3) model has the desirable characteristic of containing all the physical observables, i.e., quadrupole moments, standard angular momentum, vortex spin, and infinitesimal generators of deformation, that appear in the expression of the collective component $\hat T_{\rm coll}$ of the many-nucleon kinetic energy.
It is also related, as considered by Rosensteel 
\cite{Rosensteel88,Rosensteel93,SparksR17}, to the Riemann model of rotating  ellipsoids \cite{Chandrasekhar69} with linear combinations of  rigid and irrotational flows and a mathematical structure in terms of Yang-Mills theory,  as given by Rosensteel and Sparks   \cite{RosensteelS15,SparksR17}.

A problem with the GCM(3) model is that it is difficult to use in a calculation of nuclear properties with a many-nucleon Hamiltonian.
This is because bases for its unitary irreps are not easily constructed in terms of many-nucleon states.
Thus,  it does not lead in a practical way, to the construction of a coupling scheme for the many-nucleon Hilbert space.
A more serious concern, is that the kinetic-energy operator and results from subsequent developments  indicate that the vortex spin  of the GCM(3) model is not conserved
in the rotational states of nuclei.

\subsection{The symplectic Sp(3,$\Rb$) model}
A resolution of the problems with the GCM(3) model is  obtained
\cite{RosensteelR77a,RosensteelR80} by simply extending it
to a symplectic  Sp$(3,\Rb)$ model which includes the 
full many-nucleon kinetic energy $\hat T$ in its Lie algebra of observables.
The Lie algebra of collective observables is thereby extended to
the Lie algebra of all  bilinear combinations of the nucleon position and momentum  coordinates
\bal
& \hat Q_{ij} = \sum_{n=1}^A \hat x_{ni} \hat x_{nj} , \quad 
\hat P_{ij} =  \sum_{n=1}^A ( \hat x_{ni} \hat p_{nj} + \hat p_{ni} \hat x_{nj}),
\label{eq:8.QPL} , \\
&\hbar \hat L_{ij} 
=  \sum_{n=1}^A \big(\hat x_{ni}\hat p_{nj}  -\hat x_{nj}\hat p_{ni}\big), \quad
\hat K_{ij}= \sum_{n=1}^A  \hat p_{ni}  \hat p_{nj} , \label{eq:8.K} 
\end{align}
that are symmetric with respect to nucleon permutations.

This Sp$(3,\Rb)$ Lie algebra (sometimes referred to as Sp$(6,\Rb)$) 
is the smallest Lie algebra that  contains both the 
nuclear quadrupole moments and the many-nucleon kinetic energy.
It nevertheless contains all the  algebras of the above  models as subalgebras.
In particular, it contains the U(3) Lie algebra of the Elliott model as a subalgebra
and has the valuable property that it defines a coupling scheme 
in a U(3) $\supset$  SU(3) basis for the many-nucleon Hilbert space in a straightforward way.
The Sp$(3,\Rb)$ Lie algebra, like that of U(3) can be augmented to include the U(4)
supermultiplet spin-isospin algebra with which it commutes.
It can also be augmented to an Sp$(6,\Rb)$ (alias Sp$(12,\Rb)$) Lie algebra so that it can more readily describe  independent collective motions of the neutrons and protons; 
cf., Ref.\ \cite{Ganev14} and references therein.

The Sp$(3,\Rb)$ symplectic group is  of fundamental importance for many reasons.
It can be defined, for example, as the group of linear canonical transformations that leave invariant the  commutation relations 
\beq 
[\hat x_{ni}, \hat p_{mj}] = {\rm i} \delta_{m,n} \delta_{i,j}
\eeq
of the Heisenberg algebra.
Moreover, whereas the U(3) group of the Elliiott model is a  symmetry group of a three-dimensional spherical harmonic oscillator, the  group Sp$(3,\Rb)$, 
which contains U(3) and SU(3) as  subgroups, is a dynamical group that leaves invariant the Hilbert space of a general (not necessarily spherical) three-dimensional harmonic oscillator of a given parity.
The  Sp$(n,\Rb)$ symplectic groups for any integer $n$  have been well studied both in mathematics and physics.
In particular, the Sp$(n,\Rb)$  irreps, on spaces of $n$-dimensional harmonic-oscillators, have  been constructed \cite{Rowe84,RoweRC84,DeenenQ84}
and the U(n) irreps contained in an Sp$(n,\Rb)$ irrep have been determined
\cite{RoweWB85}.
The GCM(n) irreps contained  in an Sp$(n,\Rb)$ irrep have also  been determined
\cite{RoweR98}; it is shown, for example, that the range of vorticities in an 
Sp$(3,\Rb)$ irrep is equal to the range of angular momenta in its lowest-grade SU(3) irrep.

Because the elements of the Sp$(3,\Rb)$ Lie algebra are symmetric with respect to nucleon permutations, they conserve the so-called space symmetry of nuclear states.
And because they are bilinear combinations of the nucleon position and momentum coordinates, they are expressible as sums of their centre-of-mass and relative components, without any coupling terms, by the simple substitutions
\beq 
\hat x_{ni} = \hat x'_{ni} + \hat X_i , \quad \hat p_{ni} = \hat p'_{ni} + \frac1A \hat P_i ,
\eeq
where $\hat X_i = \frac1A \sum_n \hat x_{ni}$ and 
$\hat P_i = \sum_n \hat p_{ni}$ are centre-of-mass coordinates for the nucleus.
With this substitution, the quadrupole moments, for example, are the sums
\beq
\hat Q_{ij} = \hat Q_{ij}({\rm rel}) + \hat Q_{ij}({\rm cm}).
\eeq
In practical applications, the spurious centre-of-mass components of these operators are generally removed.
However, in the interest of simplicity, this technicality is ignored in the following.
In applications to heavy nuclei it is, in any event, a relatively minor correction due largely to the fact that the presence of the spurious contribution of the centre-of-mass degrees of freedom  does not change the Lie algebra commutation relations in any way.

\section{Applications of the symplectic model} 
\label{sect:Sp3Rcoup}
The symplectic model is an algebraic model of the rotations and collective monopole-quadrupole vibrations of nuclei  which has the additional vital property that its representations are well-defined on  the many-nucleon Hilbert spaces of nuclei.
As an algebraic model, it can be used phenomenologically 
 with a Hamiltonian expressed in terms of its Lie algebra of observables
and an irrep that best fits the low-energy properties of a nucleus.
This provides insights into the microscopic structures of  collective states.
At a more fundamental level, use can be made  of the fact that the Hilbert space of a nucleus is a sum of irreducible symplectic-model subspaces to gain an understanding of the emergence of collective states and their properties in nuclei.
Thus, as reviewed in Section \ref{sect:CMsubspaces}, 
an energy-ordering of the symplectic-model subspaces of the many-nucleon Hilbert space assists in the identification of the  relevant Sp$(3,\Rb)$ irreps for describing the low-energy states of nuclei and for defining a  coupling scheme 
 that can  be used, for example, to explore the extent to which symplectic-model irreps in nuclei are mixed in multi-shell model calculations.

Huge multi-shell model calculations with realistic interactions have been possible for some 
time in a ${\rm U(3)} \times {\rm SU(2)}_S \times {\rm SU(2)}_T$-coupled basis in light nuclei
\cite{DytrychSDBV08b,DytrychLDMVSCSLC13,LauneyDD16,DraayerDL17}
with strong implications for their Sp$(3,\Rb)$ content. 
Recent developments \cite{Dytrych17} now enable such calculaltions to be carried out
in an $[{\rm Sp}(3,\Rb)\supset {\rm U(3)}] \times {\rm SU(2)}_S \times {\rm SU(2)}_T$ basis.
However,  and not surprisingly, there are  limitations to what is  feasible with the  available computational resources in heavy nuclei.  
We therefore  seek ways to circumvent these limitations. 
This preliminary section focuses on the properties of single irreps of the symplectic model  that are learned from the calculations that have been done.

\subsection{Unitary irreps of the Sp(3,$\Rb$) Lie algebra}
\label{sect:sp3R.basis}
The irreps of  Sp$(3,\Rb)$ of relevance to nuclear physics are defined in  many-nucleon spherical harmonic-oscillator bases by expressing its Lie algebra in terms of harmonic-oscillator raising and lowering operators for which the  nucleon coordinates are given by
\beq \hat x_{ni} = \frac{1}{\sqrt{2}\,a} (c^\dagger_{ni}+ c_{ni} ) , \quad
\hat p_{ni} = {\rm i} \hbar\frac{a}{\sqrt{2}} (c^\dagger_{ni} - c_{ni} ) ,
\label{eq:xpccdag} \eeq
where $a  = \sqrt{M\omega/\hbar}$ is a unit of inverse length.
This leads to the expansions
\bal 
\begin{split}
&  \hat Q_{ij} =
 \frac{1}{2a^2}\big(2\hat{\mathcal{Q}}_{ij}  + \hat {\mathcal{A}}_{ij} + \hat {\mathcal{B}}_{ij} \big), \\
& \hat K_{ij} = \tfrac12 a^2 \hbar^2 (2 \hat{\mathcal{Q}}_{ij} 
- \hat {\mathcal{A}}_{ij} - \hat {\mathcal{B}}_{ij} ) , \\
& \hat P_{ij} = {\rm i}\hbar (\hat {\mathcal{A}}_{ij} - \hat {\mathcal{B}}_{ij}) ,
\quad L_{ij}  = -{\rm i} ( \hat {\mathcal{C}}_{ij} - \hat {\mathcal{C}}_{ji} ) , 
\end{split}
\end{align}
in which
\bal  \begin{split}
& \hat {\mathcal{A}}_{ij} = \hat {\mathcal{A}}_{ji} = \sum_n c^\dagger_{ni} c^\dagger_{nj} ,\quad
  \hat {\mathcal{B}}_{ij} = \hat{\mathcal{B}}_{ji} =\sum_n c_{ni} c_{nj} , \\
&\hat {\mathcal{C}}_{ij} =  \sum_n \big( c^\dagger_{ni} c_{nj} 
+ \textstyle\frac12 \delta_{i,j}\big), \quad
\hat{\mathcal{Q}}_{ij} = \tfrac12\big( \hat {\mathcal{C}}_{ij} + 
\hat {\mathcal{C}}_{ji} \big) ,\end{split}
  \label{eq:8.ABCQops}
\end{align}
with $n$ summed over the effective number of $A-1$ nucleons, with exclusion of  the linear combinations that  involve the nuclear centre-of-mass degrees of freedom.
(Note that removal of the centre-of-mass contributions to the  symplectic-model observables is important, especially for applications in light nuclei. However, it is easily accomplished and  does not change the commutation relations of the Sp$(3,\Rb)$ Lie algebra in any way.
Thus, for pedagogical purposes, it can be regarded as a minor technical adjustment.)

It is seen that the Sp$(3,\Rb)$ Lie algebra contains the elements 
$\{ \hat L_{ij}\}$ and $\{ \hat{\mathcal{Q}}_{ij}\}$ of a U(3) subalgebra and
 the giant monopole/quadrupole raising and lowering operators
$\{ \hat {\mathcal{A}}_{ij}\}$ and $\{ \hat {\mathcal{B}}_{ij}\}$.
The elements of the U(3) subalgebra commute with the spherical harmonic-oscillator Hamiltonian
\beq 
\hat{\mathcal{H}}_0 = \hbar\omega_0 
\sum_n \big( c^\dagger_{ni} c_{nj} + \tfrac12 \delta_{i,j}\big),
\eeq
and  the giant-resonance raising and lowering operators satisfy the commutation relations
\beq [\hat{\mathcal{H}}_0, \hat {\mathcal{A}}_{ij}] 
= 2\hbar\omega  \hat {\mathcal{A}}_{ij} , \quad
[\hat{\mathcal{H}}_0, \hat {\mathcal{B}}_{ij}] 
= -2\hbar\omega  \hat {\mathcal{B}}_{ij} .
\eeq
It follows that an irreducible Sp$(3,\Rb)$ representation
is uniquely defined by a so-called lowest-grade U(3) irrep, 
 the states of which  are annihilated by the giant-resonance lowering operators $\{ \hat {\mathcal{B}}_{ij}\}$.
The U(3) highest-weight state $|\sigma\rangle$ of this lowest-grade U(3) irrep is then the state that satisfies the equations
\bal 
&\hat {\mathcal{C}}_{ij }|\sigma\rangle = 0 , \;\; i <j , \quad 
\hat {\mathcal{C}}_{ii }|\sigma\rangle =\sigma_i |\sigma\rangle ,
\;\; i=1,2,3, \label{eq:C_ij|0>} \\
&\hat {\mathcal{B}}_{ij }|\sigma\rangle = 0, \quad i,j =1,2,3 ,
\label{eq:B_ij|0>}
\end{align}
and is  a state of U(3) weight $\sigma = (\sigma_1,\sigma_2,\sigma_3)$ given by the triple of  integers or half-odd integers ordered such that
\beq \sigma_1\geq \sigma_2\geq \sigma_3\geq 0. \eeq
Because  the state $|\sigma\rangle$ is annihilated by the Sp$(3,\Rb)$ lowering operators in Eqn.\ (\ref{eq:B_ij|0>}), it is also conveniently regarded as the lowest-weight state for the Sp$(3,\Rb)$ irrep to which it belongs.
Thus,  $|\sigma\rangle$ is simultaneously a highest-weight state for a U(3) irrep denoted by $\{ \sigma_1,\sigma_2,\sigma_3\}$ and a lowest-weight state for an Sp$(3,\Rb)$ irrep denoted by $\langle \sigma_1,\sigma_2,\sigma_3\rangle$.
It  follows that a basis of shell-model states for an Sp$(3,\Rb)$ irrep is  constructed by adding to its lowest-grade U(3) states the infinite set of multiple giant-resonance excitations generated by the 
 $\{ \hat {\mathcal{A}}_{ij}\}$ raising operators.
 
An efficient  VCS (vector coherent state) algorithm for the construction of the states and matrix elements of an Sp$(3,\Rb)$ irrep  \cite{Rowe84,RoweRC84} is described  in some detail in a recent review \cite{RoweMC16}.
Computer codes for its implementation, have been developed by Rosensteel \cite{RoweRC84}, Bahri \cite{BahriR00} and McCoy \cite{RoweMC16}. 
The VCS algorithm  makes  use of the readily available Clebsch-Gordan and other coefficients for coupling and recoupling states of SU(3) representations  \cite{AkiyamaD73,DraayerA73,BahriRD04} and the fact that the raising and lowering operators 
$\hat {\mathcal{A}}_{ij }$ and $\hat {\mathcal{B}}_{ij }$ 
are simply related to boson creation and annihilation operators as in the Dyson representation \cite{Dyson56} of  SU(2) operators.

\subsection{Calculations with schematic algebraic interactions}
\label{sect:BahriRmodel}
\begin{figure}[phtb]
\centerline{\includegraphics[width=4 in]{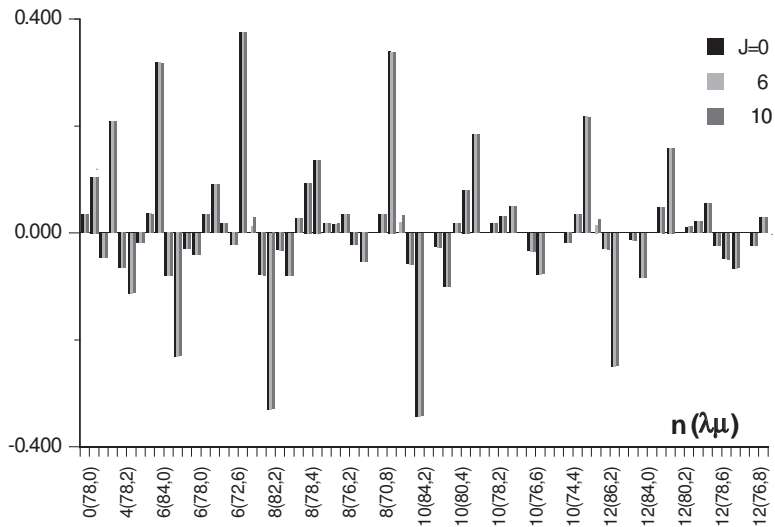}}
\caption{ \label{fig:166Erwfns} 
Amplitudes for the wave functions of the lowest-energy
$L=0$, 6, and 10, Sp$(3,\Rb)$ model states of $^{166}$Er in a U(3) basis.  
The calculation included all basis states of the irreducible 
Sp$(3,\Rb)$ representation $\langle 826.5(78,0)\rangle$ of spherical  harmonic-oscillator energy 
$\leq 12 \hbar\omega$ above and including that of its lowest U(3) states.
Particularly important is the observation that the lowest U(3) states for this representation are already at $8 \hbar\omega$, in spherical harmonic-oscillator units of energy, above the lowest available in spherical harmonic-oscillator shells.  
Note that the $+$ or $-$ signs of the amplitudes are of no significance because they are defined by the arbitrary signs of the basis states.
(The figure is adapted from one of Bahri \cite{BahriR00}.)
}
\end{figure}

Many early symplectic model calculations were reviewed in 1985 \cite{Rowe85}.
A  recent calculation of Bahri  \cite{BahriR00} shows that, 
with a simple Hamiltonian 
\beq \label{eq:BahriPot}
\hat H(\epsilon) = \hat H_0 + \chi \left( \hat Q_2 \cdot\hat Q_2 
+ \frac{\epsilon}{\hat Q_2 \cdot\hat Q_2} \right) ,
\eeq
expressed in terms of the Sp$(3,\Rb)$ Lie algebra in which
$\hat Q_2$ is the nuclear quadrupole tensor,
it is possible to obtain close-to-converged solutions for the low-energy rotational states of $^{166}$Er.
To within corrections that could be due to overly-suppressed 
centrifugal  effects by the potential-energy term of $\hat H(\epsilon)$,
remarkably good  fits were obtained with this Hamiltonian for the relative  energies of the lowest rotational states of this nucleus, up to angular momentum $J=16$, and for the E2 transitions between them.
For these calculations, a  symplectic irrep $\langle\sigma\rangle$ was 
chosen such that the quadrupole  moments and E2 transition data could be fitted without the use of an effective charge.
A  significant  result of this calculation, shown by the wave functions  in Fig.\ \ref{fig:166Erwfns}, 
is that the U(3) irreps contributing to the rotational states that emerged from the calculation were from spherical harmonic-oscillator shells  of energies ranging from $8\hbar\omega$ to $20\hbar\omega$  above that of the lowest available for this $^{166}$Er nucleus.
This is a clear indication that it is unreasonable to attempt a conventional shell model description of such rotational states in heavy deformed nuclei.
The  Bahri model  provides a simple and practical prescription for describing the rotational states of  axially symmetric nuclei.
It has two parameters, $\epsilon$ and $\chi$.
The parameter $\epsilon$ defines the deformation at which the potential is a minimum and the strength $\chi$ of the interaction can be adjusted to ensure that its eigenstates have the same deformation and  therefore give the corresponding E2 transitions between its eigenstates.

A notable  result is that the amplitudes of the U(3) basis states, displayed in the figure 
for three angular-momentum states,  are essentially identical.
An interpretation of this result is that, whereas the rotational states  of a single U(3) irrep can be obtained by angular-momentum projection from its highest-weight states, the rotational states 
of the Sp$(3,\Rb)$ irrep of the model can similarly be projected, to a high level of accuracy, 
from a single intrinsic state given by a linear combination of the U(3) highest-weight states in a manner similar to that of a quasi-dynamical symmetry  \cite{RoweRR88,RoweCancun04}.
This interpretation is derived  in the following from an algebraic mean-field perspective.
Thus, although the lowest-energy SU(3)-model components of the calculated states, which are those of the SU(3) (78,0) irrep at a harmonic-oscillator shell model excitation energy of 
$8 \hbar\omega$,  are shown in the figure to be exceedingly small, it can nevertheless be understood why the SU(3) model can work as well as it does  as an effective model, with a suitable effective interaction and effective charge.
However, it must be emphasised that the lowest U(3) irrep that contributes to the results shown belongs to a spherical harmonic-oscillator shell that is 8  shells above those that would be considered in a standard spherical shell-model calculation.  
Thus, it is an example of shape co-existence.

A limitation of the Hamiltonian (\ref{eq:BahriPot}) is that it applies only to axially symmetric representations and it makes no allowance for the possible centrifugal stretching of  a rotating nucleus with increasing angular momentum which would appear to be in evidence in the experimental spectrum of $^{166}$Er.
However, the latter deficiency could be remedied to some extent by allowing the deformation parameter to be a function of the angular momentum.

Another model with an algebraic interaction,  that is closer to one with a conventional two-nucleon interaction and  applies to triaxial as well as axially symmetric nuclei,
has a Hamiltonian of the form
\beq \label{eq:Dreyfuss}
\hat H(\gamma) = \hat H_0 + \frac{\chi}{2\gamma} 
\left( e^{-\gamma \hat Q_2 \cdot\hat Q_2 }-1 \right) ,
\eeq
which, for an axially symmetric irrep, can be diagonalised by an adjustment of Bahri's code.
It was proposed and applied by Dreyfuss \emph{et al.}\ \cite{DreyfussLDDB13} to the rotational states based on the Hoyle state of $^{12}$C and by Tobin \emph{et al.} \cite{TobinFLDDDB14} to the states of a few $sd$-shell nuclei.
The parameter $\chi$ of this Hamiltonian was fixed by a self-consistency argument and the parameter $\gamma$  was adjustable.
Good agreement for the  rotational band of states based on the Hoyle state was then obtained with an Sp$(3,\Rb)$ irrep having a (12,0) SU(3) lowest weight
of spherical harmonic-oscillator energy  $4\hbar\omega$ above that of the lowest available for the $^{12}$C nucleus and with wave functions containing components  from spherical harmonic-oscillator shells  of excitation energies up  to $16\hbar\omega$.

\begin{figure}[phbt]
\centerline{\includegraphics[width=6.0 in]{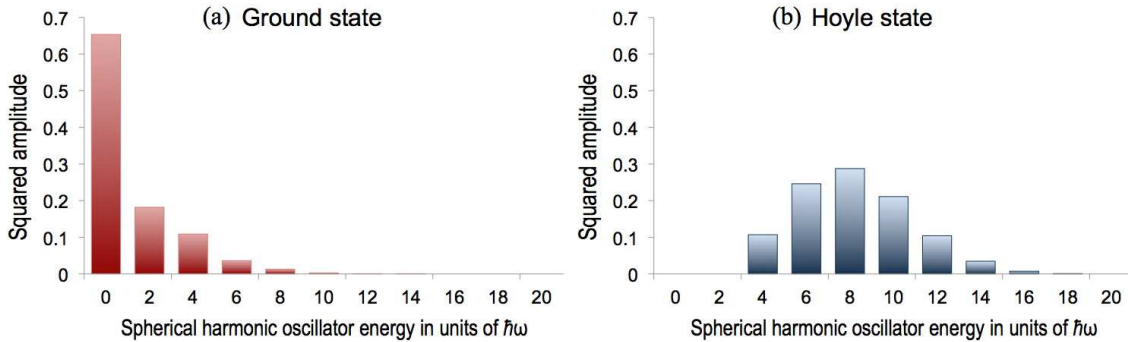}}
\caption{Squared amplitudes of the wave functions of the 
ground and  excited $0^+$ (Hoyle) state of $^{12}$C  as functions of their spherical harmonic-oscillator excitation energies relative to the lowest available for $^{12}$C.
(These results were calculated by 
Dreyfuss \emph{et al,}\ \cite{DreyfussLDDB13}
and the figure was provided by K.\ Launey.)}
\label{fig:Dreyfuss}
\end{figure}

The decomposition of the Hoyle state, calculated in terms of spherical harmonic-oscillator states and shown in figure \ref{fig:Dreyfuss}(b),
is similar to that for the rare-earth rotational states shown in figure \ref{fig:166Erwfns}.
Both  figures show the lowest-grade U(3) states of the symplectic irrep to be small components of the total.
This is in contrast to that shown for the much less deformed ground state of $^{12}$C
shown in figure \ref{fig:Dreyfuss}(a).
Another significant observation is that the Hoyle state is observed and calculated to be at an energy $\sim 8$ MeV which is very much less than the $4\hbar\omega$ spherical harmonic-oscillator energy of the  lowest-weight state that this Sp$(3,\Rb)$ calculation employed.

\subsection{ Sp$(1,\Rb)$ and Sp$(2,\Rb)$ submodels} \label{sect:ArickxHP}
In parallel with the development of the  Sp$(3,\Rb)$  model, a so-called Sp$(1,\Rb)$ model was introduced by Arickx \cite{Arickx76} founded on the observation
(subsequently confirmed in large multi-shell model calculations in U(3) bases \cite{LauneyDD16}) that the excited states which couple  most strongly
to those of a  U(3) irrep of highest weight
$( \sigma_1,\sigma_2,\sigma_3)$ are those of U(3) irreps with highest weights
\beq \label{eq:sp1.wts}
( \sigma_1 +2\nu_1,\sigma_2,\sigma_3), \quad \nu_1 = 1,2,3, \dots .
\eeq
Arickx also recognised that the highest-weight states for these U(3) irreps, 
are a basis for a unitary irrep of a group Sp$(1,\Rb)$  that is isomorphic to  SU(1,1) and  referred  to by him as Sp$(2,\Rb)$.
[Note that some physicists denote the real non-compact symplectic group Sp$(n,\Rb)$ of rank $n$ by Sp$(2n,\Rb)$ because its defining matrix representation is $2n$ dimensional.]

The next most strongly-coupled states in an Sp$(3,\Rb)$ irrep were subsequently observed  
by Peterson and Hecht \cite{PetersonH80} to be those of the U(3) irreps 
\beq \label{eq:sp2.wts}
\{ \sigma_1 +m,\sigma_2 +n,\sigma_3\}, \quad m+n = 0,2,4, \dots ,
\eeq
with $ \sigma_1 +m\geq \sigma_2 +n$. 
It was also recognised that this subset of  states of an Sp$(3,\Rb)$ irrep 
 $\langle\sigma_1,\sigma_2,\sigma_3\rangle$, which have weights with a common  
 value of $\sigma_3$, span an Sp$(2,\Rb)$ irrep.

\subsection{Applications with interacting-nucleon Hamiltonians}
Early attempts, pursued by Filippov \cite{FilippovO80,OkhrimenkoS81}, Vassanji  \cite{VassanjiR82,VassanjiR83,VassanjiR86,CarvalhoR97} and colleagues, 
at diagonalising a conventional many-nucleon Hamiltonian
within the space of an Sp$(3,\Rb)$ irrep, were based on  generator coordinate methods.
Such an approach was used by Elliott and Harvey \cite{Elliott58ab,ElliottH63} for calculations in the space of an SU(3) irrep.
The latter made use of the fact that the Hilbert space of a U(3) irrep is spanned by the states generated by SO(3) rotations of its highest-weight state 
$|\sigma\rangle \equiv |\sigma_1,\sigma_2,\sigma_3\rangle$.
Expressions for SU(3) basis states were then obtained as integrals over the rotated highest-weight state with the rotational angles serving as generator coordinates.
It is similarly observed, in Section \ref{sect:11}, that the Hilbert space of an Sp$(3,\Rb)$ irrep is spanned by the states generated by GL$(3,\Rb)$ transformations of the same state 
$|\sigma\rangle$, which is both a highest-weight state for a U(3) irrep and a lowest-weight state for an Sp$(3,\Rb)$ irrep.
This  approach was discontinued, but now 
because of its close relationship with the  algebraic mean-field theory given below, 
it appears to provide the solution not only for handling the problem of a single 
Sp$(3,\Rb)$ irrep but also that of mixed Sp$(3,\Rb)$ irreps.
Thus, it is revisited in Sections \ref{sect:GCM1} and \ref{sect:GCT}.

A different  strategy,  initiated by Escher and Draayer \cite{EscherD98} for the explicit derivation of many-nucleon wave functions for  symplectic model states, was to  start with an expression of the Sp$(3,\Rb)$ Lie algebra in terms of  nucleon creation and annihilation operators.
This potentially promising approach also does not appear to have been pursued, possibly because of the successes of the  symmetry-adapted no-core shell model (SA-NCSM).
However, it may yet prove to have important applications.

\subsection{Emergence of Sp$(3,\Rb)$ symmetry in no-core shell model
(NCSM) calculations} \label{sect:LSUprog}
The most sophisticated and most successful approach to date to understanding the emergence of collective rotational states within the framework of many-nucleon quantum mechanics,
is that of the  symmetry-adapted no-core shell model (SA-NCSM)
of the LSU research group of Draayer, Dytrych, Launey and colleagues \cite{LauneyDD16,DraayerDL17}; see also Refs.\
\cite{DytrychSBDV07a,DytrychSBDV07b,DytrychSBDV08,%
DytrychSDBV08b,DraayerDLL12}.
This model was introduced to expose the physical content of the
NCSM calculations of Navratil \emph{et al.}\
\cite{NavratilVB00b,NavratilO03,NavratilGVON07,NavratilQSB09,MarisSV09},
by expressing them in a symmetry-adapted basis.

The NCSM makes use of the massive parallel computing resources of supercomputers  and a simple $M$-scheme basis 
\cite{Whitehead72,WhiteheadWCM77}
to enable huge shell-model calculations to be performed in light nuclei with realistic interactions  derived from quantum chromodynamics \cite{MachleidtE11}
and  nucleon-nucleon interaction data \cite{ShirokovVMW07}.
These calculations are of fundamental importance in establishing the foundations of the many-nucleon quantum theory of nuclei with such interactions.
The  SA-NCSM calculations are similar, in principal, but
are carried out in  bases of states classified by 
${\rm U(3)} \supset {\rm SO(3)}$ and intrinsic neutron and proton spins.
This was  possible because of previous developments in the use of SU(3) and its tensor properties 
\cite{AkiyamaD73,DraayerA73,BahriD94}.

SA-NCSM calculations have been carried out in complete shell-model spaces of up to six  spherical harmonic-oscillator shells for the light nuclei so far considered.
A remarkable result is that  the eigenstates that  emerge contain significant contributions from only a tiny fraction of the huge number of   U(3) irreps, with corresponding neutron and proton spin wave functions,  that were included in the calculations.
Even more remarkable is the finding that most of the U(3) states that make  up the low-energy eigenstates of a nucleus in these calculations can be identified with those of a single Sp$(3,\Rb)$ irrep at the $60- 80\%$ level, while  the remaining components belong  to just one or two other Sp$(3,\Rb)$ irreps.  
For doubly even nuclei, the dominant components are 
Sp$(3,\Rb)$ states coupled to states of zero intrinsic neutron and proton spins.
Thus, the LSU results indicate the emergence of 
Sp$(3,\Rb)$ as a dominant dynamical symmetry in calculations that were 
not prejudiced by the assumption that this should be the case; i.e.,
they were carried out in a U(3)-coupled basis purely to facilitate an interpretation of the results.
The results for light nuclei are especially remarkable because the symplectic model 
\cite{RosensteelR77a,RosensteelR80} was expected to be most successful
as a microscopic collective model for heavy rotational nuclei.
 Thus,  its relevance to rotational states in light nuclei 
was  not  expected to be anywhere near as dominant as it has proved to be.

\section{Algebraic mean-field (AMF) theory} \label{sect:AMFtheory}
At present it is not feasible to carry out realistic SA-NCSM calculations in heavy deformed nuclei. The major obstacle, as Figure \ref{fig:166Erwfns} illustrates,
is that the rotational states of heavy nuclei have expansions
in a standard  shell-model basis of states with spherical harmonic-oscillator energies ranging between $\sim 6 \hbar\omega$ and $\sim 26 \hbar\omega$, in spherical  
harmonic-oscillator shell-model units  above those  that would normally be considered in a shell-model calculation.
Thus, for anything like a meaningful shell-model calculation of the usual type, the dimension of the active valence-shell space would need to be  orders of magnitude larger than could conceivably be handled in the foreseeable future.
However, there is cause for optimism because, it is now possible to do shell model calculations in bases of states restricted to just a few Sp$(3,\Rb)$ irreps \cite{Dytrych17}.
This will surely initiate a new era in the many-nucleon theory of heavy nuclei,
for which we need to be prepared.

The first challenge in attempting to apply the symplectic model to the low-lying rotational states of heavy deformed nuclei, is to select the appropriate Sp$(3,\Rb)$ irreps.
To be in a position to make a good choice from among the many available for a given nucleus, one should first consider the characteristic properties of an 
Sp$(3,\Rb)$ irrep and how it relates to the observable properties of a nucleus.
This is a primary objective of AMF theory.

When the HF and TDHF mean-field theories are expressed algebraically
in terms of equations of motion \cite{Rowe68,Rowe68b},  as in Section \ref{sect:MFtheory}, 
it becomes evident that, along with their HFB counterparts,
 these theories are special cases of a more general  AMF theory
\cite{RoweRR80,RosensteelR81b,RoweVR83}.
This section shows that, as an extension of HF theory, 
AMF theory likewise  provides an interface between the classical and the quantum representations of a variety of algebraic systems and gives an explicit realisation of the Dirac-Weyl theory of quantisation \cite{Dirac67,Weyl50}, in which the classical Poisson bracket realisation of a finite Lie algebra is quantised by construction of its unitary irreps.
Conversely, it gives a realisation of classical mechanics as constrained quantum mechanics.
Many AMF properties and applications have been developed by Rosensteel and colleagues  
\cite{DankovaR01,Rosensteel02,Rosensteel04,Rosensteel05,Rosensteel06}.
However, for present purposes, it is noted that, although AMF theory can be applied  to any algebraic model, its application is most useful for a model that has irreps defined by lowest-weight (and/or highest-weight) states.
The quantisation of its classical mean-field representation is then achieved by standard algebraic methods.

The following will show that the relationships between  AMF theory and the representation of a Lie algebra with a lowest- (and/or highest-) weight state
makes it possible to take advantage of the complementary contributions of two major approaches to nuclear structure theory: mean-field  theory and algebraic modelling.
Mean-field theory \cite{BenderHR03} is well known to be of fundamental importance in establishing the foundations of many-body theory and the nuclear shell model \cite{Barrett05}.
The power of algebraic methods  in exposing the dynamical content of a system is likewise  recognised; e.g., in Elliott's SU(3) model \cite{Elliott58ab}, 
Kerman's quasi-spin model \cite{Kerman61}, 
the Lipkin model \cite{LipkinMG65},
the Interacting Boson Models \cite{IBM87}, 
the algebraic version of the Bohr Model \cite{RoweWC09},
and the  other models discussed in Section \ref{sect:AlgModel}.
Such models are valuable for showing the many ways in which collective structures can emerge in nuclei and in providing solvable models that serve as ways to test  the validity of various approximation schemes, such as the RPA. 
Algebraic models, that define coupling schemes 
\cite{Flowers52,EdmondsF52a,Elliott58ab,FrenchHMW69} for the many-nucleon Hilbert space, are especially important.
However, apart from those of Rosensteel and colleagues, the applications of mean-field methods to algebraic models has, to date, received little attention in physics.

\subsection{AMF theory as a generalisation of HF theory}
The AMF generalisation of HF theory is valuable because of its close relationship with the structure theory of  semi-simple Lie algebras
\cite{RoweRR80}.
HF theory, is an application to the fully anti-symmetric unitary irrep of the Lie algebra of one-body operators, which  has a lowest-weight state which is a Slater determinant of so-called occupied single-particle states.
A particle-hole state is created by transferring a nucleon from an occupied state of a Slater determinant to an unoccupied state.
Thus, any Slater determinant is a particle-hole vacuum state in that it is annihilated by every one-body operator that would annihilate a nucleon in an unoccupied single-particle state and recreate it in an already-occupied single-particle state.
The significance of this observation is that  any Slater determinant of single-nucleon states can serve as the lowest-weight state for  the unique fully anti-symmetric unitary irrep, for a particular nucleus, of the Lie algebra of one-body operators.
Moreover, for any choice of lowest-weight state there are  corresponding particle-hole  creation and annihilation operators that serve as raising and lowering operators.

The important property, shown in Section \ref{sect:MFtheory},
is that the HF manifold  of all Slater determinants for a given nucleus is  a classical phase space which spans the Hilbert space of its quantisation.
It is also notable  that any Slater determinant of occupied single-nucleon states can be transformed into any other by a unitary transformation of the occupied single-nucleon states.    This means that the set of  Slater determinants for a nucleus, that form its 
HF classical phase space, can be generated by the transformations of any single determinant by the elements of the group of one-body unitary transformations.
Thus, the HF manifold, given by this classical phase space,
 is said to be an orbit of the group of one-body unitary transformations.

It is now straightforward to show that parallel relationships  
apply to any algebraic model with irreps having lowest-weight (and/or highest-weight) states 
and that, for such irreps, there is a corresponding AMF  theory.
Suppose that a state $|\phi\rangle$ is a lowest-weight state for a unitary irrep 
$\hat T$ of a Lie group $G$ and that the operators $\{ \eta^\dag_\alpha\}$ and their Hermitian adjoints $\{ \eta^\alpha\}$ are, respectively, raising and lowering operators for this irrep.
It follows immediately from the observation that the state $|\phi\rangle$ is annihilated by the 
$\{ \eta^\alpha\}$ operators that any state of the orbit
\beq 
|\phi(g)\rangle = \hat T(g) |\phi\rangle, \quad \text{for some $g\in G$} ,
\eeq
is a lowest-weight state that is annihilated by the correspondingly transformed lowering operators
\beq
\eta^\alpha(g) = \hat T(g) \eta^\alpha \hat T^{-1}(g) .
\eeq
In other words, any state $|\phi(g)\rangle$ of a $G$ orbit of a particular lowest-weight for a unitary irrep is also a lowest-weight state for that irrep.

In parallel with HF theory, AMF theory now  makes it possible to select the minimum-energy lowest-weight state for an algebraic model with a given Hamiltonian  and  determine its  normal-mode vibrational states.
The minimum-energy lowest-weight state is interpreted classically as
the  equilibrium state  of the model and, in quantum mechanics, as the closest approximation, among the lowest-weight states, to the ground state of the model.
The time-dependent normal-mode solutions of the classical equations of motion  in AMF theory likewise have an interpretation in terms of  the random-phase approximation.
The  results that emerge resemble those obtained for a system of coupled harmonic oscillators in standard coherent state theory.
Explicit examples will be given in the following for the symplectic model.

The recognition that the manifold of all Slater determinants can be identified with a classical phase space of a nucleus,  leads to many insightful results as shown in Section \ref{sect:3.A}.
For example: it takes account of the antisymmetry properties of the nucleons;
it has a classical Hamiltonian defined as a function on this phase space given
by the expectation values  of the quantum mechanical Hamiltonian; and it 
leads naturally to the Nambu-Goldstone interpretation of a broken symmetry in mean-field theory  \cite{NambuJL61,GoldstoneSW62}).
The latter follows because, 
if a minimum-energy mean-field state is not invariant with respect to SO(3) rotations, 
it represents a classical equilibrium state which, not surprisingly, has many orientations
corresponding to the energy-degenerate set of mean-field states that lie on an SO(3) orbit.

\subsection{More general AMF theories, co-adjoint orbits, and geometric quantisation}
The properties of Lie group orbits in the Hilbert spaces of their unitary irreps
have been studied widely in  mathematics
\cite{Kirillov04, Kirillov76}, as reviewed in Ref.\ \cite{Vogan98}.
In the theory of geometric quantisation \cite{Kostant70,Souriau70,Woodhouse91}, 
which relates closely to AMF theory, the orbits are  identified with coadjoint orbits, which are  orbits of a Lie group in the dual space of linear operators on its Lie algebra.

In physical terms, an element of a coadjoint orbit is  a  density operator which characterises a state of an algebraic model by the expectation values that it gives for the elements of the model's Lie algebra of observables, 
i.e., for any state $|\psi\rangle$ in a unitary irrep of an algebraic model there is a density operator 
$\hat\rho_\psi$ defined by a map in which
\beq \label{eq:psi->rho}
|\psi\rangle \to \hat\rho_\psi , \quad \hat\rho_\psi (X) = \langle \psi |\hat X|\psi\rangle , 
\eeq
for all $X$ in the Lie algebra of the model. 
For example, the dual of a Lie algebra of one-body operators, is a set of one-body density operators.
Thus, mapping a mean-field state to a point on a coadjoint orbit corresponds precisely to 
what is done in the density-matrix formulation of mean-field theory.
However, one should be aware that, in general, the map of Equation (\ref{eq:psi->rho}) is not invertable;  i.e., there may be a multiple set of states with the same density.
 The exceptional cases are then of special importance.
The states of HF theory are exceptions in that a Slater determinant is uniquely defined (to within a phase factor) by a one-body density matrix.
The essential property of such exceptional orbits of a group is that they consist of lowest-weight states that are uniquely defined by the expectation values in these states of the elements of the Lie algebra of the group.
Thus, we  restrict consideration,  in the following, to such group orbits.
These special orbits have the  advantage of being simply quantised  by the standard construction of the unitary irrep of a Lie group with a given lowest- (highest-) weight state and a corresponding system of raising and lowering operators.
They are the coadjoint orbits that are said to be `integral'.

AMF theories,  based on general coadjoint orbits of a group, have  been considered  by Rosensteel and colleagues
\cite{RosensteelR81b,DankovaR01,Rosensteel04,Rosensteel05,Rosensteel06}.
However, because the map $|\psi\rangle \to \hat\rho_\psi$,
defined by Eqn.\ (\ref{eq:psi->rho}) may be many to one,
the greater generality is obtained at a price.
For example, the expectation value of an operator,
 such as a Hamiltonian that is not an element of the Lie algebra,
 is only defined  at a point of a coadjoint orbit as the average of its values for  the states that map to that  point.
 
It will be shown in the following sections that the application of AMF methods to the irreps of an algebraic collective model of the nucleus, 
that defines a coupling scheme for the many-nucleon Hilbert space, leads to valuable  techniques for the study of nuclear collective properties.

\subsection{The symplectic model as a unified model}
A valuable property of a minimum-energy lowest-weight state for a model is its clear algebraic significance.
It has all the properties of a lowest-weight state for an irrep of a Lie algebra and the extra property of minimising the expectation value of a model Hamiltonian.

A minimal energy lowest-weight state for an Sp$(3,\Rb)$ irrep is, to within an arbitrary rotation, 
an eigenstate of a generally triaxial harmonic-oscillator Hamiltonian and can be chosen 
as a state for which  its quadrupole moments are diagonal in a Cartesian basis.
In addition, being a minimal-energy lowest-weight state, means that it has zero first-order coupling, by the Hamiltonian,  to any other states of the irrep with the exception of those generated by its rotations.
Thus, a AMF minimum-energy lowest-weight state provides a well-defined distinction between the vibrational and rotational degrees of freedom of an Sp$(3,\Rb)$ irrep.
This is a property that is characteristic of an intrinsic state of the Bohr-Mottelson unified model \cite{BohrM75}; 
in fact, the AMF expression of the symplectic model can be seen as a many-nucleon realisation of a Bohr-Mottelson unified model \cite{BohrM75}. 

In applications of the symplectic model  to the lowest-energy rotational states of heavy doubly even nuclei, the dominant symplectic model irreps are invariably those of maximal space symmetry.
Consequently, the states of these  irreps 
have vanishing intrinsic spins and minimum isospin, i.e.,  $S=0$ and $T= \frac12 (N-Z)$.
For the purposes of this review, we therefore focus primarily on such Sp$(3,\Rb)$ irreps.

It follows from Lie algebra theory that the full Hilbert space of an Sp$(3,\Rb)$ irrep is spanned by the states generated by rotations of  both its minimum-energy lowest-weight state 
$|\sigma,\omega\rangle$ and its multiple vibrational excitations.
As a result, the set of classical equilibrium states of a deformed nucleus is generated by the rotations of a lowest-weight state and a corresponding set of quantum-mechanical rotational states is obtained, in a semi-classical approximation,  by angular-momentum projection from the minimum-energy lowest-weight state (cf. Section \ref{sect:AMprojection}).
In a microscopic version of the unified model, the intrinsic one-phonon vibrational states are likewise determined by solution of the RPA equations with a minimum-energy lowest-weight state considered as the uncorrelated vacuum state.

For a generic irrep with $\sigma_1 > \sigma_2 >\sigma_2 $,
the minimum-energy lowest-weight state $|\sigma,\omega\rangle$
is an eigenstate of the three operators
\beq 
\zeta^\dag_{ii}  =  \sum_n b^\dag_{ni} b^{ni} , \quad i = 1,2,3,
\eeq
of a Cartan subalgebra.
Thus, a complete basis for the Sp$(3,\Rb)$ irrep is generated by the repeated action of the  raising operators
\bal 
&\zeta^\dag_{ij}  =  \sum_n b^\dag_{ni} b^{nj} , \quad i < j, \\
&\eta^\dag_{ij} =  \sum_n b^\dag_{ni} b^\dag_{nj} , \quad i,j = 1,2,3,
\end{align}
on the lowest-weight state $|\sigma,\omega\rangle$. 

From among the linear combinations of this set of 9 raising operators and their Hermitian-adjoint lowering operators, there are three components of angular momentum   
$\{\hat L_k, k=1,2,3\}$ relative to the axes of the lowest-weight state 
$|\sigma,\omega\rangle$ and three angular-momentum boost operators 
$\{ \hat\Theta^j, j=1,2,3\}$
that satisfy the equations
\beq
\langle  \sigma,\omega |[ \hat \Theta^j, \hat L_k ]|\sigma,\omega\rangle
= {\rm i}\delta^j_{k} .
\eeq
Also, in parallel with the Hartree-Fock RPA theory outlined in Section
\ref{sect:RPA}, there are six pairs of  excitation and de-excitation operators 
$\{ O^\dag_\lambda, O^\lambda\}$ of vibrational states that
are linear combinations of the raising and lowering operators that satisfy the  equations
 \bal \label{eq:38} 
 \begin{split}
&\langle \sigma,\omega |[ \hat X, [\hat H, O^\dag_\lambda ]]
 |\sigma,\omega\rangle =
 \hbar\omega_\lambda 
 \langle \sigma,\omega | [\hat X, O^\dag_\lambda ] |\sigma,\omega\rangle  , \\
&\langle  \sigma,\omega | [\hat X, [\hat H, O^{\lambda} ]] |\phi\rangle = 
-\hbar\omega_\lambda  
\langle  \sigma,\omega | [\hat X, O^{\lambda}]  
|\sigma,\omega,\omega\rangle , 
\end{split}  \\
&\langle \sigma,\omega | [O^{\kappa}, O^\dag_\lambda ]
 |\sigma,\omega\rangle   
    = \delta^\kappa_\lambda ,
\end{align}
for every $\hat X$ in the Sp$(3,\Rb)$ Lie algebra,
and  the orthogonality relationships 
\bal \label{eq:OLcom}
\begin{split}
&\langle  \sigma,\omega | [O^{\kappa}, \hat L_k ] 
|\sigma,\omega\rangle =
 \langle  \sigma,\omega | [O^\dag_\kappa,  \hat L_k ] 
 |\sigma,\omega\rangle = 0, \\
& \langle  \sigma,\omega | [ O^{\kappa} , \hat \Theta^j] 
|\sigma,\omega\rangle =
 \langle  \sigma,\omega | [O^\dag_\kappa ,\hat \Theta^j] 
 |\sigma,\omega\rangle = 0.
 \end{split}
 \end{align}
The operators $\{ O^{\lambda}\}$ are interpreted as one-phonon vibrational excitation
and de-excitation  operators, respectively,.
However, in doing so there is no intention to suggest that  their repeated applications generate sequences of harmonic vibrational states.
It should also be recognised that the vibrational states, excited by these $\{ O^\dag_k\}$ operators, which satisfy Eqn.\ (\ref{eq:OLcom}), are by construction intrinsic vibrational states.

For $\sigma_1 > \sigma_2> \sigma_3$,  the spectrum of states obtained 
from the  above equations will consist of  zero-phonon  rotational states of a triaxial nucleus and six sets of  rotational states corresponding to  one-phonon intrinsic vibrational excitations.
However, if $\sigma_1 > \sigma_2= \sigma_3$, the spectrum will contains a $K=0$ ground-state rotational band, plus two  $K=0$ bands and single $K=1$ and   $K=2$ rotational  bands corresponding to one-phonon vibrational excitations.
These expectations are indicated by the observation that for an SU(3) irrep 
$(\lambda\,\mu)$  the tensor product $(\lambda\,\mu)\otimes (2\,0)$ is a sum of six irreps whereas, for an axially symmetric $(\lambda. 0)$ irrep
\beq 
(\lambda,0)\otimes (2,0) =  
(\lambda\! +\!2, 0) \oplus (\lambda\! +\!1, 1)\oplus (\lambda \, 2) .
\eeq

In the $\mu=0$ case, one of the $K=0$ bands is a monopole vibrational band, 
and the other $K=0$, 1 and 2 bands are quadrupole vibrational bands.
Thus, if the $K=0$ and $K=2$ vibrational bands were to emerge at low energies, they could be interpreted as beta and gamma vibrational bands.
However, in the microscopic unified model,  corresponding to un-mixed symplectic model irreps,
they are the giant-resonance excitations of a deformed nucleus \cite{SuzukiR77b} and are not expected to lie  low in energy.
 A mixing of the rotational states arising from different symplectic model irreps could nevertheless give rise to states having some of the properties of those of low-lying beta- and gamma-vibrational bands.

For both triaxial and axially symmetric irreps, the Inglis cranking model predicts rigid-body moments of inertia for the zero-phonon rotational states generated by rotations of the shape-consistent state 
 $|\sigma,\omega\rangle$, as shown in Section \ref{sect.Inglis}, and
the Thouless-Valatin prescription for the moments of inertia gives
\beq 
 \langle \sigma,\omega |[ \hat \Theta^k [ \hat H , \hat \Theta^k] ]
|\sigma,\omega\rangle
= \frac{\hbar^2}{\scrI}_{\!\! k} .
\eeq
However, more reliable angular-momentum projection techniques 
\cite{CussonL73,Wong75,RoweBB00}, 
including those developed for wave functions in SU(3)-coupled basis 
\cite{DraayerA73,BahriRD04},
can be used, as in HF theory, first to expand the lowest-weight state as a sum 
\beq 
|\sigma,\omega\rangle = \sum_{KL} n_{KL} |KLK\rangle ,
\eeq
and then to project out an orthonormal basis of angular momentum states $\{ |KLM\rangle\}$
in term of which a nuclear Hamiltonian can then be diagonalised  to obtain bands of rotational states
\beq
|\alpha LM\rangle = \sum_K \alpha_K |KLM\rangle 
\eeq
and  their energies.
One can then also evaluate the separate contributions to these energies coming from the kinetic energy and interaction components of the Hamiltonian
\beq
\hat H = \hat T + \hat V,
\eeq
and determine the extent to which the potential energies of symplectic-model rotational states  take  a constant value for adiabatically small values of the rotational angular momentum.

\section{Low-energy collective states of symplectic model irreps}
\label{sect:11}
It is important to recognise  that an application of  AMF theory is not necessarily an approximation.
It could simply be a first step in the construction of an irrep of a semi-simple  Lie algebra. 
The choice of a minimum-energy lowest-weight  is nevertheless a useful starting point for good approximations such as given by angular-momentum projection and RPA methods.
A higher-order approximation for the low-energy rotational states known as  'variation after projection' \cite{CussonL73},
is to determine the optimal lowest-weight for each value of the projected angular momentum
independently; this enables account to be taken of the changing structure of the intrinsic state with increasing angular momentum.

Methods that make use of  the algebraic properties of  Sp$(3,\Rb)$ irreps are of particular interest, especially when they give results that are valid for any choice of Hamiltonian.
It is shown, for example, that angular-momentum projection can be achieved  explicitly and analytically for a shape-consistent lowest-weight state of an axially symmetric Sp$(3,\Rb)$ irrep.
This makes it possible to calculate the properties of the projected states of an Sp$(3,\Rb)$ irrep as determined by the matrix elements between these states of any observable in the Sp$(3,\Rb)$ Lie algebra, such as quadrupole matrix elements, and E2 transition rates.
As an example, the calculated expectation values of the many-nucleon kinetic energies of axially symmetric Sp$(3,\Rb)$ states have been calculated. 
Similar algebraic projection methods are possible, in principle, for the generic triaxial Sp$(3,\Rb)$ irreps, but have  yet to be developed.
However, the first steps of projecting a minimum-energy lowest-weight state onto a U(3) basis are considered in this section.

Also discussed are approximations based on the restriction  of an Sp$(3,\Rb)$ calculation to subsets of basis states and on a generator-coordinate approach. 
These  approaches, are shown to be related to the AMF methods and provide intermediate steps in assessing the accuracy of the AMF-RPA, as a microscopic unified model, in comparison with full Sp$(3,\Rb)$ model calculations.

\subsection{Angular-momentum-projected rotational states
of an axially symmetric Sp$(3,\Rb)$ irrep} \label{sect:axialSCstate}
Having determined a minimum-energy lowest-weight state for an Sp$(3,\Rb)$ irrep,  
angular-momentum projection methods, such as developed for HF theory (see Section \ref{sect:MFrotor}), can be used to determine its low-energy rotational states.
Likewise, standard RPA methods can be used to determine its one-phonon intrinsic vibrational states.
In fact, explicit expressions have already been derived  \cite{RoweBB00} for the multiplicity-free angular-momentum states projected  from a  lowest-weight state $ |\sigma,\omega\rangle$ of an axially symmetric
Sp$(3,\Rb)$ irrep $\langle\sigma\rangle = \langle\sigma_1,\sigma_2,\sigma_2\rangle$
which is an eigenstate of energy $\sigma_1\hbar\omega_1+2\sigma_2\hbar\omega_2$ of
an axially symmetric harmonic oscillator.
This derivation, 
starts with the expression of the state $ |\sigma,\omega\rangle$ 
 as an SU(1,1) scale transformation 
 \beq \label{eq:transf}
 |\sigma,\omega\rangle = \hat S(\epsilon) |(\sigma_1,\sigma_2,\sigma_2),\omega_2\rangle
 \eeq	
 of an eigenstate $ |(\sigma_1,\sigma_2,\sigma_2),\omega_2\rangle$ 
 of energy $(\sigma_1+2\sigma_2)\hbar\omega_2$
 of a spherical harmonic oscillator of frequency $\omega_2$, 
 and determines that
 \beq \label{eq:S1transf}
  \hat S(\epsilon) |(\sigma_1,\sigma_2,\sigma_2),\omega_2\rangle  =  \sum_{\nu \geq 0}
C_\nu(\sigma_1,\epsilon) |(\sigma_1 + 2\nu,\sigma_2,\sigma_2),\omega_2\rangle ,
\eeq
with
\beq \label{eq:S(e)a}
C_\nu(\sigma_1,\epsilon) =  \frac{1}{(\cosh \epsilon )^{\sigma_1}}
(\tanh{\epsilon})^\nu 
\sqrt{\frac{(\sigma_1+\nu-1)!}{(\sigma_-1)! \,\nu!}} 
\eeq
and $e^{2\epsilon} = \omega_2/\omega_1$.
This expression of a scale transformation is derived in Appendix \ref{sect:SU11Scale}.
This gives the state 
$|\sigma,\omega\rangle$ 
as a sum, 
\beq \label{eq:transf2}
|\sigma,\omega\rangle = \sum_{\nu \geq 0}
C_\nu(\sigma_1,\epsilon) |N_\nu (\lambda+2\nu,0),\omega_2\rangle,
\eeq
of U(3)$\,\supset\,$SU(3) highest-weight states
with $N_\nu = \sigma_1+2\sigma_2 + 2\nu$ and $\lambda = \sigma_1-\sigma_2$.
(Note that  a single scale transformation is sufficient, for an axially symmetric irrep, if the frequency $\omega_2$ is chosen  to give the required volume of the shape-consistent lowest-weight state.)

Use is then made of the expansion \cite{RoweB00}
\beq
|N(\lambda, 0),\omega_2\rangle 
= \sum_{L\geq 0} a^\lambda_L |N(\lambda ,0)L0,\omega_2\rangle
\eeq
with
\beq 
a^\lambda_L= \tfrac12  \big(1+(-1)^{\lambda+L}\big)
\left[\frac{(2L+1)\lambda!}{(\lambda-L)!! (\lambda+L+1)!!} \right]^\frac12 ,
\eeq
to obtain the expansion
\beq
|\sigma,\omega\rangle = \sum_L N_L(\epsilon) |\epsilon  L0\rangle ,
\eeq
with
\beq
 N_L(\epsilon)|\epsilon  L0\rangle = \sum_{\nu\geq 0} 
C_\nu(\sigma_1,\epsilon) a^{\lambda+2\nu}_L |N_\nu (\lambda+2\nu,0) L0\rangle.
\eeq 
Then, with the $N_L(\epsilon)$ coefficients chosen to be real and the observation that,
with
\beq
 | N_L(\epsilon)|^2 =
 \sum_{\nu\geq 0} |C_\nu(\sigma_1,\epsilon)|^2\, |a^{\lambda+2\nu}_L|^2 ,
\eeq
they  satisfy the  identity $\sum_L | N_L(\epsilon)|^2 $, 
it  follows that
\beq  \label{eq:amexpansion}
|\epsilon  L0\rangle = \sum_{\nu\geq 0}
f_{\nu L}(\epsilon) |N_\nu (\lambda+2\nu,0) L0\rangle ,
\eeq
with $f_{\nu L}(\epsilon) =C_\nu(\sigma_1,\epsilon) a^{\lambda+2\nu}_L /
 N_L(\epsilon)$.

\begin{figure}[htb]
\centerline{\includegraphics[width=4 in]{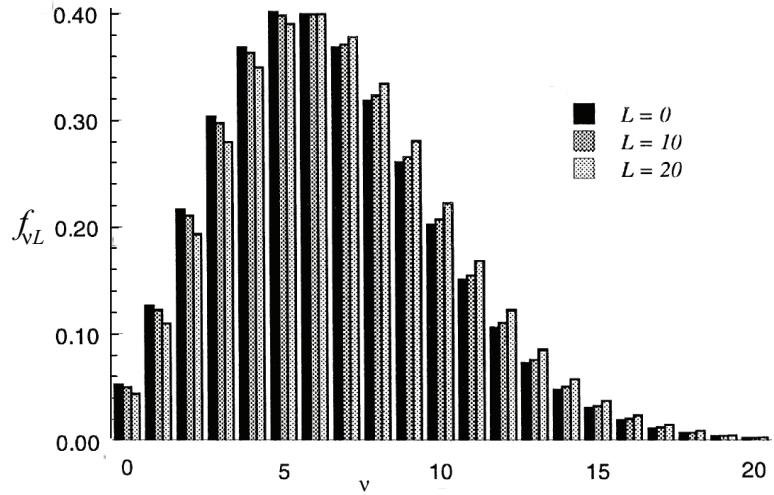}}
\caption{ \label{fig:Ams166Erwfns} 
Amplitudes for the wave functions of the lowest $^{166}$Er states of angular momentum
$L=0$, 10, and 20  projected from a shape-consistent lowest-weight state
(as defined in the text) for an Sp$(3,\Rb)$ irrep $\langle 836.5(78,0)\rangle$. 
(The figure is adapted from one of Ref.\ \cite{RoweBB00} for which $N_0$ was given the value $N_0=826.5$.)}
\end{figure}

 These amplitudes are shown in Fig.\ \ref{fig:Ams166Erwfns}
for the $L=0$, 10 and 20 states of the ground-state rotational band of $^{168}$Er, 
for the shape-consistent lowest-weight state
of the Sp$(3,\Rb)$ irrep  $\langle N(\lambda,0)\rangle = \langle 836.5(78,0)\rangle$.
The irrep $\langle 836.5(78,0)\rangle$ was chosen to give  experimentally observed E2 transition rates \cite{JarrioWR91}.
The figure shows that the wave functions for the states of this irrep  are spread over  $\gtrsim 15$ spherical harmonic-oscillator shells.

From the known representations  of the Sp$(3,\Rb)$ Lie algebra in a U(3) $\supset$ SO(3) basis \cite{Rowe84} (reviewed in Ref. \cite{RoweMC16}), it is now straightforward to calculate the matrix elements of any
Sp$(3,\Rb)$ observable between the states obtained by angular-momentum-projection 
from the shape-consistent state.
As an example,  Table \ref{tab:KEengies} gives the expectation values of the nuclear  kinetic energy
 of the angular-momentum-projected states, relative to that of the $L=0$ ground state, for the states of angular momentum ranging from 2 to 16.
\begin{table}[ht]
\caption{\label{tab:KEengies} The kinetic  energies, relative to that of the $L=0$ ground state, of the angular-momentum states projected from  the shape-consistent lowest-weight state
(defined in Section \ref{sect:min.l.wt}) of the axially symmetric Sp$(3,\Rb)$ irrep 
$\langle N_0(\lambda,0)\rangle = \langle 836.5(78,0)\rangle$.
The kinetic energies in the second column are given in keV and the ratios
$\frac12 L(L+1)/K.E.$ in the third column are given in inverse MeV. 
(These are previously unpublished results.)}
$ \begin{array}{|r|r|c|c|c|c|c|c| } \hline
\;L\; & \;K.E. & \frac12 L(L+1)/K.E.  \\ \hline
2\; & 30 & 99.5 \\
4\; & 101 & 99.3 \\
6\; & 212 & 99.1 \\
8\; & 364 & 98.9 \\
10\; & 568 & 98.5 \\
12\; & 795 & 98.1 \\
14\; & 1,076 & 97.6 \\
16\; & 1,401 & 97.1 \\
\hline   
\end{array}$ \\ %
\end{table} 
They show that to within $\sim 2\%$, the expectation values of the nuclear kinetic-energy, relative to that for the $L=0$ state, are proportional to $L(L+1)$ for states of angular momentum ranging from $L=2$ to 16. 
Thus, if it is assumed that the potential energy of a rotor depends only on its intrinsic state and is independent of its rotational angular momentum, then the moment of inertia of the rotor can only depend on its kinetic energy and should be expressible in the form
$K.E. =K_0 + \hbar^2 L(L+1)/2\scrI_{\rm KE}$.
This moment of inertia  and that obtained from the  experimentally observed energy for the  $L=2$ state, have the respective values 
\beq \label{eq:keMofI}
\scrI_{KE}/\hbar^2 = 99.5 \; {\rm MeV}^{-1}, \quad
\scrI_{\rm expt}/\hbar^2 = 37.6 \; {\rm MeV}^{-1} .
\eeq
For comparison the rigid and irrotational-flow moments of inertia for the
minimum-energy lowest-weight state have the values
\beq 
\scrI_{\rm rig}/\hbar^2 = 84.4 \; {\rm MeV}^{-1} , \quad
\scrI_{\rm irr}/\hbar^2 = 5.8 \; {\rm MeV}^{-1} .
\eeq
A conclusion of these results is that the kinetic-energy components of the excitation energy 
for the $2^+$ state of this $^{168}$Er Sp$(3,\Rb)$   irrep is less  than half  the experimentally observed value. 
Obvious interpretations of these results, assuming no mistakes have been made, are  that the shape-consistent lowest-weight state for the chosen Sp$(3,\Rb)$ irrep 
is a poor approximation to the intrinsic state of the ground-state rotational band of $^{166}$Er and/or that a large fraction of  rotational energies are potential energies.
However, regardless of whether or not the chosen irrep is the most appropriate for this nucleus,
which is certainly questionable, the results show that the kinetic-energy component of the rotational energies of a symplectic model irrep is close to that of a rigid rotor.
Note that, in Bahri's symplectic-model calculations \cite{BahriR00} for  $^{166}$Er  with a Davidson-like potential, good fits were obtained  in which the rotational energies were more than half potential energies.

According to conventional wisdom, a large mixing of symplectic model irreps, brought about by pairing interactions, is needed to obtain observed  moments of inertia.
However, the  results of Bahri's calculations suggest that the potential-energy contributions to rotational energies might be sufficient for this purpose.
In any event, it is important to determine the potential and kinetic energy components of the 
rotational energies of the decoupled collective model states of the symplectic model because,
as observed in Section \ref{sect:CMsubspaces}, the states of different symplectic model irreps are pure uncoupled collective states in the sense that there can be no isoscalar E2 transitions between them.
Thus, it is clear that
\emph{an evaluation of the separate contributions  of the kinetic and potential energies to the rotational energies of  axially symmetric nuclei is essential for understanding the dynamics of nuclear rotations.}

\subsection{The  rotational states of a triaxial Sp$(3,\Rb)$ irrep}
Angular-momentum projection methods for the low-energy rotational states of a nucleus and RPA methods for its intrinsic excitations, as developed for use in HF theory (see Section \ref{sect:MFrotor}), can similarly be applied to the minimum-energy lowest-weight states of  triaxial Sp$(3,\Rb)$ irreps of maximal space symmetry.
It is also likely that if a minimum-energy lowest-weight state of a triaxial Sp$(3,\Rb)$ irrep is expanded in a U(3) basis,  it can be further expanded in an SO(3) basis 
by precise algebraic methods, as shown above for an axially symmetric irrep and as
employed for the calculation of SU(3) Clebsch-Gordan coefficients in an SO(3) basis \cite{DraayerW69,RoweB00}.
Thus, an outline is given in this section of the expansion of a shape-consistent lowest-weight of a triaxial Sp$(3,\Rb)$ irrep in a U(3) basis.

If $|\sigma,\omega_3\rangle$ is a lowest-weight state of a triaxial Sp$(3,\Rb)$ irrep
$\langle \sigma\rangle \equiv \langle \sigma_1,\sigma_2,\sigma_3 \rangle$
that is an eigenstate of a spherical harmonic-oscillator Hamiltonian of frequency $\omega_3$,
then the corresponding shape-consistent lowest-weight state is given, 
in parallel with that of Eqn.\ (\ref{eq:transf}), by the 
SU(1,1)$\times$SU(1,1) transformation
\beq \label{eq:scaletransf}
|\sigma, \omega_3\rangle  \to |\sigma,\omega\rangle = 
 \hat S_1(\epsilon_1) \hat S_2(\epsilon_2) |\sigma,\omega_3\rangle
\eeq
to an eigenstate  of a triaxial harmonic oscillator with a triple of frequencies 
$\omega \equiv (\omega_1,\omega_2,\omega_3)$,
with
\beq
e^{2\epsilon_1}= \frac{\omega_3}{\omega_1}  , \quad
e^{2\epsilon_2}=\frac{ \omega_3}{\omega_2}.
\eeq
(Note that a third scale transformation is avoided by setting $\omega_3$ such that the state
$|\sigma,\omega_3\rangle$ has the desired volume.)
The scaled state  
is then the sum of states
\beq \label{eq:U2wts}
|\sigma,\omega\rangle
= \sum_{\nu_1,_3\nu_2} C_{\nu_1}(\sigma_1,\epsilon_1)  
C_{\nu_2}(\sigma_2,\epsilon_2) 
|(2\nu_1,2\nu_2)\rangle ,
\eeq
where 
\beq 
|(2\nu_1,2\nu_2)\rangle \equiv 
|(\sigma_1+2\nu_1,\sigma_2+2\nu_2, \sigma_3,),\omega_3\rangle . 
\eeq

\begin{figure}[htb]
 \centerline{\includegraphics[width=5 in]{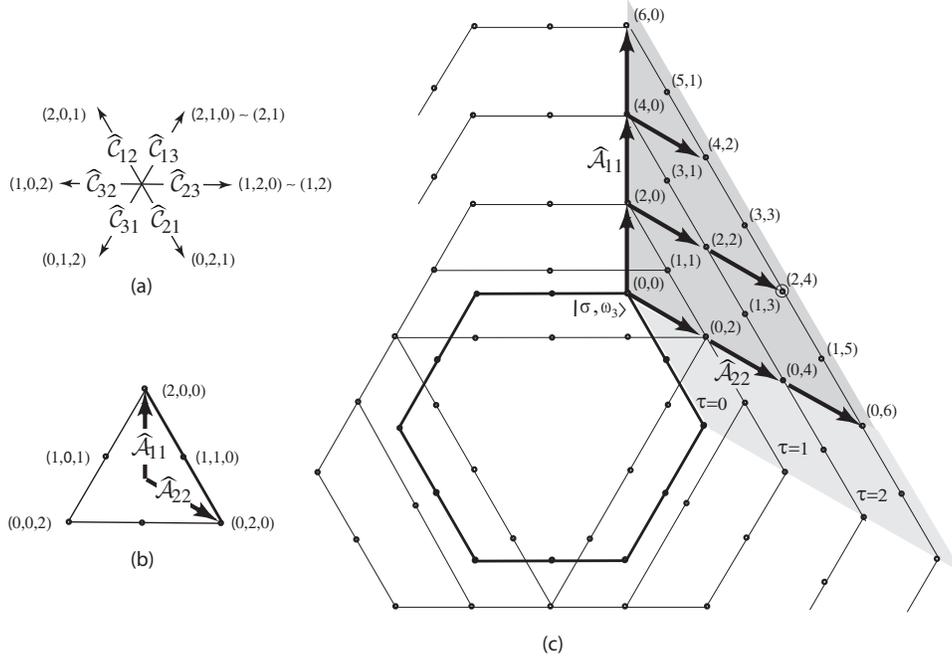}}
\caption{\label{fig:weights}
The figure  shows weight diagrams: 
(a) for the SU(3) Lie algebra;
(b) for the Sp$(3,\Rb)$  raising operators
$\hat{\mathcal A}_{11}$ and $\hat{\mathcal A}_{22}$
 as components of a U(3) tensor of highest weight (2,0,0), 
and (c) for the U(3) weights of states 
in the expansion of the scaled lowest-weight state 
$ \hat S_1(\epsilon_1)\hat S_2(\epsilon_2)|\sigma,\omega_3\rangle$
in a basis of spherical harmonic-oscillator eigenstates. 
The weight diagram of figure (c) is two-dimensional in the grey shaded areas in which the third component of the weights takes the constant value $\sigma_3$.
The state  $|\sigma,\omega_3\rangle$, corresponding to the point  (0,0) in the figure, 
is, by definition, both a lowest-weight state  for an Sp$(3,\Rb)$ irrep with 
$\sigma = (\sigma_1, \sigma_2, \sigma_3)$ and a  highest-weight state for a U(3) irrep. 
All the states with weights that lie on the vertical line above that of the U(3) highest-weight state 
$|\sigma,\omega_3\rangle$ and  are connected to this state by $\hat{\mathcal A}_{11}$ arrows are basis states for an ${\rm SU(1,1)} \equiv {\rm Sp}(1,\Rb)$ irrep.  
All the states with weights in the 
combined light- and dark-grey areas  are a basis for an Sp$(2,\Rb)$ irrep.  
The subset of these states with weights that are connected 
by $\hat{\mathcal A}_{11}$ and $\hat{\mathcal A}_{22}$ arrows to the state
$|\sigma,\omega_3\rangle$ are a basis for an SU(1,1) $\times$ SU(1,1) irrep.
As  shown in the text, a minimum-energy Sp$(3,\Rb)$ lowest-weight state is a linear combination of the basis states for the above-defined SU(1,1) $\times$ SU(1.1) irrep.
}
\end{figure}

The states $ |(2\nu_1,0)\rangle$ in the expansion of  
$\hat S_1(\epsilon_1) |\sigma,\omega_3\rangle$
have the useful property of being U(3) highest-weight states.
They are the states illustrated in Fig.\,\ref{fig:weights} that have U(3) weights on the vertical line above the weight of the  Sp$(3,\Rb)$ lowest-weight state 
$|(0,0)\rangle \equiv |\sigma,\omega_3\rangle$ 
and are generated from this lowest-weight state by the $\hat{\mathcal A}_{11}$ operators
of Eqn. (\ref{eq:8.ABCQops}).
It is clear that that these states are all of U(3) highest weight.

The states $|(0,2\nu_2)\rangle$ in the expansion of 
$\hat S_2(\epsilon_2) |\sigma,\omega_3\rangle$
 are similarly generated from the lowest-weight state  $|(0,0)\rangle$   
 by the  $\hat{\mathcal A}_{22}$ operators.  
Thus, the states  in the expansion of
$ \hat S_1(\epsilon_1) \hat S_2(\epsilon_2) |\sigma,\omega_3\rangle$
have U(3) weights defined by  all the points 
in Fig.\,\ref{fig:weights}(c) that are connected to the state  $|(0,0)\rangle$ by  
$\hat{\mathcal A}_{11}$ and $\hat{\mathcal A}_{22}$ arrows.
However, although a state $|(2\nu_1,2\nu_2)\rangle$ has well-defined U(3) weight
$(\sigma_1+2\nu_1, \sigma_2+2\nu_2 , \sigma_3)$,
it is only a state of U(3)  highest weight if $\nu_2  = 0$.
As a result, the expansion of the shape-consistent lowest-weight state of a triaxial 
Sp$(3,\Rb)$ irrep  as a sum of states of good angular-momentum quantum numbers is not as simple as for an axially symmetric irrep.
It can no doubt be determined by use of the transformations between SU(3) states in SU(2) and SO(3) bases \cite{DraayerW69,BahriD94,RoweB00}.
As  discussed in the following section,  the information gained from the content of Figure \ref{fig:weights}, when considered in the light of the Sp$(1,\Rb)$ and Sp$(2,\Rb)$ approximations and the general coordinate  method, also leads to alternative ways of proceeding to get reliable results.

\subsection{Relationship of the AMF approximations to the Sp$(1,\Rb)$ and Sp$(2,\Rb)$ submodels}
Approximate Sp$(3,\Rb)$ calculations, which relate to the 
AMF-RPA methods, are the so-called
Sp$(1,\Rb)$ and Sp$(2,\Rb)$ submodels of Sp$(3,\Rb)$ proposed, respectively, 
by Arickx \cite{Arickx76} and by Peterson and Hecht \cite{PetersonH80}, as outlined in Section 
 \ref{sect:ArickxHP}.

 A first observation is that, in truncating the Hilbert space of a full Sp$(3,\Rb)$ model 
 to a subspace, these models both maintain a constant value  of 
 $\sigma_3$ for every U(3) irrep $\{\sigma'_1,\sigma'_2,\sigma_3\}$ in the truncated space.
 Thus, both models can be improved  by choosing  the  value of $\omega_3$ to be that 
 for which the  shape-consistent lowest-weight state has the desired volume of essentially incompressible nuclear matter, as opposed to choosing it for  the spherical harmonic-oscillator lowest-weight state.
 With this adjustment, they provide  sub-models that are intermediate between those of the AMF-RPA and the full Sp$(3,\Rb)$ model.
A comparison of their results for  particular Sp$(3,\Rb)$ irreps,
for which each of them is able to give converged solutions to an acceptable level of accuracy, could then be very informative.
Hopefully, they will show that the simple AMF-RPA model, once programmed, is able to give quick and sufficiently accurate results that could be used to obtain a first understanding of nuclear data in terms of unmixed symplectic model irreps.
If this proves to be the case, it then become worthwhile to take it to the next level of approximation given by the generator-coordinate theory of single and mixed Sp$(3,\Rb)$ irreps.

\subsection{Generator-coordinate theory for single  Sp$(3,\Rb)$ many-nucleon irreps}
\label{sect:GCM1}
It is known  that in the Elliott model, the  states of an SU(3) irrep can be expressed as linear combinations of the states obtained by rotations of its highest-weight state.
Likewise,  the state of an Sp$(3,\Rb)$ irrep can all be expressed as linear combinations of the states obtained by general-linear transformations of its lowest-weight state \cite{VassanjiR86}.
 This observation underlies the  generator coordinate approach of 
Filippov, Vassanji and colleagues
\cite{FilippovO80,OkhrimenkoS81,VassanjiR82,VassanjiR83,VassanjiR84},
which is now seen to have a potentially powerful relationship with the AMF developments.

From an algebraic perspective, the  states of an Sp$(3,\Rb)$ irrep  are linear combinations of the states generated by the rotations and deformations of a lowest-weight state.
Thus, if we start with a minimum-energy lowest-weight state and augment it to a set  of locally deformed states generated by infinitesimal generators of the Sp$(3,\Rb)$ algebra, other than those that generate rotations, the result is a space corresponding to the intrinsic small-amplitude  vibrational states of a rotational nucleus; the rotational states are then generated by the relatively adiabatic rotation of all the intrinsic states, as understood in the Bohr-Mottelson unified model.

This perspective  leads naturally to an extension of the Sp$(3,\Rb)$ model to a model of mixed Sp$(3,\Rb)$ irreps in which the mixing takes places in the intrinsic frame of an adiabatic rotor in the manner of a quasi-dynamical symmetry as discussed in the following Section, \ref{sect:GCT}.

\section{The many-nucleon Hilbert space as a sum of collective model subspaces}
\label{sect:CMsubspaces}
A significant property of the nuclear shell model \cite{Mayer49,HaxelJS49,MayerJ55}
is that it defines a decomposition of the many-nucleon Hilbert spaces of nuclei into sums of energy-ordered  subspaces of  spherical harmonic-oscillator eigenstates coupled  to spin and isospin states.
This energy-ordered decomposition has been used widely to select truncated subspaces for shell-model calculations.
It has also been useful for separating nuclei into classes of  magic (closed-shell) nuclei, singly-closed shell nuclei, and doubly-open shell nuclei, each of which has  characteristic properties.
However, in spite of its successes, it is now apparent that the spherical shell model, when truncated to a manageable size, is unable to describe the rotational states of strongly deformed nuclei in anything like a realistic manner; it can do so only with the use of  fitted effective interactions and huge effective charges for electric quadrupole (E2) transition operators.
The problem is that such effective shell model states have essentially zero overlaps with
 realistic many-nucleon states in heavy nuclei.

\subsection{Coupling schemes for nuclear Hilbert spaces} \label{sect:CMspaces}
As Nilsson model \cite{Nilsson55,Lamm69}, mean-field \cite{BenderHR03} 
and symplectic model \cite{ParkCVRR84,BahriR00} calculations with schematic interactions 
(see Fig.\ \ref{fig:166Erwfns}) testify,
the spherical harmonic-oscillator-based decomposition and energy-ordering of the standard shell model is seriously inappropriate for realistic descriptions of the rotational states of heavy deformed  nuclei.
They  show that significant components of the states needed do not begin to occur in  standard shell-model spaces of spherical harmonic-oscillator energies of less that $\sim 10 \hbar\omega$, in spherical harmonic-oscillator units, above those normally considered.
Thus, for a realistic shell-model theory of heavy deformed nuclei,
it is necessary  to consider decompositions of their many-nucleon Hilbert spaces  into more appropriate sequences of subspaces.

An appropriate decomposition 
is given by a coupling scheme, defined by an algebraic model of nuclear rotations and quadrupole vibrations whose irreps can be ordered in a physically relevant manner.
Without prejudice as to the nature of 
 the rotational dynamics of nuclei, the Lie algebra of the model should include the angular momentum, the quadrupole moment operators which define the orientation of a rotor and isoscalar E2 transitions,  and the  nuclear kinetic energy.
It is then inferred, see Section \ref{sect:Sp3Rcoup}, that the smallest Lie algebra that contains these observables is that  of the Sp$(3,\Rb)$ symplectic model which, in addition contains infinitesimal generators of quadrupole deformation.
The essential property of the decomposition into irreducible symplectic model subspaces is that, if the whole many-nucleon Hilbert space of a nucleus is expressed as a sum of  such subspaces, there is no iso-scalar E2 transitions between the states of these subspaces.
A finer  decomposition, which distinguishes equivalent subspaces but which retains this property with fewer multiplicities, is given by the irreps of the direct product group 
${\rm Sp}(3,\Rb)\times {\rm U}(4)$, where U(4) is  Wigner's supermultiplet group which contains, as subgroups, the U(2) spin and isospin groups.
Thus, the irreps of the group ${\rm Sp}(3,\Rb)\times {\rm U}(4)$ can be said to define ideal collective states for the study of  rotational dynamics.

\subsection{Energy-ordered symplectic-model subspaces}
\label{sect:min.l.wt}
It is apparent from Fig.\ \ref{fig:166Erwfns} and other symplectic model calculations,  
that highly deformed rotational states which lie low in energy  invariably have their dominant components  in much higher spherical harmonic-oscillator shells than the lowest available to them.
Thus, instead of ordering symplectic model irreps by the energies of their lowest spherical harmonic-oscillator components, 
it is  more meaningful to order them by the eigenvalues of the generally triaxial harmonic-oscillator Hamiltonians of which their minimum-energy lowest-weight states are eigenstates.
As this section shows, the minimum-energy lowest-weight states of strongly deformed irreps then have much lower  energies than those of their  spherical harmonic-oscillator counterparts.

Such an energy-ordered sequence of Sp$(3,\Rb)$ irreps
is derived below by mean-field shape-consistency methods.
It must be recognised, however, that although the sequence of irreps obtained is much more meaningful than the partially ordered sets (with huge degeneracies) given by the energies of  spherical harmonic-oscillator lowest-weight states, it is only intended to provide a list of candidates.
A better ordering of irreps  would be given by the energies of their lowest-energy eigenstates calculated for a suitable model Hamiltonian restricted to the spaces of single irreps.
However, given the huge number of possible irreps,
such an  approach is only  feasible for refining the order of a small set of irreps prepared by other, e.g., shape-consistency, methods.
It is also possible to select the most relevant irreps from the shape-consistency list on the basis of  their ability  to describe experimental observations  as in Ref. \cite{JarrioWR91}.

To within a rotation, a lowest-weight state for an Sp$(3,\Rb)$ irrep is 
an eigenstate  $|\sigma,\omega\rangle$ of a generally triaxial harmonic oscillator     
\beq \label{eq:THO}  
\hat{\mathcal{H}}(\omega) =  \sum_{n=1}^A \sum^3_{i=1} \hbar\omega_i
(b^\dag_{ni} b_{ni}+ \tfrac12),
\eeq
with a triple of frequencies $\omega =(\omega_1, \omega_2 , \omega_3)$,
for which the harmonic-oscillator raising and lowering $b^\dag_{ni}$ and $b_{ni}$ operators are defined in terms of the nucleon position and momentum coordinates by
\beq 
\hat x_{ni} = \frac{1}{\sqrt{2}\,a_i}(b_{ni}^\dag + b_{ni}) , \quad
\hat p_{ni} = {\rm i} \hbar \frac{a_i}{\sqrt{2}} (b_{ni}^\dag - b_{ni}), 
\eeq
with generally unequal units of inverse length $a_i  = \sqrt{M\omega_i/\hbar}\,$.
The objective is then to determine the  lowest-weight state for which the energy 
$\langle \sigma,\omega|\hat H | \sigma,\omega \rangle$ 
of a chosen many-nucleon nuclear Hamiltonian $\hat H$ is minimised.

Based on experience gained from standard Hartree-Fock theory, one can be sure that, for a  many-nucleon Hamiltonian with predominantly attractive short-range interactions between the nucleons, the energy
$\langle \sigma,\omega | \hat H |\sigma,\omega\rangle$ will be minimised
when, to a good approximation, the ellipsoid defined by the quadrupole moments of the state 
$|\sigma,\omega\rangle$ has the same shape as that of the potential-energy component of 
$\hat{\mathcal{H}}(\omega)$.  
Thus, if the lowest-weight state $|\sigma\rangle$ is an eigenstate of the triaxial harmonic-oscillator Hamiltonian  
$\hat{\mathcal{H}}(\omega)$ with potential energy
\beq
\hat V =  \frac12 \sum_{ni} \omega_i^2 \hat x_{ni}^2 ,
\eeq
then the single-particle density of the lowest-weight state 
has expectation values  given by
\beq
\langle \sum_n \hat x^2_{ni} \rangle 
= \frac{\hbar}{2M \omega_i} \sum_n \langle  (b^\dag_{ni} b_{ni} +b_{ni}b^\dag_{ni})\rangle
= \frac{\hbar \sigma_i}{M\omega_i },
\eeq
and defines an $n$-independent ellipsoidal surface given by
\beq 
\sum_i \frac{x_i^2}{ \langle \sum_n\hat x^2_{ni}\rangle} 
= \sum_i \frac{M}{ \hbar\, \sigma_i \omega_i} \omega^2_i x_i^2
=\text{const.}
\eeq
It follows that the density of the lowest-weight state  has the same  ellipsoidal shape as the potential energy $\hat V$ if and only if
\beq \label{eq:shape.consistency.b}
\sigma_1\omega_1 = \sigma_2\omega_2 = \sigma_3\omega_3 .
 \eeq
This shape-consistency relationship has been used for other purposes by Bohr, Mottelson  \cite{BohrM55,BohrM75} and others \cite{CastelRZ90}.

Equation (\ref{eq:shape.consistency.b})  defines the relative values of 
 $\omega_1$, $\omega_2$ and $\omega_3$ 
 for a given Sp$(3,\Rb)$ irrep $\langle\sigma\rangle$.
 The strength of the interaction component of $\hat H$ should also be such that their magnitudes give the  volume of the deformed nucleus consistent with that expected for near-incompressible nuclear matter.
 This then determines the absolute magnitudes of the frequencies.
Given that a symplectic-model irrep is defined  by its lowest weight, 
$\sigma = (\sigma_1,\sigma_2,\sigma_3)$, it follows that the set of symplectic model irreps for a nucleus can be ordered by the energies of the corresponding shape-consistent lowest-weight states.
 
Among a set of Sp$(3,\Rb)$ irreps with a  common value of 
$N_0 = \sigma_1 + \sigma_2 + \sigma_3$, which are irreps having lowest-weight states   of  spherical harmonic-oscillator energy $N_0 \hbar\omega_0$,
it is a simple matter to   identify those of  maximal space symmetry and observe that they are the maximally deformed irreps; they have the largest values of the SU(3) Casimiar invariant $\lambda^2 + \lambda\mu + \mu^2 + 3(\lambda+\mu)$
and correspondingly the  largest values of $2\lambda+\mu$ or $\lambda+2\mu$.
The energy expectation values 
\beq \label{eq:LWtHOstate}
E_\sigma =\langle \sigma,\omega |\hat{\mathcal H}(\omega) | \sigma,\omega\rangle 
= \sum_i \hbar \omega_i \sigma_i  
= 3 (\sigma_1\sigma_2\sigma_3)^\frac13 \hbar \omega_0 , 
\quad \text{with $\omega_0^3= \omega_1\omega_2\omega_3$},
\eeq
for the corresponding shape-consistent lowest-weight states
for these irreps can then be evaluated for increasing values of $N_0$, starting from the lowest values allowed by the  Pauli exclusion principle.
Table \ref{tab:order.reps} shows the lowest energies determined in this way for three 
\begin{table}[th]  
\caption{\label{tab:order.reps} Comparison of minimum values
$E_\sigma =\langle \sigma,\omega |\hat{\mathcal H} | \sigma,\omega\rangle$
in units of  $\hbar\omega_0$  for values of $N_0 = \sigma_1+\sigma_2+\sigma_3$ increasing from the minimum value allowed by the Pauli exclusion principle.
The quantum numbers $\lambda = \sigma_1-\sigma_2$, 
and $\mu = \sigma_2-\sigma_3$ are defined as usual.
The contribution of the centre-of-mass to the energies shown has been removed.
 \vspace{0.2cm} }
  \begin{minipage}[t]{1.6in}
\centerline{${}^{12}${\rm C} }\vspace{0.1cm}
$\begin{array}{|c|c|c|c|c|c|}\hline
{N_0} &  \lambda &  \mu & 2\lambda+ \mu &  E_\sigma   \\ \hline
24.5 &   0 & 4  &  4  & 23.75 \\ 
28.5 & 12 & 0  & 24 & 24.27 \\ 
26.5 &   6 & 2  & 14 & 24.68 \\ 
30.5 & 10 & 2  & 22 & 26.91 \\
32.5 & 12 & 2  & 26 & 27.90 \\
\hline
\end{array}$
\end{minipage}   
\begin{minipage}[t]{1.7in}
\centerline{${}^{16}${\rm O} }\vspace{0.1cm}
$\begin{array}{|c|c|c|c|c|}\hline
{N_0} &  \lambda &  \mu & 2\lambda+ \mu  & E_\sigma   \\ \hline
34.5 &    0 &  0  & 0   & 34.50 \\ 
38.5 &    8 &  4  & 20 & 35.68  \\ 
36.5 &    4 &  2  & 10 & 35.78 \\ 
46.5 &  24 &  0  & 48 & 36.30 \\
42.5 &  16 &  2  & 34 & 36.61\\ 
40.5 &  10 &  4  & 28 & 36.86 \\ 
\hline
\end{array}$
\end{minipage} 
\begin{minipage}[t]{1.7in}
\centerline{${}^{168}${\rm Er} }\vspace{0.1cm}
$\begin{array}{|c|c|c|c|c|}\hline
{N_0} &  \lambda &  \mu & 2\lambda+\mu &  E_\sigma  \\ \hline
812.5 &  30  &   8   &  68  & 811.11 \\ 
824.5 &  96  &  20  & 212 &  811.38 \\
822.5 &  82  &  26  & 190 & 811.47 \\
826.5 & 104 &  20  & 228 &  811.49 \\
814.5 &  40  &  16  &  96  &  811.51 \\
820.5 &  70  &  28  & 168 &  811.53 \\
816.5  &  52 &  20  & 124 &  811.58 \\
818.5  &  60 &  26  & 146 &  811.59 \\ 
828.5  & 114 & 16  & 244 &  811.66 \\ 
\hline	
\end{array}$
\end{minipage} 
\end{table}
doubly-even nuclei, in order of increasing energy, for a range of $N_0$ values.
An immediate observation is that the order by increasing values of $E_\sigma$ differs markedly from the ordering by increasing values of $N_0$ as  would be obtained if the 
minimum-energy lowest-weight states were  eigenstates  of spherical harmonic-oscillator Hamiltonians.
As expected \cite{CastelRZ90,Rowe16}, it is found that the minimum  energies of the shape-consistent lowest-weight states obtained in this way are almost invariably those with
$\lambda >\mu$, consistent with the observed dominance of prolate over oblate deformations.

A  satisfying property of these simply derived results is that the lowest three irreps obtained for  $^{12}$C  and $^{16}$O are precisely those determined to be appropriate for describing the low-energy positive-parity states of these nuclei; cf.\ 
\cite{DreyfussLDDB13} for $^{12}$C and \cite{RoweTW06} for $^{16}$O.
It is  remarkable  that  the  irreps shown for $^{168}$Er 
have lowest-weight states  that range over 16 spherical harmonic oscillator 
$\hbar\omega_0$ shells but have  triaxial harmonic-oscillator energies 
with a spread of only $\sim 0.5 \hbar\omega_0$.
These results are particularly significant because they explain the emergence of shape coexistence \cite{HeydeW11,WoodH16} in a dramatic way  without the use of adjustable parameters.
Another  characteristic is that the sets of lowest-weight states of nearby energies tend to have significantly different values of $N_0$ and substantially different deformation shapes.
As a result, the  Sp$(3,\Rb)$ irreps of similar energies will  mix much less than would occur for irreps of similar deformations.
This is an indication that Sp$(3,\Rb)$ is likely to be a considerably better dynamical symmetry than might otherwise have been expected.
It also  indicates that it is hopeless to give a thought to ever being able to perform realistic many-nucleon shell-model calculations in a conventional spherical harmonic-oscillator basis for heavy rotational nuclei.

The above shape-consistency results suggest that it would be profitable to explore the more reliable ordering of minimum-energy lowest-weight states for realistic Hamiltonians  restricted to the triaxial harmonic-oscillator lowest-weight states of each Sp$(3,\Rb)$ irrep.
In identifying the appropriate irrep for the lowest-rotational states of a given nucleus, 
one should also be aware 
that the states of lowest angular momentum of a symplectic model irrep are expected to have energies significantly below that of its minimum-energy lowest-weight state and to be lowered most for the most-strongly deformed irreps. 
For example,   although $E_\sigma$ for the less-deformed  shape-consistent lowest-weight state for the irrep with $N_0=812.5$ lies slightly lower in energy than 
that of the next lowest irrep with $N_0=824.5$, it is not surprising to find that the observed low-energy rotational states of $^{168}$Er are closer to those of the 
$N_0 =824.5$ irrep \cite{JarrioWR91}.
This lowering could be estimated from an evaluation of the expectation value of 
$\hat L \cdot \hat L$ in the minimum-energy lowest-weight state and the expected moment of inertia for the nucleus.

  Note also that the weights $\langle \sigma\rangle$ of  Sp$(3,\Rb)$ irreps needed to describe the observed E2 transitions and quadrupole moments of a number of heavy rotational nuclei,  have been estimated by Jarrio \emph{et al.}\ \cite{JarrioWR91}  and, from a theoretical perspective of what is possible, by Carvalho \cite{CarvalhoR92}, and are qualitatively in agreement with the above results.

\subsection{Generator-coordinate theory for  mixed Sp$(3,\Rb)$ many-nucleon irreps}
\label{sect:GCT}
The  potential for a dramatic reduction in the computational complexity of calculations for  microscopic Hamiltonians in spaces of multiple symplectic model irreps
by generator coordinate methods, coupled with the insights gained from the AMF perspective, raises expectations for the achievement of accurate results from rapidly converging sequences of calculations in sub-spaces of relatively low dimension.
As illustrated in a simple application of the generator coordinate method to the $L=0$ states of  mixed Sp$(3,\Rb)$ irreps by Carvalho \emph{et al.}\ \cite{CarvalhoVR87},
it is profitable to proceed by diagonalising a nuclear Hamiltonian in  spaces 
of states, that are angular-momentum projected from an increasing number of optimally selected lowest-weight states from  a set of Sp$(3,\Rb)$ irreps, until converged results are obtained to  the desired level of accuracy for the chosen set of irreps.

Such a procedure could be a solution to the  challenge of
handling the mixing of states of different
${\rm Sp}(3,\Rb) \times {\rm U(4)}$ irreps by those of a competing dynamical symmetry.
The concept of quasi-dynamical symmetry was introduced in a study of such systems with competing dynamical systems \cite{RoweRR88} to understand the apparent persistence of the sub-dynamics of a system when it is adiabatic relative to the competition.
Such quasi-dynamical symmetries,  reviewed in Ref.\ \cite{RoweCancun04},  have  been
shown to emerge in many examples
\cite{RochfordR88,BahriRW98,BahriR00,HessAHC02,Rowe04,Rowe04b,RosensteelR05}.

For a rotor model, quasi-dynamical symmetry  is realised as a coherent linear sum of many
similar irreps,  all of which vary in a similar way with respect to their angular-momentum content so that the result is essentially their average.
Thus, it is apparent that, when there are  competing dynamical symmetries, 
the dynamical symmetry of an adiabatic rotational model
becomes a quasi-dynamical symmetry and is easily handled.
Within the spirit of the Bohr-Mottelson-Nilsson unified model, one has only to 
replace the shape-consistent lowest weight of a single Sp$(3,\Rb)$ irrep by an eigenstate of the nuclear Hamiltonian in the space of the shape-consistent lowest-weight states of the several mixed Sp$(3.\Rb)$ irreps which then becomes the intrinsic state for one of several mixed collective-model representations.
However, the possibilities of such a potentially powerful approach to the symmetry mixing of symplectic irreps requires  investigation.


\section{Concluding remarks}     
For many years, the Nilsson model \cite{Nilsson55} has been understood as providing the intrinsic states for the unified model  of rotational nuclei.
In doing so, it has been interpreted as a phenomenological approximation to HF theory.
This perspective is now given a more fundamental foundation in which a Hartree-Fock solution is identified as a minimum-energy lowest-weght state for the unique totally anti-symmetric irreps of the Lie group of one-body unitary transformation of nuclei.
There is, however, a problem with the underlying Hartree-Fock theory in that, although it is in principle a complete theory, in the sense that a set of Slater determinants spans the whole Hilbert space of a nucleus and although the Hartree-Fock manifold of Slater determinants has many local minimum-energy states, it is not clear how these local minimum energy states
relate to the excited rotational bands of a nucleus.
This problem is now resolved by extending Hartree-Fock  theory to an algebraic mean-field theory and applying it to the irreps of the symplectic model, each of which has a single minimum-energy lowest-weight state and which together form a complete set of intrinsic states for the collective model states of a nucleus.

The irreps of the symplectic model for a nucleus have many invaluable properties.
In particular, there are no isoscalar E2 transitions between the states of its different irreps
 and all matrix elements of the nuclear kinetic energy and of other elements of the Sp$(3,\Rb)$ Lie algebra have vanishing values between the states of diferent irreps.   
 Thus, the irreducible Sp$(3,\Rb)$ subspaces are completely independent
collective-model subspaces of the many-nucleon Hilbert space.
Their study is then a profitable first step towards understanding the rotational dynamics of nuclei.
At the same time, it is  recognised that coherent mixtures of states from the different collective subspaces can result in the  emergence of new and unexpected properties.
For example, the correlations of nucleon dynamics induced by pairing interactions could no doubt result in more superfluid-like flows, especially in states of small deformation.
Thus, in reality one can expect there to be a competition between  different dynamical symmetries and the emergence of rotational states characterised by quasi-dynamical symmetries as discussed in Section \ref{sect:GCT}.

Several approximations for the  calculation of nuclear rotational properties with many-nucleon Hamiltonians in  spaces of  single or a few Sp$(3,\Rb)$ irreps have been outlined in this review, although in most cases further developments of the technical tools for such calculations are needed.
Moreover, to assess the accuracy of these approaches, comparisons with precise calculations
are required for at least a  few Sp$(3,\Rb)$ irreps of medium-mass nuclei of modest deformation.
Fortunately, it appears that such calculations are now becoming possible with the supercomputer program of the LSU group as discussed in Section \ref{sect:LSUprog}.

The remarkable successes of the Bohr-Mottelson collective and unified models leave little doubt that nuclei  have states with the essential properties of the symplectic models.
However, in seeking to understand the   dynamics of nuclear rotations many questions  arise of which the following are a few that emerge.

\begin{enumerate}

\item
What is the significance of the cranking model moments of inertia, which in one situation gives  moments of inertia corresponding to the kinetic energies of a rigid rotor and, in another, the irrotational-flow moments of inertia of a quantum fluid?

It has traditionally been supposed that an observed moment of inertia is a mass parameter in the expression of the kinetic energy of a rotating nucleus.
However, this is clearly inappropriate if the so-called rotational energies are not predominantly kinetic energies.  
Questions also arise regarding the interpretation of nuclear moments of inertia as being  linear combinations of those for irrotational and rigid flows and regarding the magnitude  of the inertial (Coriolis and centrifugal) forces, if rotational energies are only fractionally kinetic energies.

\item
To what extent are the symplectic model irreps, that are best able to describe the low-energy rotational states of doubly-even nuclei, among the leading irreps  of maximal space symmetry given by shape-consistency criteria?
How should the criteria be modified for the ${\rm Sp}(3,\Rb) \times {\rm U}(4)$  irreps of other than doubly-even nuclei?

\item
Is there a better  algorithm, than that of computing the energies of shape-consistent lowest-weight states, for a practical  ordering of symplectic irreps and for identifying those that contribute most strongly to the observable lower-energy states of nuclei?

\item
What evidence is there for  low-energy  beta and gamma vibrational bands?
And why is  the axially symmetric unified model  as successful as it is,
when the minimum-energy shape-consistent lowest-weight states available to a nucleus,
as defined by its maximally decoupled  irreducible Sp$(3,\Rb)$ subspaces,
have triaxial shapes much more frequently than axially symmetric shapes?

The existence of beta vibrations  has been questioned in recent years
\cite{Garrett01,GarrettW10} in the light of experimental observations of the decay properties of  possible candidates.  
One can also question the existence of gamma vibrations.
From a symplectic model perspective, it is expected that with no mixing of its irreps, the vibrations of the intrinsic states of rotational nuclei would only have vibrational states at giant-resonance energies.
It is also expected that the larger the intrinsic deformation of a nucleus, the more rigid it will be and the less it will mix with other irreps,
Of particular note is the observations that, in listing and ordering the symplectic irreps available to a nucleus by their generally triaxial harmonic-oscillator energies, as in Table \ref{tab:order.reps}, 
there are relatively few axially symmetric irreps.
More precisely, if  the length scales of the shape-consistent lowest-weight state are defined by their Sp$(3,\Rb)$ quantum numbers $(\sigma_1,\sigma_2,\sigma_3)$, it is found that
 $\lambda =\sigma_1-\sigma_2$ is generally substantially larger than 
 $\mu=\sigma_2-\sigma_3$, consistent with the observation that nuclear deformations are  more  prolate than oblate \cite{CastelRZ90,Rowe16}.
However,  it appears that $\mu$, which is a measure of  axial asymmetry, is rarely zero.
Nevertheless, the description of rotational states with an axially symmetric rotor model has been remarkably successful.
 This suggests that there might be a restoration of axial symmetry about the major axis which could be due to the mixing of states from other Sp$(3,\Rb)$ irreps, e.g., by pairing interactions.
 However, the coherent mixing of irreps in the restoration of axial symmetry implies  that axially symmetric nuclei can have low-energy gamma vibrational excited states as suggested  in Ref. \cite{CarvalhoLeBVRMcG86}.
Maybe the standard HF and HFB models \cite{BenderHR03,MatsuyanagiMNHS10,MatsuyanagiMNYHS16} are already able to address  this question!

\item
  A complementary question is; what is the experimental evidence for rotational states of triaxial nuclei nuclei?

The general triaxial rotor model for an even-even  nucleus  has mixed
$K=0,2,4,\cdots$ sequences of rotational bands whereas the axially symmetric rotor model predicts a $K=0$  band, a one-phonon $K=2$ gamma-vibrational band, two-phonon $K=0$ and 4 bands, and so on.  
However, there appears to be little experimental evidence of either two-phonon gamma bands or of  $K\geq 4$ states of a triaxial rotor.
Maybe they exist but are not  observed.
Recall also that there is little unequivocal evidence of multi-phonon excitations of any kind in nuclei.
Thus, it is of fundamental  interest to explore the extent to which a triaxial rotor model is able to fit the observed data of rotational nuclei \cite{AllmondW17}  
and to explore the  existence or non-existence of  $K=0$ and $K=4$ rotational bands that are strongly coupled to the first excited $K=2$ bands of even-even rotational nuclei.

\item
To what extent is it possible to determine the goodness of the $K$ quantum number and hence the mixing of rotational bands in the rotor-model interpretation of nuclear data?

As shown in  Appendix \ref{sect:app.B}, $K$ is defined as an integer-valued quantum number for the basis states of the rotor model and, because SU(3) states can be expressed, 
in a vector coherent-state representation, as  linear combinations of rigid rotor-model states, it also has  remarkably well-defined essentially integer values for appropriately defined basis states of  a U(3) irrep.
However, the $K$ quantum number is conserved only for the states of an axially symmetric rotor.
Thus, measurements of the mean values of $K$ for the rotational states of a nucleus provide a measure of triaxiality.

\end{enumerate}

In addressing  these questions, there is much to be gained by comparing applications of the HF, HFB and AMF-symplectic models with one another and experimental data.
It must be emphasised,  however, that there is no reason to expect that the AMF-symplectic model calculations, when restricted to a single Sp$(3,\Rb)$ irrep will fit  experimental data any better than standard HF and HFB calculations for the same Hamiltonian when applied to  a ground-state rotational band.  
On the contrary, given that  the minimum-energy lowest-weight states of symplectic model irreps are most likely to be single Slater determinants when combined with appropriate spin and isospin wave functions,  the standard HF and HFB models have the potential, to give better results,  as far as the ground-state band of collective states is concerned, and also provide information on the  mixing of irreps and  some of the questions posed above.
Thus, it must be emphasised that  the  objective of this review is not simply  to fit  rotational data but to develop tools for understanding the nature of rotational dynamics in nuclei.
Thus, comparisons of HF, HFB, AMF and experimental results are expected to provide more insights and answers to important questions  than either approach on its own.
The most important contribution of the AMF-symplectic model is that it facilitates the analyses of the dynamics of idealised (that is decoupled) collective states and, hence, a means to interpret experimental data in terms of a mixing of such states.
If everything works out in a logically consistent way, we can then be satisfied.
However, it is perhaps more likely that unexplained phenomena will emerge which will present new and exciting challenges

The appendices to this review include brief summaries of some developments of importance in the applications of the symplectic model for which details can be found in the literature.

\acknowledgments{The author is pleased to acknowledge many helpful discussions of this review with J. L. Wood and his proof reading of the manuscript.}

\begin{appendix}

\section{SU(1,1) scale transformations of an Sp$(3,\Rb)$ lowest-weight state} \label{sect:SU11Scale}
A pair of position and momentum operators for a particle satisfies the Heisenberg commutation relations $[\hat x,\hat p] = {\rm i} \hbar$ and can be expressed in terms of the raising and lowering operators of a simple harmonic-oscillator Hamiltonian 
\beq 
\hat h(\omega_0) =  \frac{1}{2M} \hat p^2 
+ \frac12  M\omega_0^2 \hat x^2 
\eeq
of frequency $\omega_0$ with
\beq  \label{eq:xpc}
\hat x = \frac{1}{\sqrt{2}\,a}(c^\dag + c) , \quad
\hat p = {\rm i} \hbar \frac{a}{\sqrt{2}}(c^\dag - c), 
\eeq
where $a  = \sqrt{M\omega_0/\hbar}$ is a unit of inverse length, and  
\beq
[c^\dag, c^\dag] =[c, c] = 0, \quad
[c, c^\dag] = 1 .
\eeq
An SU(1,1) scale transformation  of these operators, that preserves their Heisenberg commutation relation, is then given by
\beq
  \hat x \to \hat S(\epsilon) \hat x \hat S(-\epsilon)
 = e^{-\epsilon} \hat x, \quad  
 \hat p \to \hat S(\epsilon) \hat p \hat S(-\epsilon) = e^{\epsilon} \hat p , 
\eeq 
with 
\beq
\hat S(\epsilon) = e^{\epsilon (\hat S_+ -\hat S_-)}
\eeq
and
\beq
\hat S_+ = \tfrac12 c^\dag c^\dag, \quad
\hat S_- = \tfrac12 c\, c, \quad
\hat S_0 = \tfrac14 (c^\dag c + c\, c^\dag) .
\eeq
Thus, a harmonic-oscillator Hamiltonian with a scaled
potential is  given by the transformation
\beq
\hat h(\omega_0) \to \hat h(\omega) 
=  e^{-2\epsilon} \hat S(\epsilon) \hat h(\omega_0)\hat S(-\epsilon)
= \frac{1}{2M} \hat p^2 + \frac12  M\omega^2 \hat x^2 ,
\eeq
with $e^{-2\epsilon}= \omega/\omega_0$.

A parallel expression gives the  many-particle Hamiltonian
\beq 
\hat h(\omega) =  \frac{1}{2M}\sum_n \hat p_n^2 
+ \frac12  M\omega^2 \sum_n\hat x_n^2  
\eeq 
as a transformation
\beq
\hat h(\omega) 
=  e^{-2\epsilon} \hat S(\epsilon) \hat h(\omega_0)\hat S(-\epsilon)
\eeq
of the  Hamiltonian  $\hat h(\omega_0)$, with
\beq 
\hat S(\epsilon) = \exp \Big[ \tfrac12 \epsilon  
 \sum_n\big(c^\dag_n c^\dag_n - c_n c_n \big) \Big].
\eeq
It follows that, if $|0,\omega_0\rangle$ is the ground state of  
$\hat h(\omega_0)$, the ground state of  $\hat h(\omega)$ is given by
 \beq \label{eq:SU11transf}
|0, \omega\rangle =
\hat S(\epsilon) |0,\omega_0\rangle .
\eeq
The $\{ C_\nu(\epsilon)\}$ coefficients in the expansion
\beq \label{eq:expansionS(e)}
\hat S(\epsilon) |\sigma,\omega_0\rangle
= \sum_{\nu \geq 0}
C_\nu(\epsilon) |\sigma + 2\nu,\omega_0\rangle
\eeq
can now be derived.  

In its fundamental matrix representations, an SU(1,1) group element is a complex matrix of the form
\beq 
g = \left(\begin{matrix}  a& b \\ b^*&a^*  \end{matrix} \right) , \quad
\text{with} \;\; |a|^2-|b|^2 = 1,
\eeq
and its Lie algebra is  expressed in terms of the matrices
\beq \label{eq:su11mats}
S_+\! =\! \left(\begin{matrix}  0&-1 \\ 0&0  \end{matrix} \right) , \;\;
S_-\! =\! \left(\begin{matrix}  0&0 \\ 1&0  \end{matrix} \right) , \;\;
S_0\! =\! \left(\begin{matrix}  \tfrac12& 0 \\ 0&-\tfrac12  \end{matrix} \right) ,
\eeq 
which satisfy the commutation relations
\beq
[S_0, S_\pm ] = \pm S_\pm , \quad [S_-, S_+] = 2S_0 .
\eeq

An SU(1,1) matrix has a so-called Gauss factorisation
\beq 
\left(\begin{matrix}  a& b \\ b^*&a^*  \end{matrix} \right)
 = \left(\begin{matrix}  1& b/a^* \\ 0 &1  \end{matrix} \right)
 \left(\begin{matrix}  1/a^*& 0\\ 0 &a^*  \end{matrix} \right)
 \left(\begin{matrix}  1& 0 \\ (b/a)^* &1  \end{matrix} \right) .
 \eeq
The Lie algebra element $S(\epsilon) = \exp{[\epsilon  (S_+-S_-)]} \in$ SU(1,1)  
is then the matrix
\beq
S(\epsilon)  = 
 \exp{\left[ -\epsilon
\left(\begin{matrix} 0&1\\1&0 \end{matrix} \right)\right]} 
=\left(\begin{matrix}\, \cosh{\epsilon}&-\sinh{\epsilon} \\
- \sinh{\epsilon}&\,\cosh{\epsilon} \end{matrix} \right) 
\eeq 
and has the factored form
\beq 
S(\epsilon) \!=\!
\left(\begin{matrix} 1&\! - \tanh{\epsilon}\\ 0&1 \end{matrix} \right)
\left(\begin{matrix} 
(\cosh{\epsilon})^{-1} &\!\!0\\\!\! 0 &\!\!\!\!\!\! \cosh{\epsilon} \end{matrix} \right)
\left(\begin{matrix} 1 &0\\
- \tanh{\epsilon}&1 \end{matrix} \right). 
\eeq 
Hence, with  $\gamma(\epsilon)$ defined such that
$e^{\gamma(\epsilon)/2} \!=\!  (\cosh{\epsilon})^{-1}$ and
\beq \label{eq:expS0}
 \left(\begin{matrix}   
(\cosh{\epsilon})^{-1} &\!\!0\\\!\! 0 &\!\!\!\!\!\! \cosh{\epsilon} \end{matrix} \right)
=\left(\begin{matrix}  e^{\gamma(\epsilon)/2} & \!\!0 \\ 
                                 0 &\!\!\!\!e^{-\gamma(\epsilon)/2}  \end{matrix} \right) 
=e^{\gamma(\epsilon) S_0}    ,
 \eeq   
 and  the observation that
 \beq
 \left(\begin{matrix}   1 &x \\ 0 & 1\end{matrix} \right) 
= \exp  \left(\begin{matrix}   0 &x \\ 0 & 0\end{matrix} \right) ,
\eeq
it follows that  $S(\epsilon)$ is represented as an operator
\beq
\hat S(\epsilon)\! = \!
\exp\!\big[(\tanh{\epsilon)\, \hat S_+}\big]
\exp\!\big[\gamma(\epsilon) \hat S_0 \big]
\exp\!\big[\! -\! ( \tanh{\epsilon)\, \hat S_-} \big] .
\eeq
It also follows from the expansion
\beq \label{eq:exptS+}
\exp\!\big[(\tanh{\epsilon})\,\hat S_+\big] 
= \sum_\nu \frac{(\tanh{\epsilon})^\nu \big(\hat S_+\big)^\nu}{\nu!}, 
\eeq           
that, for an SU(1,1) lowest-weight state 
$|\sigma\rangle$ for which
$ \hat S_-  |\sigma\rangle = 0$ and 
$\hat S_0 |\sigma\rangle = \tfrac12 \sigma |\sigma\rangle$,
\beq \label{eq:2exp.tS+}
\hat S(\epsilon) |\sigma\rangle = 
e^{\sigma \gamma/2} \exp\!\big[(\tanh{\epsilon})\hat S_+\big] |\sigma\rangle =
\frac{1}{(\cosh \epsilon)^\sigma}
\sum_\nu \frac{(\tanh \epsilon)^\nu}{\nu!} 
\big(\hat S_+\big)^\nu |\sigma\rangle .
\eeq

Finally, the $C_\nu(\epsilon)$   coefficients in 
Eqn.\,(\ref{eq:expansionS(e)})  are obtained from the known expression of a unitary SU(1,1) irrep \cite{VanderJeugtJ98}, with lowest-weight state
$|\sigma\rangle$ and
basis states $ \{ |\sigma +2\nu\rangle, \nu = 0,1,2,\dots\}$, given by
\bal
&\hat S_0 |\sigma +2\nu\rangle = 
       \tfrac12 (\sigma + 2\nu) |\sigma +2\nu\rangle , \label{eq:S0}\\
&\hat S_+ |\sigma +2\nu\rangle = 
       \sqrt{(\sigma+\nu)(\nu+1)}  |\sigma +2\nu+2\rangle , \\
&\hat S_- |\sigma +2\nu\rangle = 
       \sqrt{(\sigma+\nu-1)\nu}  |\sigma +2\nu-2\rangle.
\end{align}
It follows that  
\beq \label{eq:S(e)1}
\exp\!\big[(\tanh{\epsilon})\hat S_+\big] |\sigma\rangle =
\sum_\nu (\tanh{\epsilon})^\nu 
\sqrt{\frac{(\sigma+\nu-1)!}{(\sigma-1)! \,\nu!}}\,
|\sigma + 2\nu\rangle 
\eeq
and, from Eqn.\ (\ref{eq:2exp.tS+}), that 
\beq \label{eq:S(e)}
\hat S(\epsilon) |\sigma\rangle = \sum_{\nu \geq 0}
C_\nu(\sigma,\epsilon) |\sigma + 2\nu\rangle ,
\eeq
with 
\beq \label{eq:S(e)3}
C_\nu(\sigma,\epsilon) =  \frac{1}{(\cosh \epsilon )^\sigma}
(\tanh{\epsilon})^\nu 
\sqrt{\frac{(\sigma+\nu-1)!}{(\sigma-1)! \,\nu!}} .
\eeq

\section{Vector-coherent-state (VCS) representations of SU(3) in an SO(3) basis}
\label{sect:app.B}
In the notations of Section \ref{sect:sp3R.basis}, the U(3) Lie algebra is expressed as linear combinations of  operators
\beq 
\hat{\mathcal C}_{ij} = \sum_n c^\dag_{ni} c_{nj}, \quad i,j = 1,2,3,
\eeq
 that satisfy the commutation relations
\beq [\hat{\mc{C}}_{ij}, \hat{\mc{C}}_{kl}] 
= \delta_{j,k} \hat{\mc{C}}_{il} - \delta_{i,l} \hat{\mc{C}}_{jk} .
\label{eq:Cijcoms}
\eeq
It is also expressed in terms of the angular-momentum and quadrupole operators defined in a spherical-tensor basis by
\begin{subequations}\label{eq:6.L_k2}
\bal 
&\hat L_0 = \hat L_{23}  ,\quad 
\hat L_{\pm1} =\mp\tfrac{1}{\sqrt{2}}( \hat  L_{31} \pm {\rm i}  L_{12}),  \\
&\hat{\mc{Q}}_{2,0}  = 
    2  \hat{\mc{Q}}_{11} -  \hat{\mc{Q}}_{22}-  \hat{\mc{Q}}_{33}, \\
&   \hat{\mc{Q}}_{2,\pm 1} = 
  \mp \sqrt{6}\, ( \hat{\mc{Q}}_{12} \pm {\rm i}  \hat{\mc{Q}}_{13}), \\
&\hat{\mc{Q}}_{2,\pm 2} =  \textstyle\sqrt{\frac{3}{2}}\, 
 ( \hat{\mc{Q}}_{22}-  \hat{\mc{Q}}_{33} \pm 2{\rm i}  \hat{\mc{Q}}_{23}),
\end{align}
\end{subequations}
in which   
\beq
\hat L_{ij}  =  -{\rm i} ( \hat {\mathcal{C}}_{ij} - \hat {\mathcal{C}}_{ji} ) ,
\quad \hat{\mathcal{Q}}_{ij} = 
\textstyle\frac12\big( \hat {\mathcal{C}}_{ij} 
+ \hat {\mathcal{C}}_{ji} \big) . \label{eq:LijQij}
\eeq

 In a  VCS expression \cite{RoweLeBR89,Rowe12,RoweMC16}
of an arbitrary SU(3) representation $(\lambda\,\mu)$,
in which  $\lambda$ and $\mu$ are  positive integers  \cite{Elliott58ab}, 
the SU(3) quadrupole operators are represented as operators
 \bal \label{eq:GammaQ}
 \hat\Gamma({\cal Q}_{2\nu})  
 = (2\lambda+\mu+3) \hat\scrD^2_{0\nu} -
\tfrac12\,\big[ \hat{\bf L}\cdot \hat{\bf L}, \hat\scrD^2_{0\nu}\big]
+ \sqrt{6}\big( \hat \sigma_+\hat\scrD^2_{2\nu} +
 \hat \sigma_-\hat\scrD^2_{-2,\nu}\big) , 
\end{align}
that act on linear combinations of rotor-model wave functions of the form
 \bal 
 \Phi^{(\lambda\mu)}_{KLM}&(\Omega) 
 =   \sqrt{ \frac{2L+1}{16\pi^2(1+\delta_{K,0})}}  \label{eq:6.phiKLM}  
\big( \xi_{K}\scrD^L_{KM}(\Omega) +
(-1)^{\lambda+L+K} \xi_{-K} \scrD^L_{-K,M}(\Omega)\big) .
\end{align}
In this representation the operator  $\hat\scrD^2_{\mu\nu}$  acts
multiplicatively, i.e.,
\beq \hat\scrD^2_{\mu\nu} \Psi(\Omega)
= \scrD^2_{\mu\nu}(\Omega)\,\Psi(\Omega) ,
\eeq
and $\hat \sigma_{0}, \hat \sigma_{\pm}$  operate on  intrinsic-spin states 
$\{ \xi_K, K= -\mu, -\mu+2, \dots , \mu\}$
according to the SU(2) equations
\begin{subequations}\label{eq:S+-ops.1}
\bal 
\hat \sigma_0 \xi_K = \tfrac12 K \,\xi_K ,\quad
\hat \sigma_\pm \xi_K = \tfrac12 \sqrt{(\mu \mp K)(\mu\pm K+2)}\,\xi_{K\pm 2} ,
\end{align}
\end{subequations}
with $K$ taking integer  values in the range $-\mu, -\mu+2, \dots , \mu$.

In spite of the close similarity between the above VCS representations of SU(3) and the standard irreps of the rigid-rotor model, there is a fundamental difference; 
for, whereas the rotor-model irreps are of infinite dimension, those of SU(3) are finite.
This emerges explicitly in the VCS representation $\hat\Gamma({\cal Q}_{2\nu})$
of Eqn.\ (\ref{eq:GammaQ}) for which the matrix elements of the quadrupole operators  between states of increasing angular momentum vanish at some value of $L$ at which a band of states terminates  because of the increasingly negative term 
$-\tfrac12\,\big[ \hat{\bf L}\cdot \hat{\bf L}, \hat\scrD^2_{0\nu}\big]$
in this equation.
However, it is apparent that, for values of $ 2\lambda+\mu+3 \gg L$ and $\mu\gg K$,
 the matrix elements of the quadrupole  operators of SU(3) approach those of a corresponding rigid-rotor irrep with
 \beq \label{eq:contract}
 \hat\Gamma({\cal Q}_{2\nu})  \to (2\lambda+\mu+3) \hat\scrD^2_{0\nu} 
+ \sqrt{\tfrac32\mu(\mu+1)} \big( \hat\scrD^2_{2\nu} + \hat\scrD^2_{-2,\nu}\big) .
 \eeq
and, therefore,  that $K$ should become a good quantum number in the  
$\lambda/L\to\infty$ and $\mu/K \to\infty$ limit.

\section{The $K$ quantum number in the SU(3) model}
\label{sect:app.B}
The presence of the 
$\tfrac12\big[ \hat{\bf L}\cdot \hat{\bf L}, \hat\scrD^2_{0\nu}\big]$
term in the VCS expression of the SU(3) quadrupole operators has the result that 
the VCS quadrupole operators are not Hermitian relative to the rotor-model inner product for which the wave functions $\{ \Phi^{(\lambda\mu)}_{KLM}(\Omega)\}$
are an orthonormal basis.
The adjustment of the inner product to that of a finite-dimensional SU(3) irrep is straightforward \cite{RoweLeBR89,Rowe12},
but has the result that $K$ is no longer a precise integer-valued quantum number.  
Nevertheless, it transpires that an orthonormal SO(3) basis for an SU(3) irrep can be defined \cite{RoweT08} as eigenstates of a linear combination of SO(3)-invariant operators that have expectation values of $K$ that take integer values to an extraordinarily high degree of accuracy.

The $K$ quantum number for a rigid-rotor model is a  component of angular momentum relative to an axis of an intrinsic  body-fixed frame in which the Cartesian quadrupole moments of the nucleus are diagonal.
This is a frame in which
 the $L=2$ spherical tensor components of the Cartesian quadrupole moments
\bal \label{eq:2.Qmoments}
\hat q_0 =\sqrt{\tfrac{5}{16\pi}}\, (2\hat Q_{11}-\hat Q_{22}-\hat Q_{33}), \quad 
\hat q_{\pm 1} =\mp \sqrt{\tfrac{15}{8\pi}}\, (\hat Q_{12} \pm {\rm i}\hat Q_{13}),
\quad \hat q_{\pm 2} =\sqrt{\tfrac{15}{32\pi}}\, 
(\hat Q_{22}-\hat Q_{33} \pm 2{\rm i}\hat Q_{23}),
\nonumber
\end{align}
can be expressed in the familiar form
\beq
\bar q_0 = \beta\cos\gamma, \quad \bar q_{\pm 1} = 0,
\quad \bar q_{\pm 2} = \frac{1}{\sqrt2} \beta\sin\gamma .
\eeq
The $K$ quantum number is then defined in this intrinsic frame as the  $\hat L_0$ component of the rotor's angular momentum relative to the 1-axis.
This definition is  model dependent.
However, it has been determined \cite{RoweT08} that,  
for the states of an arbitrary rigid  rotor classified by $KLM$ quantum numbers, 
the ratios of reduced matrix elements
\bal \label{eq:RLK}
R(L,K)= \frac{\langle K+2,L 
\|\big(\hat L\otimes [\hat Q_2\otimes Q_2]_2 \otimes\hat L\big)_0\|KL\rangle}
{\langle K+2,L \|\big(\hat L\otimes Q_2 \otimes\hat L\big)_0\|KL\rangle}
= \frac{
\langle K+2,L \| [\hat Q_2\otimes Q_2]_2  \|KL\rangle}
{\langle K+2,L \| Q_2 \|KL\rangle}
\end{align}
take the precise value
\beq 
R(L,K) =\sqrt{\tfrac87}\, \bar q_0.
\eeq
This means that rotor-model states with good $K$ quantum numbers are eigenstates of the SO(3) scalar operator
\beq \label{eq:Rotor_SU3.Zrot}
 \hat {\mathcal{Z}}^{({\rm rot})} =
[\hat L \otimes[\hat{{Q}}_2 \otimes\hat{{Q}}_2 ]_2 \otimes \hat L]_0 
-  \sqrt{\tfrac87}\, \bar q_0 
[\hat L \otimes \hat{{Q}}_2 \otimes \hat L]_0  .
\eeq
A parallel definition is now adopted for an SU(3) irrep.

For large values of   $2\lambda +\mu$ and $L\ll 2\lambda +\mu$,  the term 
 $\tfrac12\,\big[ \hat{\bf L}\cdot \hat{\bf L}, \hat\scrD^2_{0\nu}\big]$ 
 in Eqn.\ (\ref{eq:GammaQ}) is comparatively negligible and, for values of $K\ll \mu$,
 the VCS expression $\hat\Gamma({\cal Q}_{2\nu})$ becomes identical
  to its rotor-model counterpart with intrinsic quadrupole moments
 \beq \label{eq:su3.rot}
 q_0 = 2\lambda+\mu+3,  \quad  q_{\pm 1} =0,
\quad  q_{\pm 2} = \sqrt{\tfrac32}\, (\mu+1) .
\eeq
This is a so-called contraction or macroscopic limit.
Thus, a resolution of the multiplicity of SO(3) states of the same angular momentum in an SU(3) irrep that will give states of good $K$ quantum numbers in this  contraction limit is obtained by requiring the SU(3) basis states
to be eigenstates of both $\hat{\bf L}\cdot \hat{\bf L}$ and the linear combination
 \beq \label{eq:ZSU3op}
\hat {\mathcal{Z}}^{(\lambda\mu)} 
=[\hat L \otimes[\hat{\mathcal{Q}}\otimes\hat{\mathcal{Q}}]_2 
\otimes \hat L]_0  - \sqrt{\tfrac87}\, (2\lambda+\mu +3)
 [\hat L \otimes \hat{\mathcal{Q}}_2 \otimes \hat L]_0  
\eeq
of the SO(3)-invariant  operators.
The eigenstates $\{ |\alpha LM\rangle\}$ obtained are then expressed in a VCS representation as linear combinations of rotor-model states with good $K$ quantum numbers
\beq
 |\alpha LM\rangle = \sum_K  |KLM\rangle {\mathcal K}_{K\alpha}.
 \eeq
Thus, it is simple to determine the mean values of $K$ in this orthonormal basis.
 
Some mean values of $K$, taken from Ref.\ \cite{RoweMC16}, are shown in Table \ref{tab:Kqno}.
\begin{table}[pth]
\caption{\label{tab:Kqno} 
Comparison of the   $K$ quantum number and diagonal quadrupole matrix element 
$ \langle (\lambda\mu)\alpha L \| \hat{\mathcal{Q}}_2 
\|(\lambda\mu)\alpha L\rangle$ computed  in their SU(3) and rotor-model limits.
The $K$ quantum number is denoted by 
$K^{(\rm rot)}_{\alpha L}$ in the rotor-model limit for a state 
$|(\lambda\mu)\alpha LM\rangle$ and its mean value in the corresponding SU(3) state is denoted by $\langle K\rangle_{\alpha L}$.
The quadrupole matrix elements
are denoted by $\langle \hat Q\rangle^{({\rm rot})}_{\alpha L}$
for the rotor limit  and by $\langle \hat Q\rangle^{({\rm su3})}_{\alpha L}$
when computed precisely.  }\vspace{0.2cm}
$ \begin{array}{|c|c|c|r|r|c|c|} \hline
(\lambda\;\mu)& \alpha &L& 
\langle \hat Q\rangle^{({\rm rot.})}_{\alpha L}\;
& 
\langle \hat Q\rangle^{({\rm su3})}_{\alpha L}\;
& K^{({\rm rot})}_{\alpha L} &  \langle K\rangle_{\alpha L}\;
\\   \hline
(70\, 6) & 1  & 0 & 0   \;\;\;\;\;\;  & 0 \;\;\;\;\;\;  &  0    &  0.000\\
             & 2  & 2 & 178.089     &     178.089 &  2    & 2.000\\
             & 3 &  4 & 318.937     &     318.937 & 4     &  4.000\\
             & 4 &  6 & 425.927     &     425.927 & 6     &  6.000\\
             & 1 &  30 &-582.098 & -582.193  &0        & 0.000\\
             & 2 &  30 &-574.587 &-574.609  & 2        & 2.000\\
             & 3 &  30 &-552.055 &-552.050  & 4        & 4.000\\
             & 4 &  30 &-514.500 &-514.388  & 6        & 6.000\\
  \hline
(70\, 7) & 1 & 1 &     95.304 &     95.304 & 1   &  1.000\\
             & 2 & 3 &   256.174 &  256.174 & 3   &  3.000\\
             & 3 & 5 &   377.874 &  377.874 & 5   &  5.000\\
             & 4 & 7 &   475.213 &  475.213 & 7   &  7.000\\
             & 1 & 31 & -546.077 &  -545.784 & 1   &  1.000\\
             & 2 & 31 & -579.311 &  -545.784 & 3   &  3.000\\
             & 3 & 31 & -550.495 &  -550.462 & 5   &  5.000\\
             & 4 & 31 & -507.272 &  -507.108 & 7   &  7.000\\
\hline
(10\, 4)  & 1 & 2 &  -32.271 &  -32.281  & 0  & 0.000\\
             & 2 & 2 &    32.271 &   32.281 &  2  & 2.000\\
             & 3 & 4 &    57.794 &   57.811 &  4  & 4.000\\
             & 1 & 10 & -62.077   &  - 63.297  & 0   &  0.003\\
             & 2 & 10 & -55.305   &   -54.896  & 2   &  2.001\\
             & 3 & 10 & -34.989    &  -34.177  &  4   &  3.996\\
             & 1 & 12 & -67.663     &  -39.409 &  -   &  1.012\\
             & 2 & 12 & -62.458    &  -44.447  & -    & 3.872\\
             & 1 & 14 & -72.830    &  -35.837  & -    & 2.010\\
\hline   
\end{array}$%
\end{table} 
The remarkable result is how extraordinarily close they are to the corresponding integer rotor-model values, both for even and odd values of $\mu$, with the exception of the few values of the angular momentum close to those at which the finite-dimensional SU(3) band terminates.

\end{appendix}


\end{document}